\documentclass[final,3p]{elsarticle}

\usepackage[T1]{fontenc}
\usepackage[british]{babel}
\usepackage[utf8]{inputenc}
\usepackage{caption}
\usepackage{subcaption}
\usepackage{wrapfig}
\usepackage{makeidx}
\usepackage{multicol}
\usepackage{footmisc}
\usepackage{amsmath}
\usepackage{amsthm}
\usepackage{amssymb}
\usepackage{amsbsy}
\usepackage{tikz}
\usepackage{color}
\usepackage{url}
\usepackage[firstpage]{draftwatermark}
\usepackage{graphicx}
\usepackage{cleveref}
\usepackage{todonotes}
\usepackage{booktabs}   
\usepackage{pgfplots}
\usepackage{centernot}
\usepackage{xspace}
\usepackage{nomencl} 
\usepackage{stmaryrd} 
\usepackage{xfrac} 
\usepackage{units} 
\usepackage{multirow} 
\usepackage{epstopdf}
\allowdisplaybreaks

\crefname{chapter}{Chapter}{Chapters}
\crefname{section}{Section}{Sections}
\crefname{appendix}{Appendix}{Appendices}
\crefname{subsection}{Section}{Sections}
\crefname{subsubsection}{Section}{Sections}
\crefname{equation}{Equation}{Equations}
\crefname{figure}{Figure}{Figures}
\crefname{table}{Table}{Tables}
\crefname{subfigure}{Figure}{Figures}
\crefname{listing}{Listing}{Listings}

\graphicspath{{figures/}}

\usepackage{tgcursor}  
\usepackage[activate={true,nocompatibility},final,kerning=true,tracking=true,spacing=true]{microtype}
\microtypecontext{spacing=nonfrench}

\usetikzlibrary{calc} 

\makenomenclature

\usetikzlibrary{plotmarks,shapes,arrows}
\tikzstyle{decision} = [diamond, draw, fill=blue!20,
    text width=4.5em, text badly centered, node distance=3cm, inner sep=0pt]
\tikzstyle{block} = [rectangle, draw, fill=blue!20,
    text width=13em, text centered, rounded corners, minimum height=4em]
\tikzstyle{line} = [draw, -latex']
\tikzstyle{cloud} = [draw, ellipse,fill=red!20, node distance=3cm,
    minimum height=2em]

\let\baraccent=\= 
\renewcommand{\=}[1]{\stackrel{#1}{=}} 

\begin{document}

\begin{frontmatter}

\title{Shape deformation of a vesicle under axisymmetric non-uniform alternating electric field}
\author[ntnu]{Kumari Priti Sinha}
\cortext[cor1]{Corresponding author}
\author[ntnumat]{Rochish M Thaokar\corref{cor1}}
\ead{rochish@che.iitb.ac.in}

\address[ntnu]{Department of Chemical Engineering, Indian Institute of Technology Bombay, Mumbai-400076, India}
\address[ntnumat]{Department of Chemical Engineering, Indian Institute of Technology Bombay, Mumbai-400076, India}

\begin{abstract}

Non-uniform fields are commonly used to study vesicle dielectrophoresis and can be used to hitherto relatively unexplored areas of vesicle deformation and electroporation. A common but perplexing problem in vesicle dynamics is the cross over from the entropic to enthalpic (stretching) tension during vesicle deformation. A lucid demonstration of this concept is provided by the study of vesicle deformation and dielectrophoresis under axisymmetric quadrupole electric field. Small deformation theory incorporating the Maxwell stress approach is used (employing area and volume conservation constraints) to estimate the dielectrophoretic velocity. The entropic and enthalpic tensions are implemented to understand vesicle electrohydrodynamics in low and high tension limits. The shapes obtained using the entropic and the enthalpic approaches, show significant differences. A strong dependence of the final vesicle shapes on the ratio of electrical conductivities of the fluids inside and outside the vesicle as well as on the frequency of the applied quadrupole electric field is observed which could be used to estimate electromechanical properties of the vesicle. Moreover, an excess area dependent transition between the entropic and enthalpic regimes is observed. The Maxwell stress approach, used in this work, indicates that Clausius-Mossotti factor obtained by the dipole moment method together with the drag on a rigid sphere explains vesicle dielectrophoresis. Interestingly, the coupling of hydrodynamic and electric stress, important in drops is absent in vesicle dielectrophoresis to linear order.
\end{abstract}


\end{frontmatter}
\section{Introduction}
 Non-uniform fields are commonly used to study dielectrophoresis (DEP), which is the movement of an uncharged particle under a spatially non-uniform electric field. DEP results from the interaction of an electric field gradient and an induced dipole in the particle. A particle in a non-uniform field is termed to undergo positive DEP if particles migrate towards a region of high electric field while in negative DEP the particles migrate towards region of lower electric field. An understanding of the DEP behavior of bioparticles \cite{guido2009, li2011}  has importance in several biotechnological and biomedical applications \cite{weigl2003,khoshmanesh2011,jubery2014,dey2015}. First stuggested for yeast cells, DEP \cite{pohl1971, crane1971}, has subsequently been widely studied for other bioparticles such as RBCs,\cite{gascoyne1997, leonard2008} bacteria,\cite{nakano2016} DNAs,\cite{chou2002, tuukkanen2005} proteins\cite{nakano2011} etc. The technique has potential applications in cell manipulation, \cite{gagnon2011, jaka2013} separation, \cite{kang2008,meighan2009,nuttawut2011,gagnon2011}  sorting, \cite{stefan1998,brian2005,thomas2007} to study electrorotation, \cite{reichle1999, han2013} electrofusion, \cite{zimmermann1982,cavallaro2012,yang2012} cell-cell interaction \cite{yeli2011} as well as in characterizing their physical properties \cite{patel2008} etc as well as for for diagnosis of cancer \cite{alshareef2013,peter2014,sonnenberg2014}. On the other hand, single cell studies \cite{jang2009, wang2013, huang2014} have important engineering applications in drug delivery, gene introduction, cloning technology etc, apart from the fundamental insights into the mechanism of dielectrophoresis that such studies provide. Excellent controllability, high efficiency and small damage to cells make DEP technique appropriate for contact-free trapping of a cell in a region of low electric field \cite{jang2009}. This has led to research in designing effective non-uniform electric fields by judicious design of electrodes, often in  microfluidic/nanofluidic chips by microfabrication techniques \cite{hadi2010,jaka2013,huang2014} has gained prominence.\\  
  
  Giant Unilamellar Vesicles (Liposomes) (GUVs) have emerged as a very reliable bio-memetic system and has been used to understand the DEP response of cells \cite{stoicheva1994, hadady2015}. Unlike biological cells, there are very few experimental  \cite{stoicheva1994, victoria2009, kodama2013} and theoretical \cite{jones1990} investigations on the DEP of vesicles . Korlach et al., \cite{webb2005} created a 3D electric field cage to study vesicle deformation and electro-rotation by trapping a vesicle using optical tweezers. Studies on the modification of the electrical properties of GUVs to serve them as test particle for DEP study\cite{desai2009}, high-frequency DEP response of vesicles to estimate upper and lower crossover frequency at different interior conductivity and membrane electric properties \cite{hadady2015} and DEP studies on surface-modified liposomes in AC fields\cite{victoria2009}, have also been reported .\\ 
A vesicle under non-uniform, axisymmetric quadrupole electric field, not only exhibits dielectrophoresis, but can also deform. Although several experimental and theoretical papers have demonstrated vesicle deformation under AC\cite{Vla2009, Dimova2010,Antonova10, Peterlin10}, pulsed DC\cite{Vla2014}, DC fields\cite{Mc2013, Vlah2015}, these fields are mostly uniform. A uniform field leads to prolate and oblate spheroidal (dipolar) deformations, and these have been summarized into a phase diagram\cite{RDimova2009, Dimova2010}. Moreover non-uniform fields have also shown promise in more efficient electroporation as compared to uniform fields \cite{issadore2010}.  \\
  
Interestingly, very few such experimental studies on deformation of a vesicle have been conducted \cite{webb2005,issadore2010} in non-uniform fields whereas there is hardly any theoretical study reported. The experiments \cite{webb2005,issadore2010} indicate fascinating shapes (prolate, oblate, pear, diamond, square) due to action of quadrupolar and higher order potentials and associated Maxwell stress. The resulting shape is clearly a balance of electric, hydrodynamic and membrane stress. Amongst the different membrane stresses, there is a good understanding of the bending stress as well as the non-uniform tension that arises on account of local membrane incompressibility. However, to describe the uniform tension, two approaches have been used. The entropic approach, wherein, the tension arises due to the thermal undulations of the excess area. The assumption here is that under an external force, the excess area present in the thermal undulations is reduced, leading to a tension (hereafter called as the entropic tension). The membrane is then assumed to have enough excess area not to cause stretching at a molecular level and was employed  to describe vesicle deformation under a uniform electric field \cite{Vla2009,PS2017}. On the other hand, when a membrane is completely stretched, the uniform tension arises because the excess area of a vesicle can not increase during the shape deformation process (we call this the enthalpic tension). This approach has been used to describe shape deformation for vesicles in shear flow\cite{shear2007},wherein shapes are described by the $2^{nd}$ Legendre mode. On the other hand, the quadrupolar field provides two degrees of freedom for shape deformation, namely the $2^{nd}$ and $4^{th}$ Legendre modes. \\
  
  These concepts form the basis for micropipette experiments which were initially proposed by \cite{evan1990}, wherein it was showed that for tensions lower than $0.5 mN/m$, the aspiration can be considered entropic and the area change is logarithmic in tension. On the other hand for higher values of tension, a membrane stretches proportional to the tension, and inverse to the area incompressibility modulus (typically of the order $100-200 mN/m$). 
   
  Motivated by these issues  we ask the following questions,
  \begin{enumerate}
  \item What is the vesicle deformation in pure quadrupolar field (as well as a mix of uniform and quadrupole fields), and can the prolate, oblate, pear, diamond and square shapes, seen in experiments be explained?
  \item When is the deformation dominated by entropy and enthalpy or how does the deformation differ from the uniform field case?
  \end{enumerate} 
 
\section{Mathematical formulation}

\subsection{Model description}
The system considered consists of a spherical vesicle of radius $R_o$, surrounded by a non-conducting bilayer membrane of thickness $\delta_m$ that has an electrical conductivity ($\sigma_m$) and a finite permittivity ($\epsilon_m$). This bilayer membrane which separates the inner fluid from the suspending medium is characterized by an interfacial tension $\gamma_{init,0}$, bending rigidity $\kappa_b$ and the dilatational viscosity of the membrane, $\mu_m$. The Newtonian fluid enclosed within a vesicle has permittivity $\epsilon_{in}$, conductivity $\sigma_{in}$, and viscosity $\mu_{in}$, the suspending Newtonian medium has permittivity $\epsilon_{ex}$, conductivity $\sigma_{ex}$, and viscosity $\mu_{ex}$. Gravity effects are neglected on account of their small size ($R_o=5-10 \mu m$). We define the ratios of fluid physical properties as
  $\sigma_r=\sigma_{in}/\sigma_{ex}, \epsilon_r=\epsilon_{in}/\epsilon_{ex}, \mu_r=\mu_{in}/\mu_{ex}$.  Note that subscript 'in' and 'ex' represent quantities associated with the inner and the outer fluid, respectively.
  
\begin{figure}[tbp]
\centering
\includegraphics[width=.7\textwidth]{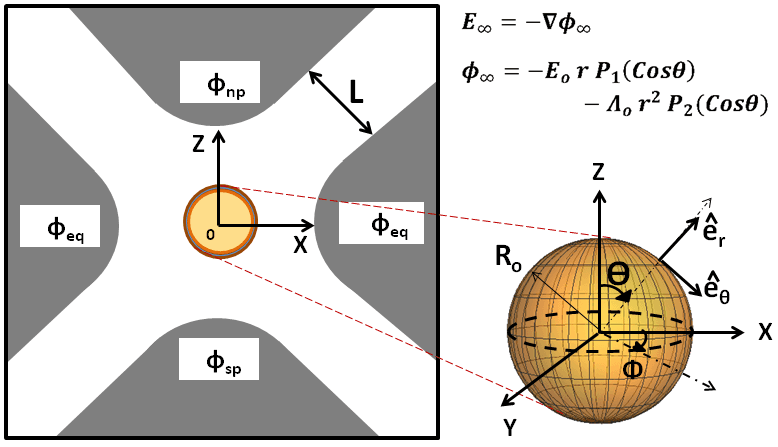}
\caption{Schematic presentation of a vesicle subjected to an axisymmetric, non-uniform AC electric field. $\boldsymbol{\hat{e}_r}$ and $\boldsymbol{\hat{e}_{\theta}}$ are unit normal and tangent vectors, respectively.}
\label{SCHEMATIC}
\end{figure}

  To generate a non-uniform electric field, axisymmetric quadrupole electrodes (symmetric about the z axis) are used in this work (Figure\ref{SCHEMATIC}). The geometric center of the axisymmetric electrode setup is the region of minimum electric field, whereas the electric field is maximum at the electrode edges. A spherical coordinate system $(r, \theta, \Phi)$ is assumed such that the origin of the coordinate system is at the geometric center of the electrode system. A time-periodic, non-uniform, axisymmetric, AC electric field is externally applied to a vesicle placed at the center of the electrode system. The applied electric potential is expressed as a sum of uniform and quadrupole electric potentials as $\phi_{\infty}=(-E_o r P_1 (\cos\theta)-\Lambda_o r^2 P_2 (\cos\theta))\cos\omega t$, where $E_o$ and $\Lambda_o$ are the intensities of the uniform and the quadrupole electric fields, respectively.  Here $P_1 (\cos\theta)$ and $P_2 (\cos\theta)$ denote Legendre Polynomials of first and second degree, respectively. The electric field generated due to this applied electric potential can be expressed as $E_{\infty}=- \nabla  \phi_{\infty}$. In this work, most of the equations are expressed in their dimensional form (with no over-bar) but the results are primarily presented in non-dimensional form (represented with an over-bar) using appropriate dimensionless parameters.
\subsection{Governing equations and boundary conditions} 
\subsubsection{Electrodynamics}
The inner and outer fluids are assumed to be leaky dielectrics. The solution of Laplace equation $\nabla^2 \phi_j=0$ (where $j=ex,in$) in spherical coordinate system results in the electric potential outside ($\phi_{ex}$) and inside ($\phi_{in}$) the vesicle to be of the form,
\begin{align}
&\ \phi_{ex}=\phi_{\infty}+\frac{A_1}{r^2} P_1 (\cos\theta)+\frac{A_2}{r^3} P_2 (\cos\theta) &\\
&\ \phi_{in}=B_1 r P_1 (\cos\theta)+B_2 r^2 P_2 (\cos\theta)&
\end{align}
where coefficients $A_1, A_2, B_1, B_2$ (provided in Appendix-A), are obtained by solving the following electrostatic boundary conditions at the membrane interface ($r=R_o$) and using orthogonality of Legendre polynomials,
\begin{align}
&\ \phi_{in}-\phi_{ex}-V_{m1} P_1 (\cos\theta)-V_{m2} P_2 (\cos\theta)=0 &\\
&\ (\sigma_{in}+i \omega \epsilon_{in}) \frac{d\phi_{in}}{dr}-(\sigma_{ex}+i \omega \epsilon_{ex}) \frac{d\phi_{ex}}{dr}=0 &\\
&\ (\sigma_{ex}+i \omega \epsilon_{ex}) \frac{d\phi_{ex}}{dr}+C_m i \omega (V_{m1} P_1 (\cos\theta)+V_{m2} P_2 (\cos\theta))+G_m (V_{m1} P_1 (\cos\theta)+V_{m2} P_2 (\cos\theta))=0&
\end{align} 
Here $V_{m1}$ and $V_{m2}$ are the transmembrane potentials (Appendix-B) across the membrane associated with the $P_1$ and $P_2$ Legendre modes, respectively. $C_m$ and $G_m$, which can be modelled as  $C_m=\epsilon_m/\delta_m$ and $G_m=\sigma_m/\delta_m$, are the membrane capacitance and conductance respectively. \\
The normal and tangential electric fields are obtained from the electric potentials using the definitions
$E_{r \ ex,in}=-\frac{\partial \phi_{ex,in}}{\partial r}$, 
$E_{\theta \ ex,in}=-\frac{1}{r}\frac{\partial \phi_{ex,in}}{\partial \theta}$.

Using the Maxwell's stress tensor $\textbf{T}=\epsilon \epsilon_0 (\textbf{E} \textbf{E}-\frac{1}{2} \textbf{I} E^2)$ where $E^2=\textbf{E}.\textbf{E}$ and $\textbf{I}$ is the identity tensor, the normal (${\hat{\textbf{e}}_r.\textbf{T}.\hat{\textbf{e}}_r}$) and tangential (${\hat{\textbf{e}}_r.\textbf{T}.\hat{\textbf{e}}_\theta}$) electric stresses acting at the vesicle surface can be estimated. Here $\hat{\textbf{e}}_r$ and $\hat{\textbf{e}}_\theta$ represents the normal and tangent unit vectors to an undeformed sphere. The non-oscillatory part of the Maxwell stresses can be further expressed as\cite{PS2017, RT2016PRE}.
\begin{align}
\tau_{r \ ex,in}^E=\frac{\epsilon \epsilon_0}{4}(E_{r \ ex,in} E_{r \ ex,in}^*-E_{\theta \ ex,in} E_{\theta \ ex,in}^*) &\\
\tau_{\theta \ ex,in}^E=\frac{\epsilon \epsilon_0}{4} (E_{\theta \ ex,in} E_{r \ ex,in}^*+E_{r \ ex,in} E_{\theta \ ex,in}^*)
\end{align} 
where * represents complex conjugate of the respective physical quantity and $E_r$ and $E_{\theta}$ are the radial and tangential electric fields respectively. 
The net normal and tangential electric stresses on the vesicle are  
$\tau_r^E= \tau_{r \ ex}^E-\tau_{r \ in}^E, \ \tau_\theta^E=\tau_{\theta \ ex}^E-\tau_{\theta \ in}^E$.
Full expressions for these quantities in a simplified form are provided in the Appendix-C.

\subsubsection{Hydrodynamics}
The velocity and pressure fields corresponding to each fluid region (inner or outer) are given by the Stokes equation and the continuity equation ($\nabla p_{in,ex}=\mu_{in,ex} \nabla^2 \boldsymbol{u_{in,ex}}, \ \
\nabla.\boldsymbol{u_{in,ex}}=0$). Here, the inertial effects are ignored, thereby addressing small Reynolds number conditions. Assuming axisymmetry and adopting a stream function approach, the stream functions for the outer and inner regions are of the generalized form
\begin{align}
&\ \psi_{ex}=\left(\frac{C_{1e}}{r}+C_{2e}r\right)G_2+\left(\frac{C_{3e}}{r^2}+C_{4e}\right)G_3+\left(\frac{C_{5e}}{r^3}+\frac{C_{6e}}{r}\right)G_4+\left(\frac{C_{7e}}{r^4}+\frac{C_{8e}}{r^2}\right)G_5 &\\
&\ \psi_{in}=\left(C_{1i}r^4+C_{2i}r^2\right)G_2+\left(C_{3i}r^5+C_{4i}r^3\right)G_3+\left(C_{5i}r^6+C_{6i}r^4\right)G_4+\left(C_{7i}r^7+C_{8i}r^5\right)G_5&
\end{align}
where $G_1-G_5$ are the Gegenbauer's function of first kind (Appendix-D). The velocity fields can be expressed in terms of stream functions ($\psi$) as
$\ v_{r \ ex,in}=\frac{1}{r^2 \sin\theta}\frac{\partial \psi_{ex,in}}{\partial\theta}, \  v_{\theta \ ex,in}=-\frac{1}{r \sin\theta}\frac{\partial\psi_{ex,in}}{\partial r}$,
where $v_r$ and $v_\theta$ are the normal and tangential velocity components, respectively. 

The pressure is governed by the solution of the Laplace equation, $\nabla^2 P=0$. For the outer and inner regions it is considered to be of the form 
\begin{align}
&\ p_{ex}=\frac{C_{9e}}{r^2} P_1 (\cos\theta)+\frac{C_{10e}}{r^3} P_2 (\cos\theta)+\frac{C_{11e}}{r^4} P_3 (\cos\theta)+\frac{C_{12e}}{r^5} P_4 (\cos\theta) &\\
&\ p_{in}=p_0+C_{9i} r P_1 (\cos\theta)+C_{10i} r^2 P_2 (\cos\theta)+C_{11i} r^3 P_3 (\cos\theta)+C_{12i} r^4 P_4 (\cos\theta)&
\end{align}
here $C_{1i}-C_{12i}$ and $C_{1e}-C_{12e}$ are unknown coefficients to be determined (Appendix-E).\\
The normal and tangential hydrodynamic stresses at the outer and inner surfaces of the membrane are given by
\begin{align}
&\tau_{r \ ex,in}^H=-p_{ex,in}+2 \mu_{ex,in} \frac{\partial v_{r \ ex,in}}{\partial r} &\\
&\tau_{\theta \ ex,in}^H=\mu_{ex,in} \left(\frac{1}{r}\frac{\partial v_{r \ ex,in}}{\partial \theta}-\frac{v_{\theta \ ex,in}}{r}+\frac{\partial v_{\theta  ex,in}}{\partial r} \right)&
\end{align}
\subsubsection{Membrane mechanics}
The surface of a deformed vesicle is described by 
\begin{align}
\ r_s=\alpha+s_1 P_1  (\cos\theta)+s_2 P_2  (\cos\theta)+s_3 P_3  (\cos\theta)+s_4 P_4  (\cos\theta)
\end{align}
where $r_s$ is the radial position of a slightly deformed vesicle surface from it's center, $s_1, s_2, s_3, s_4$ are deformation amplitudes associated with respective Legendre modes. $\alpha$ is obtained by the constraint of volume conservation $\int_{\phi=0}^{2\pi} d\phi \int_{\theta=0}^{ \pi}d\theta (1/3) r_s^3 \sin\theta=4\pi R_o^3/3$, to yield
\begin{align}
\alpha=R_o \left(1-\left(\frac{s_1^2}{3}+\frac{s_2^2}{5}+\frac{s_3^2}{7}+\frac{s_4^2}{9}\right)\frac{1}{R_o^2}\right)
\end{align}
The deviation of the vesicle shape from a sphere (of fixed volume $4\pi R_o^3/3$) is defined by a shape function $F=r-r_s(\theta)$, therefore the unit normal at $r=r_s$ is $\textbf{n}=\nabla F/|\nabla F|$. The constraint of area conservation of the bilayer membrane results in a relation between deformation amplitudes and excess area of the form $\int_{\phi=0}^{2\pi} \int_{\theta=0}^{ \pi} r_s^2/(\hat{\textbf{e}}_r.\textbf{n}) \sin\theta d\theta d\phi=4\pi R_o^2+\Delta$. This gives the non-dimensional excess area as
\begin{align}
\bar{\Delta}=\left(\frac{2 s_2^2}{5}+\frac{5 s_3^2}{7}+s_4 ^2 \right)\frac{1}{R_o^2} \label{ExcessA}
\end{align}
 
\textbf{Membrane stress}: 
The stress due membrane bending ($\tau^B$), due to both uniform ($\tau_u^T$) and nonuniform tensions ($\tau_{r,nu}^T$, $\tau_{\theta,nu}^T$) as well as due to normal and tangential interfacial membrane stresses ($\tau_r^\nu$, $\tau_\theta^\nu$) are given by
\begin{align}
&\tau^B=\frac{6\kappa_b}{R_o^4} \left(\sum_{l=2}^{4} (l(l+1)-2) s_l P_l (\cos\theta)\right)&\\
&\tau_u^T= \gamma_u \left(\frac{2}{r}+\frac{1}{r^2}\sum_{l=2}^{4} (l(l+1)-2) s_l P_l (\cos\theta)\right) &\\
&\tau_{r,nu}^T=\frac{2}{r}\left(\sum_{l=1}^{4}\gamma_{nul} P_l (\cos\theta) \right) &\\
&\tau_{\theta,nu}^T=\frac{1}{R_o} \left(\sum_{l=1}^{4}\frac{\partial (\gamma_{nul} P_l (\cos\theta))}{\partial\theta} \right) &\\
&\tau_r^\nu=\frac{Bq \mu_{ex}}{R_o}\frac{\partial}{\partial \theta}\left(\frac{1}{\sin\theta}\frac{\partial (v_\theta \sin\theta)}{d\theta}\right)&\\
&\tau_\theta^\nu=\frac{2 Bq \mu_{ex}}{R_o \sin\theta}\frac{\partial (v_\theta \sin\theta)}{d\theta}&
\end{align}
here $Bq=\mu_m/(\mu_{ex}R_o)$ is the Boussinesq number and $\mu_m$ is the membrane viscosity associated with dilatational deformation of the membrane. The normal membrane stress component is  $\tau_r^M=(\tau^B+\tau_{u}^T+\tau_{r, nu}^T+\tau_r^\nu)$, 
the tangential membrane stress component is $\tau_\theta^M=(\tau_{\theta, nu}^T+\tau_{\theta}^\nu)$.  Note that the stress associated with the membrane tension has contributions from both uniform tension, which could be entropic or enthalpic and discussed later (equation \ref{UnifEntTens} and Appendix-E)  as well as the non-uniform tension which varies with position $\theta$ along the surface and is obtained  by applying membrane incompressibility condition ($\nabla_s.v_j=0$), where $\nabla_s$ is surface gradient operator. This conforms local area conservation and yields the non-uniform tension associated with each mode namely $\gamma_{nu1}, \gamma_{nu2}, \gamma_{nu3}, \gamma_{nu4}$  (Appendix-F).

\subsubsection{Boundary conditions}
The balance of membrane and fluid stresses and continuity of their velocity fields across the membrane interface are given by Taylor expanding the following boundary conditions at the undeformed membrane surface $r=R_o$\\ 
 (b1) Normal stress balance: $[[\tau_r^E+\tau_r^H]]+\tau_r^M=0$, 
 where [[.]] represents the difference in properties of outer and inner fluid across the interface\\
 (b2) Tangential velocity continuity: $v_{\theta \ in}=v_{\theta \ ex}$\\
 (b3) Membrane incompressibility condition:
 \begin{align}
 \frac{1}{R_o}\frac{\partial (v_{\theta \ ex, in}\sin\theta) }{\partial\theta}+\frac{2}{R_o}v_{r \ ex, in} \sin\theta=0 
  \end{align}
 (b4) Tangential stress balance:$[[\tau_{\theta}^E+\tau_{\theta}^H]]+\tau_\theta^M=0$\\ 
 (b5) Kinematic condition:
 \begin{align}
\ v_{r \ in}=v_{r \ ex}=\sum_{l=1}^{4}\frac{ds_l}{dt} P_l (\cos\theta) \label{b5}
\end{align}
The above boundary conditions (b1-b5) are solved using the orthogonality condition for Legendre polynomials. The boundary conditions are integrated at each order of the Legendre polynomials in order to get the unknown constants associated with the velocity and pressure fields for both the inner and the outer fluids (Appendix-G). 

\section{Entropic and enthalpic tension approach for uniform tension}
In the present work, the electric field induced shape deformations are estimated by two different ways of describing the uniform tension in the membrane. These are based on the two regimes for tension as given by \cite{evan1990}. In the first case, when the induced tension is low, the tension is estimated by using the entropic theory. At low tension a membrane is in a highly fluctuating state where in the excess area of a vesicle is present in various deformation modes. Therefore an applied stress leads to a change in area that is described by the amplitudes $s_2, s_3, s_4$ of modes $P_2, P_3, P_4$ (equation\ref{ExcessA}) due to straightening of the fluctuations (wiggles), resulting in membrane tension that is given by
 \begin{align}
 \gamma_u^{ent} = \gamma_{init,0} \  e^{\left(
 \frac{8 \pi \kappa_b \left(2 s_2^2/5 + 5 s_3^2/7 + s_4^2\right)}{R_o^2 K_B T}\right)}
 \end{align}
Therefore, the uniform tension in the membrane because of excess area is
\begin{align}
\gamma_u^{ent}=\gamma_{init, 0} \ e^{\left(\frac{8\pi  \kappa_b \Delta}{K_B T}\right)} \label{UnifEntTens}
\end{align}
where $\gamma_{init, 0}$ is the initial tension in the membrane. Here, $\Delta$ contains contribution from all the three modes ($P_2, P_3$ and $P_4$).\\ 

In the second case, when the induced tension is high, a vesicle is said to be in the enthalpic regime. In this state, the deformation modes are related by the constraint of the vesicle area remaining constant during deformation (equation\ref{ExcessA}), $\dot{\Delta}=0$, leading to  
 \begin{align}
 \frac{4}{5} s_2\frac{ds_2}{dt}+\frac{10}{7} s_3 \frac{ds_3}{dt}+\frac{28}{11} s_4 \frac{ds_4}{dt}=0
 \end{align}
 By substituting the solutions from kinematic conditions (equation \ref{b5}) for $\frac{ds_2}{dt}, \frac{ds_3}{dt}, \frac{ds_4}{dt} $ the enthalpic tension, $\gamma_u^{enth}$, can be determined and is provided in Appendix-E.

The overall shapes of a vesicle obtained in the entropic and the enthalpic cases are described by the positions of their interface, that is $r_s=\alpha+s_2 P_2(\cos\theta)+s_3 P_3(\cos\theta)+s_4 P_4(\cos\theta)$. Here $s_2$ deformation mode deforms a vesicle into a prolate/oblate ellipsoids, the $s_3$ mode makes the shapes non-axisymmetric about the z-axis, while the $s_4$ mode imparts higher order shapes (square, diamond, pear etc).

 \section{Dimensionless parameters}
 All length scales are non dimensionalized by the radius of vesicle $R_o$, the potential by either $E_o R_o$ or $\Lambda_o R_o^2$, the tension by $\kappa_b/R_o^2$, and the velocity and the stress by $\epsilon_{ex} R_o^3 \Lambda_o^2/\mu_{ex}$ and $\epsilon_{ex} R_o^2 \Lambda_o^2$, respectively. A dimensionless factor $\bar{f}=E_o/(\Lambda_o R_o)$ is introduced, which compares the relative strength of the applied uniform and quadrupole electric field, such that $\bar{f}=0$ represents a pure quadrupole field. Another dimensionless quantity which compares the relative strength of the shape deforming electric stress and shape resisting bending stress is the capillary number  $Ca=\epsilon_{ex} R_o^5 \Lambda_o^2/\kappa_b$. There are several time scales in the problem, the hydrodynamic time scale $t_H=\mu_{ex}/(\epsilon_{ex} R_o^2 \Lambda_o^2)$, the charge relaxation time scale of outer fluid $t_{ex}=\epsilon_{ex}/\sigma_{ex}$, the Maxwell Wagner relaxation time $t_{MW}=(2\epsilon_{ex}+\epsilon_{in})/(2\sigma_{ex}+\sigma_{in})$ and the membrane charging time $t_{cap}=\bar{C}_m (\frac{1}{2}+\frac{1}{\sigma_r})$, where $\bar{C}_m= C_m R_o/\epsilon_{ex}$. Here we use $t_{ex}$ for non-dimensionalizing the time and $\bar{\omega}=\omega \epsilon_{ex}/\sigma_{ex}$ represents the non-dimensional frequency. For simplicity we present results for the case $t_H=t_{ex}$, which is valid for low conductivity fluids. \\
 
 The non-dimensional equations for the time evolution of the various deformation modes are given by,\\

 \textbf{Rate of shape deformation}
 \begin{align}
 &\ \frac{d\bar{s}_1}{d\bar{t}}=\frac{2 \bar{f} (-30 \bar{Y}_0 + 18 \bar{Y}_2 + \bar{X}_1 \bar{Y})}{9 \bar{Y}} &\\ 
 &\  \frac{d\bar{s}_2}{d\bar{t}}= \frac{6 (-28 \bar{s}_2 \bar{Y} (6 + \bar{\gamma}_u)/\bar{Ca} + (14 \bar{Y}_{1a} + 90 \bar{Y}_3 + 7 \bar{X}_{2a} \bar{Y} + 7 \bar{f}^2 (2 \bar{Y}_{1b} + \bar{X}_{2b} \bar{Y})))}{7 \bar{Y} (32 + 23 \mu_r)}&\\ 
 &\ \frac{d\bar{s}_3}{d\bar{t}}=\frac{12 (-10 \bar{s}_3 \bar{Y} (6 + \bar{\gamma}_u)/\bar{Ca} + \bar{f} (-8 \bar{Y}_2 + \bar{X}_3 \bar{Y}))}{\bar{Y} (85 + 76 \mu_r)} &\\
 &\ \frac{d\bar{s}_4}{d\bar{t}}=\frac{20 (-126 \bar{s}_4 \bar{Y} (6 + \bar{\gamma}_u)/\bar{Ca} + (-48 \bar{Y}_3 + 7 \bar{X}_4 \bar{Y}))}{63 \bar{Y} (20 + 19 \mu_r)}&
 \label{evoleqn}
 \end{align}
 where constants terms ($\bar{X}'s$ and $\bar{Y}'s$) are parts of the normal and tangential electric stresses, respectively. These are lengthy expressions and therefore not provided in the manuscript.

 \section{Results and discussion}
  Although the results are presented in non-dimensional parameters (represented by an over-bar) it is important to mention the typical experimental parameters of relevance\cite{victoria2009, webb2005}. These are mostly borrowed from the values reported in \cite{victoria2009, webb2005} and \cite{issadore2010}, and have been used to determine the range of  non-dimensional parameters used in this work. An isolated spherical vesicle of $R_o=5\mu m$, motivated by typical size of a biological cell, membrane conductivity $\sigma_m=4\times10^{-9} S/m$, membrane permittivity $\epsilon_m=10 \epsilon_0$, initial membrane tension $\gamma_{init,0}=2.75\times10^{-6}N/m$, and membrane thickness $\delta_m=5nm$, containing a fluid of $\sigma_{in}=1.4\times10^{-2} S/m, \epsilon_{in}=80 \epsilon_0$ is assumed to be suspended in a medium of $\sigma_{ex}=5\times10^{-5} S/m$, $\epsilon_{ex}=80 \epsilon_0$. Here $\epsilon_0$ is the permittivity of free space.  
   
   The axisymmetric quadrupole electrodes in experiments typically consist of two end cap electrodes maintained at a certain voltage and a ring electrode which is often ground.  An AC electric field can be generated by a  peak-to-peak voltage of $\phi_{\infty,np,sp}$ = 40 $V_{pp}$ applied to the end-cap (live) and the ring electrode ($\phi_{\infty,eq}$ = 0 $V_{pp}$) is grounded. An additional 10 $V_{pp}$ potential can be superimposed "between" two end cap electrodes to produce a simultaneous uniform and quadrupole field between the electrodes. The frequency can be varied from around 500 Hz to more than 10 $MHz$. A separation of $L=20\mu m$ and $r_{o}=\sqrt{2} z_o$, can be maintained between the electrodes (see figure\ref{SCHEMATIC}), where $z_{o}, r_{o}$  are the distances of the end caps and the ring respectively, from the center of quadrupole system. These trap parameters are similar to the experimental work of \cite{victoria2009} except while their electrode arrangement is 2D planar quadrupole, the present work deals with axisymmetric quadrupolar system. Similar to \cite{victoria2009} the  conductivity  ratio $\sigma_r=280$ is considered, and  additionally,  $\sigma_r=0.003$ is also used to investigate the $\sigma_r<1$ case.
   
   A vesicle, initially positioned at the center of electrode system, deforms under applied electric field. Additionally, it undergoes dielectrophoresis for a non-zero value of the parameter $\bar{f}$.  Therefore, the study is divided into three parts.  The first and second parts address the problem of  vesicle deformation under  pure quadrupole field ($\bar{f}=0$) as well as  under simultaneous uniform and quadrupole fields (non-zero $\bar{f}$), respectively. Both entropic and enthalpic tension approaches are used in the deformation studies and a variety of shapes are reported. A qualitative comparison with  the   vesicle shapes reported in recent experimental papers \cite{webb2005, issadore2010} is also made. Lastly,  the dielectrophoretic motion of a vesicle under non-uniform electric field is presented when a combination of uniform and quadrupole electric field is applied (non-zero $\bar{f}$) and the details are provided in the supplementary material.
   
   \subsection{Transmembrane potential}
\begin{figure}[tp]
        \centering
       \includegraphics[width=0.6\linewidth]{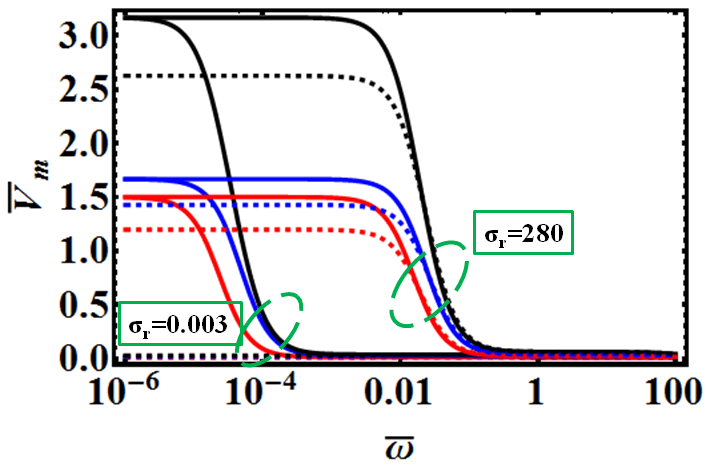}
       \caption{Role membrane conductance on transmembrane potential when $\bar{G}_m=0$(solid) and $\bar{G}_m=0.5$(dashed) when  $\sigma_r$=280 and $\sigma_r$=0.003 for $\bar{f}=1$ ($\bar{C}_m=125, \epsilon_r=1$)} 
        \label{TP}
     \end{figure}

   The transmembrane potential (TMP) generated by a  uniform electric field is equal and opposite at the north and south poles of the vesicle. On the contrary the transmembrane potential due to the quadrupole field is symmetric about the equator and therefore identical at the north and south poles. Therefore for membranes with a finite resting potential (in the absence of field), while uniform field would lead to a difference in poration tendency at the north or south pole, a quadrupole field will cause symmetric poration at the two poles. The variation of the amplitude of the transmembrane potential  (Appendix B: equation \ref{Vm1} and \ref{Vm2}), plotted in figure \ref{TP} for $\bar{f}=1$, shows that the TMP falls over a frequency equivalent to  the reciprocal of the capacitor charging time $(t_{cap}^{-1}=0.015$ for $\sigma_r=280$ and $t_{cap}^{-1}=2.4\times10^{-5}$ for $\sigma_r=0.003$) for  both uniform and quadrupolar parts of the applied fields. 
     The transmembrane potential due to quadrupolar part is greater than that due to the uniform part. 
     Increasing the membrane conductance from 0 to 0.5 reduces the transmembrane potential slightly for $\sigma_r\gg1$. However, the reduction is precipitous for  $\sigma_r<1$. The transmembrane potential in the low frequency limit in a non-conducting membrane is a result of the charges built on the membrane to reduce the normal electric field in the outer region to zero. This potential is independent of the conductivity ratio of the two fluids for a non-conducting membrane. In a conducting membrane, the outer electric field need not be zero, since the membrane can allow current to flow through it. This leads to a drop in the transmembrane potential. A reduction in $\sigma_r$ means a lower electrical conductivity of the inner fluid for the same conductivity of the outer fluid (used in non-dimensionalization). This leads to lower transmembrane current (ohmic) which results in a very small transmembrane potential being able to drive ohmic current through the conducting membrane when $\sigma_r<1$.\\
     
\subsection{Deformation in pure quadrupole electric field}
  A vesicle deforms in an applied electric field on account of the Maxwell stress, proportional to the square of electric field, acting on it. Thus while a uniform electric field ($P_1$) produces shapes described by the $P_2$ mode, when a pure quadrupolar ($P_2$) field is applied, the Maxwell stresses and thereby the resulting shapes are expected to have the $P_2,P_4$ modes. Atleast two experimental results are reported in the literature on vesicle deformation in multipolar fields \cite{webb2005,issadore2010}. Unlike uniform electric fields which typically result in prolate or oblate spheroids, higher order multipolar fields (e.g. quadrupole and octupole) lead to interesting shapes such as square, diamond and even hexagonal. We therefore understand the deformation of a vesicle in quadrupole potentials using the Maxwell stress approach.
  
  To determine realistic parameters for quadrupolar fields that can cause deformation, one can consider around $16 V_{pp}$ potential applied between the end cap electrodes and the ring electrodes, with $r_o=\sqrt{2} z_o$  and $z_o$=10 $\mu m$. The applied potential generates an axisymmetric quadrupole electric field of strength $\Lambda_o=-2.12\times10^{10} V/m^2$. A vesicle of size $R_o=5\mu m$ and excess area $\bar{\Delta}=0.2$ can be considered to be at the center where the DEP force on the vesicle is zero on account of zero electric field, thereby preventing its  translation. This allows a systematic analysis of shape deformations due to non-zero electric stress at the vesicle surface. In calculations presented $\kappa_b=25 K_B T$, that gives a capillary number of $Ca=\epsilon_{ex} R_o^5 \Lambda_o^2/\kappa_b=9684$. These numbers serve as a reference for the choice of non-dimensional parameters used in the present analysis. \\
  
  \begin{figure}[tbp]
               \centering
  \includegraphics[width=0.7\linewidth]{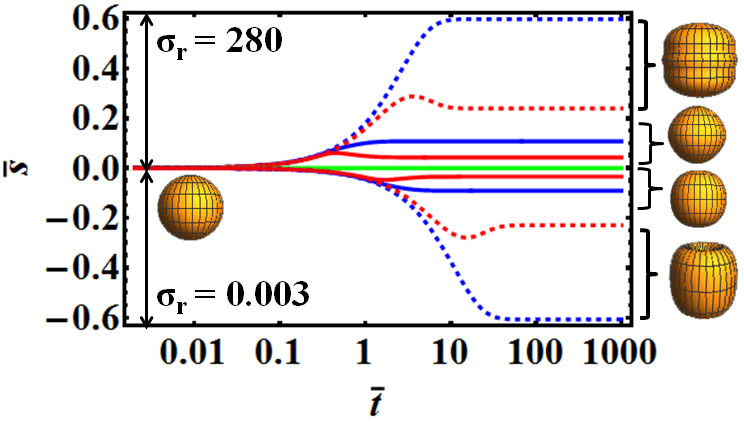}        
  \caption{Variation of the amplitudes of shape deformation modes with time in pure quardratic AC electric field, entropic (solid) and enthalpic approach (dashed), $\sigma_r=0.003, 280$ ($\bar{\gamma}_{init, 0}=668$, Ca=9864, $\bar{\Delta}$=0.2, $\bar{C}_m$=125, $\bar{G}_m$=0, $\mu_r=\epsilon_r$=1, $\bar{\omega}=1$). \textcolor{blue}{\textemdash} ($\bar{s}_2$), \textcolor{green}{\textemdash} ($\bar{s}_3$), \textcolor{red}{\textemdash} ($\bar{s}_4$)}  
                \label{DefVsTimePureQ}
  \end{figure}
  Figure \ref{DefVsTimePureQ} shows the evolution of amplitude of deformation modes, $\bar{s}_2$ and $\bar{s}_4$, with time for very high and low $\sigma_r$ at an intermediate frequency ($\bar{\omega}=1$) in both entropic and enthalpic regimes. For $\sigma_r>1$ both $\bar{s}_2$ and $\bar{s}_4$ are positive, while for $\sigma_r<1$, $\bar{s}_2$ and $\bar{s}_4$ are negative. The $P_3$ deformation mode is not admitted  in the pure quadrupole case ($\bar{s}_3=0$). Thus starting with an initial spherical shape, both the enthalpic and entropic regimes show that a final, non-spherical, equilibrium shape is reached. The deformations in the enthalpic regime are higher than those in the entropic regime.\\
  
  \begin{figure}[tbp]
            \centering
        \begin{subfigure}[b]{0.48\linewidth}
    \includegraphics[width=\linewidth]{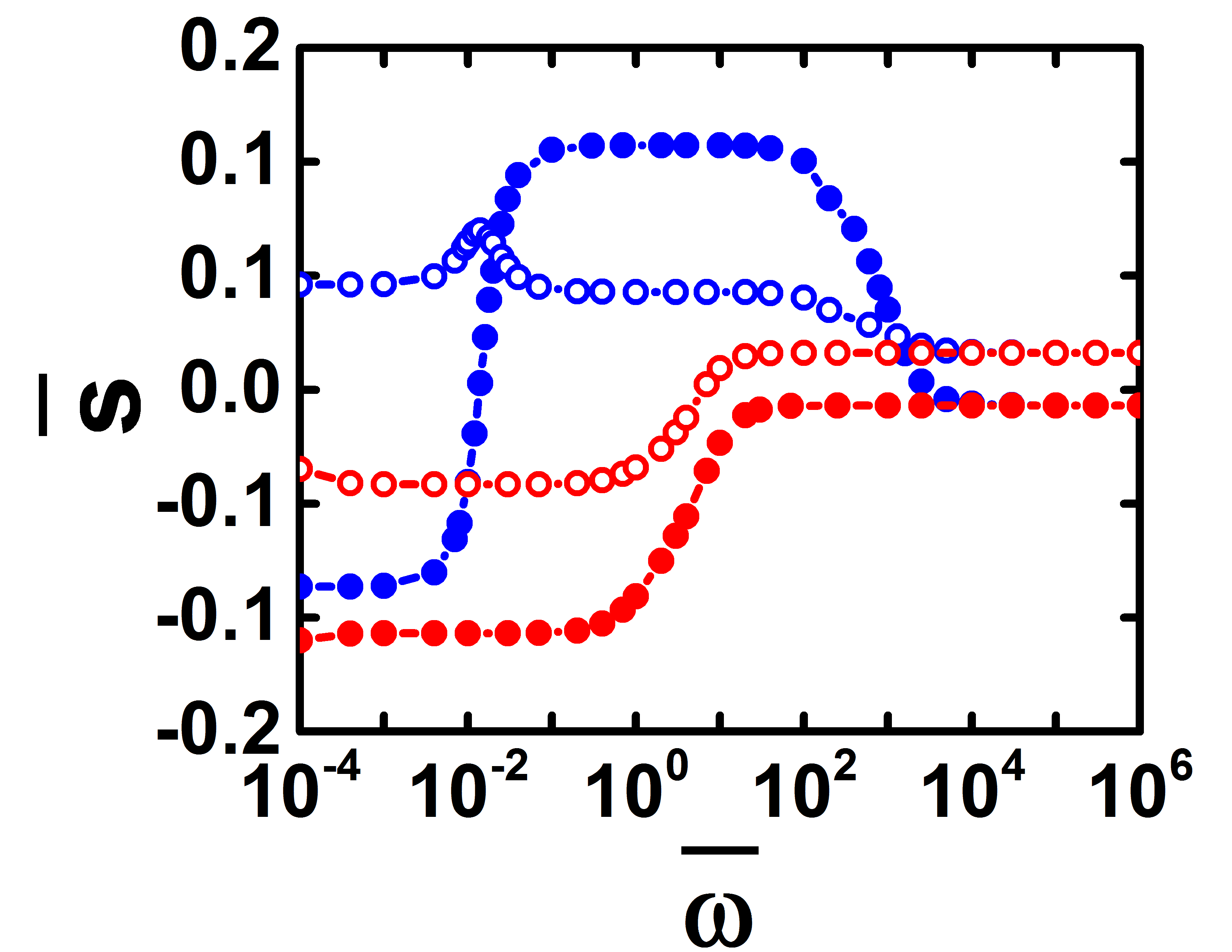}
             \caption{}
         \end{subfigure}
     \hspace{0.12cm}
    \begin{subfigure}[b]{0.48\linewidth}
    \includegraphics[width=\linewidth]{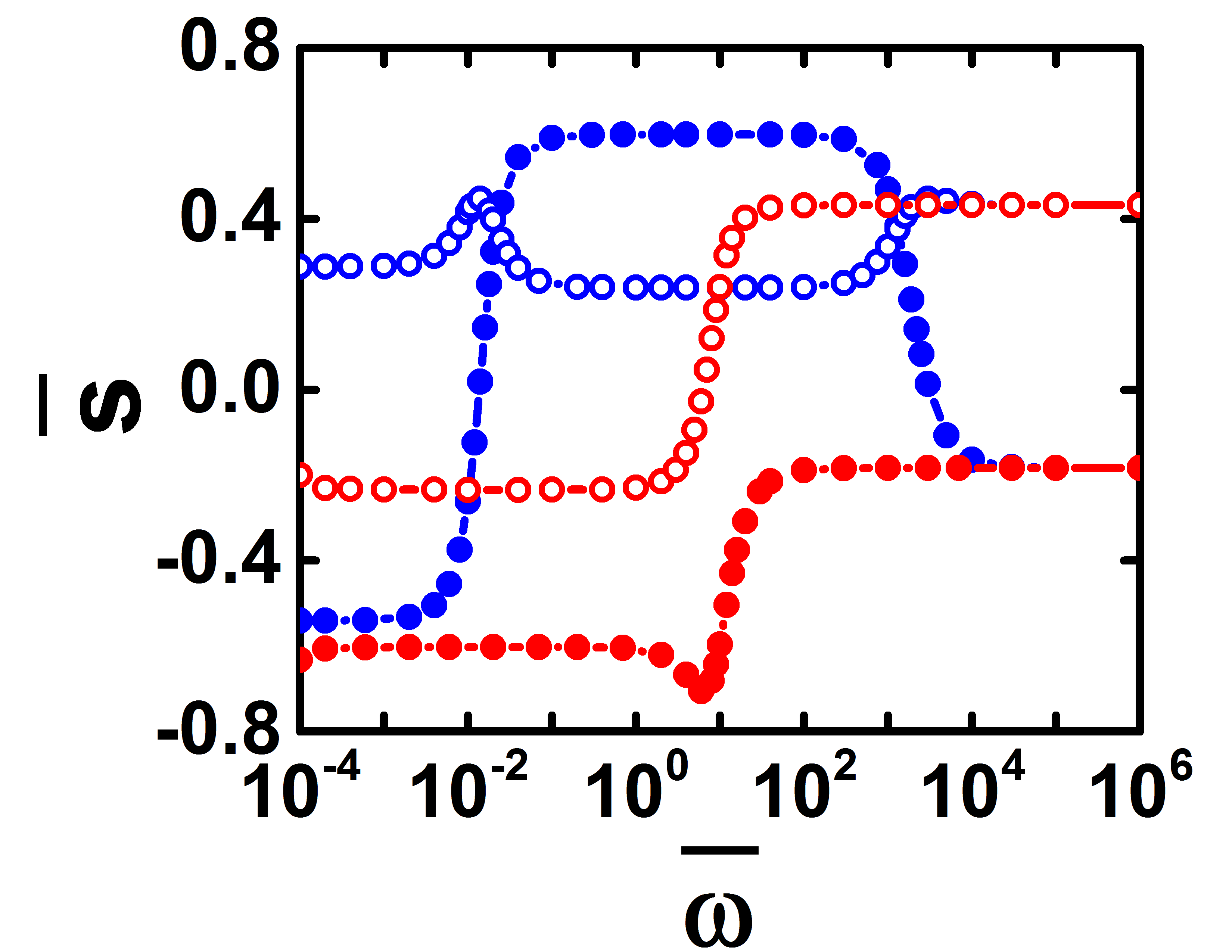}
       \caption{}
     \end{subfigure} 
  \caption{Variation of amplitudes of shape deformation modes with frequency in pure quadrupole AC electric field in (a) entropic and (b) enthalpic tension regime for Ca=9684, $\bar{\Delta}$=0.2, $\bar{C}_m=125, \bar{G}_m=0, \epsilon_r=\mu_r=1, \bar{\gamma}_{init,0}=668$. (\textcolor{blue}{\textemdash} ($\bar{s}_2$), \textcolor{red}{\textemdash} ($\bar{s}_4$), Filled circle: $\sigma_r=280$, Hollow circles: $\sigma_r=0.003$)} 
    \label{DefVsFreqPureQ}
  \end{figure}
  \begin{figure}[tbp]
       \centering
        \begin{subfigure}[b]{0.49\linewidth}
    \includegraphics[width=\linewidth]{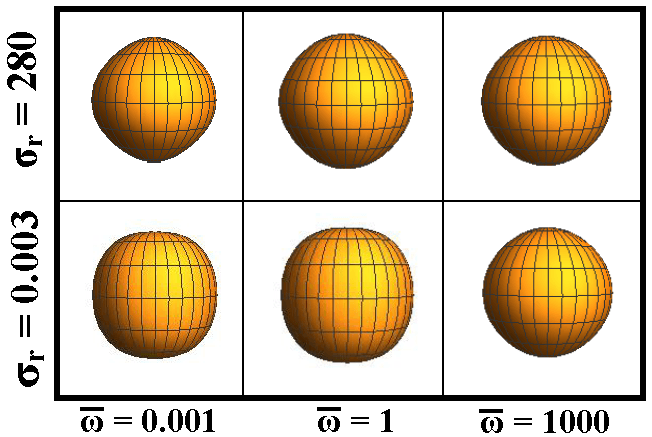}
            \caption{}
        \end{subfigure}
        \hspace{0.12cm}
      \begin{subfigure}[b]{0.45\linewidth}
   \includegraphics[width=\linewidth]{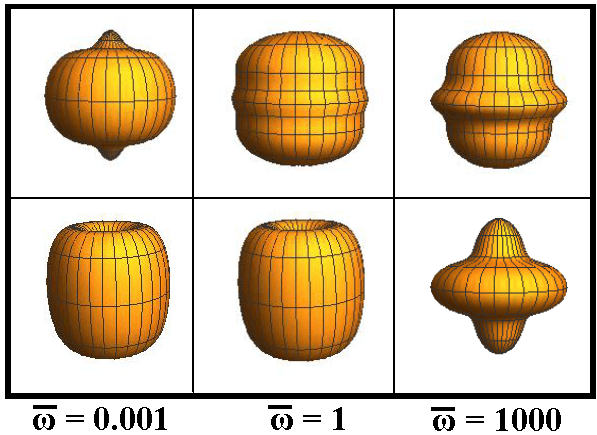}
            \caption{}
        \end{subfigure} 
  \caption{Vesicle shape transition in pure quadrupole AC electric field with frequency by: (a) entropic and (b) enthalpic approach, $\sigma_r =$ 280 and 0.003 ($\bar{\Delta}$=0.2, $\bar{C}_m$=125, $\bar{G}_m$=0, $\epsilon_r=\mu_r=1,\bar{\gamma}_{init,0}=668, Ca=9684$).} 
        \label{PureShapeEnt}
   \end{figure}
\begin{figure}[tbp]
   \centering
   \includegraphics[width=0.6\linewidth]{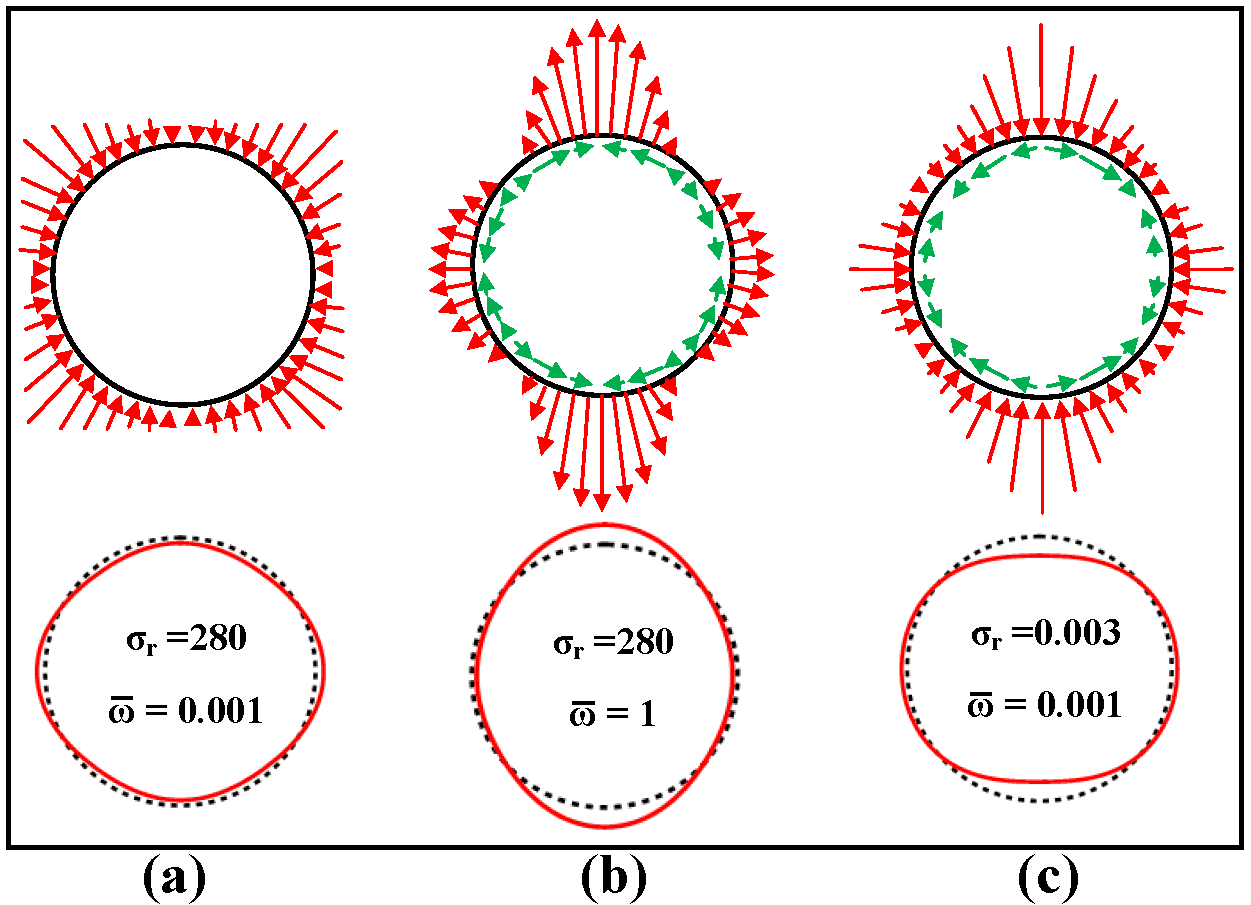}
   \caption{Schematic of normal and tangential electric stress distribution on an undeformed vesicle in pure quadrupole AC electric field ($\bar{C}_m$=125, $\bar{G}_m$=0).} 
   \label{stressdist}
  \end{figure}

  Figures \ref{DefVsFreqPureQ} and \ref{PureShapeEnt} show the variation of the deformation ($\bar{s}_2$ and $\bar{s}_4$) and the  final equilibrium shapes, respectively. The variation is shown with frequency, at different conductivity ratios, in the entropic and enthalpic regimes. Unlike the case of uniform fields, where prolate and oblate spheroids imply that $\bar{s}_2$ deformation suffices to explain the shape of the vesicles, in quadrupole fields, occurrence of $\bar{s}_2$ and $\bar{s}_4$ amplitudes suggest that the actual shape should be discussed (figure \ref{PureShapeEnt}). A distinctive feature of pure quadrupole fields is the observation of oblate deformation for the $P_4$ ($-\bar{s}_4$) mode even at very low frequencies $\omega<t_{cap}^{-1}$ in agreement with similar results for a drop \cite{SD2012} (Figure \ref{stressdist}a). This is essentially due to the fact that pure normal stresses in quadrupole field favor oblate shapes even in the absence tangential stresses \cite{SD2012}, unlike uniform fields wherein vesicles are always prolate in the low frequency regime. 
  
  In the entropic regime, in the intermediate frequency range, $t_{cap}^{-1}<\omega<t_{MW}^{-1}$, a vesicle behaves like a drop. At higher inner conductivity ($\sigma_r>1$) the deformation is predominantly prolate for all frequencies (see figure \ref{stressdist}b), whereas for $\sigma_r<1$, oblate deformations are observed at low and intermediate frequencies  (see figure \ref{stressdist}c). A feature of these plots (compared to uniform field which admits only the $P_2$ mode) is the different qualitative behaviour of the $\bar{s}_2$ and $\bar{s}_4$ amplitudes of the $P_2$, $P_4$ modes with frequency. In the high frequency regime, a nearly spherical shape is seen since the Maxwell stresses become small although do not disappear. Thus unlike a drop even at high frequencies, a vesicle does show non-zero, although small, tangential and normal electric fields and stresses, on account of the capacitance of the membrane. 
  
  In the enthalpic regime, at low and intermediate frequencies, the variation of the shapes with frequency is qualitatively similar to that observed in the entropic case. However, unlike the entropic case, low deformation (that is nearly spherical shape) is not observed at very high frequencies, since the modes have to conserve the total area and volume simultaneously. The distribution of this area between $\bar{s}_2$ and $\bar{s}_4$ is clearly seen in the figures \ref{DefVsFreqPureQ},\ref{PureShapeEnt}. A variety of shapes are therefore observed which are a competition between minimization of the bending, tension and the electrostatic energy. The role of electrostatics is indicated by prolate shapes for $\sigma_r>1$ and oblate shapes at intermediate frequencies for $\sigma_r<1$ and also by the fact that the the shapes depend upon the conductivity ratio. Both the entropic and enthalpic regimes show significant deviation from ellipsoidal shapes, exhibiting rhomboidal, cuboidal shapes on account of $\bar{s}_4$. The transition frequencies can be used to estimate electromechanical properties of a vesicle.\\
  
\begin{figure}[tbp]
     \centering
      \begin{subfigure}[b]{0.47\linewidth}
   \includegraphics[width=\linewidth]{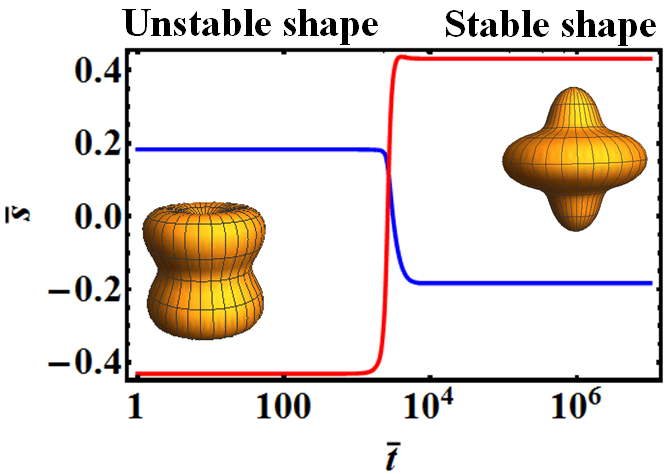}
        \caption{}
      \end{subfigure}
      \hspace{0.12cm}
      \begin{subfigure}[b]{0.48\linewidth}
   \includegraphics[width=\linewidth]{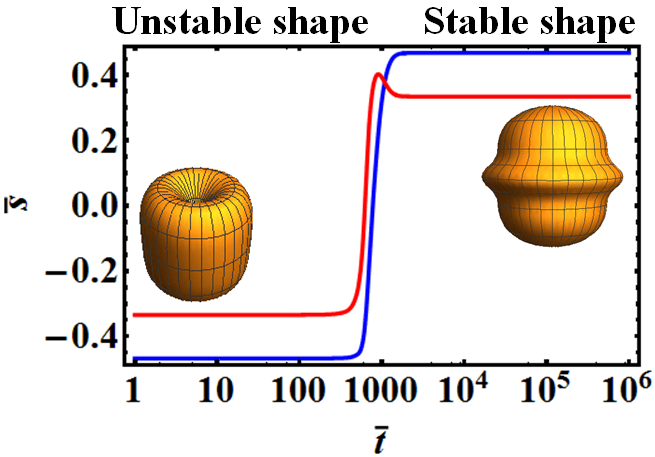}
       \caption{}
     \end{subfigure}
  \caption{Unstable and stable vesicle shapes during vesicle deformation in enthalpic tension regime under pure quadrupole AC electic field for (a)$\sigma_r=0.003$, and (b)$\sigma_r=280$ ($\bar{\omega}=1000$, $\bar{\Delta}$=0.2, $\bar{C}_m$=125, $\bar{G}_m$=0, $\epsilon_r=\mu_r=1,\bar{\gamma}_{init,0}=668, Ca=9684$). \textcolor{blue}{\textemdash} ($\bar{s}_2$), \textcolor{red}{\textemdash} ($\bar{s}_4$)} 
             \label{PureStability1}
  \end{figure}

  Figure \ref{PureShapeEnt}a shows the stable shapes for high and low $\sigma_r$ in the entropic regime.  It should be noted that while the shapes in the pure entropic regime are obtained by solving the dynamical equations for $\bar{s}_2$ and $\bar{s}_4$, calculations in the enthalpic regime lead to different nature of the equations. The  steady state solution to the evolution equations (33-\ref{evoleqn})  for $\bar{s}_2$ and $\bar{s}_4$ in the enthalpic regime gives multiple roots (as pairs in $\bar{s}_2,\bar{s}_4$) of which only one root (and thereby shape) is expected to be stable. It should be noted that each root corresponds to a different shape. To find the stable root, the eigenvalues of the linearized coefficient matrix resulting from the the two non-linear differential equations  for $\bar{s}_2$ and $\bar{s}_4$ (33-\ref{evoleqn}) are determined for each of the roots, and the root is deemed stable if the eigenvalues are negative (and unstable if at least one of the eigen value is positive). This is also demonstrated by solving the evolution equations (33-\ref{evoleqn}) with respect to time, with the initial shape ($\bar{s}_2$ and $\bar{s}_4$ values) being a perturbation around the stable or unstable roots. When the initial condition corresponds to the unstable roots, the system is seen to evolve to the stable roots (and thereby shapes).  Figure \ref{PureStability1} shows these plots for $\sigma_r>1$ and $\sigma_r<1$ respectively. A systematic evolution to the stable shapes is seen, with an interesting period of relative quiescence before the transition.
             
  \begin{figure}[tbp]
       \centering
        \begin{subfigure}[b]{0.48\linewidth}
    \includegraphics[width=\linewidth]{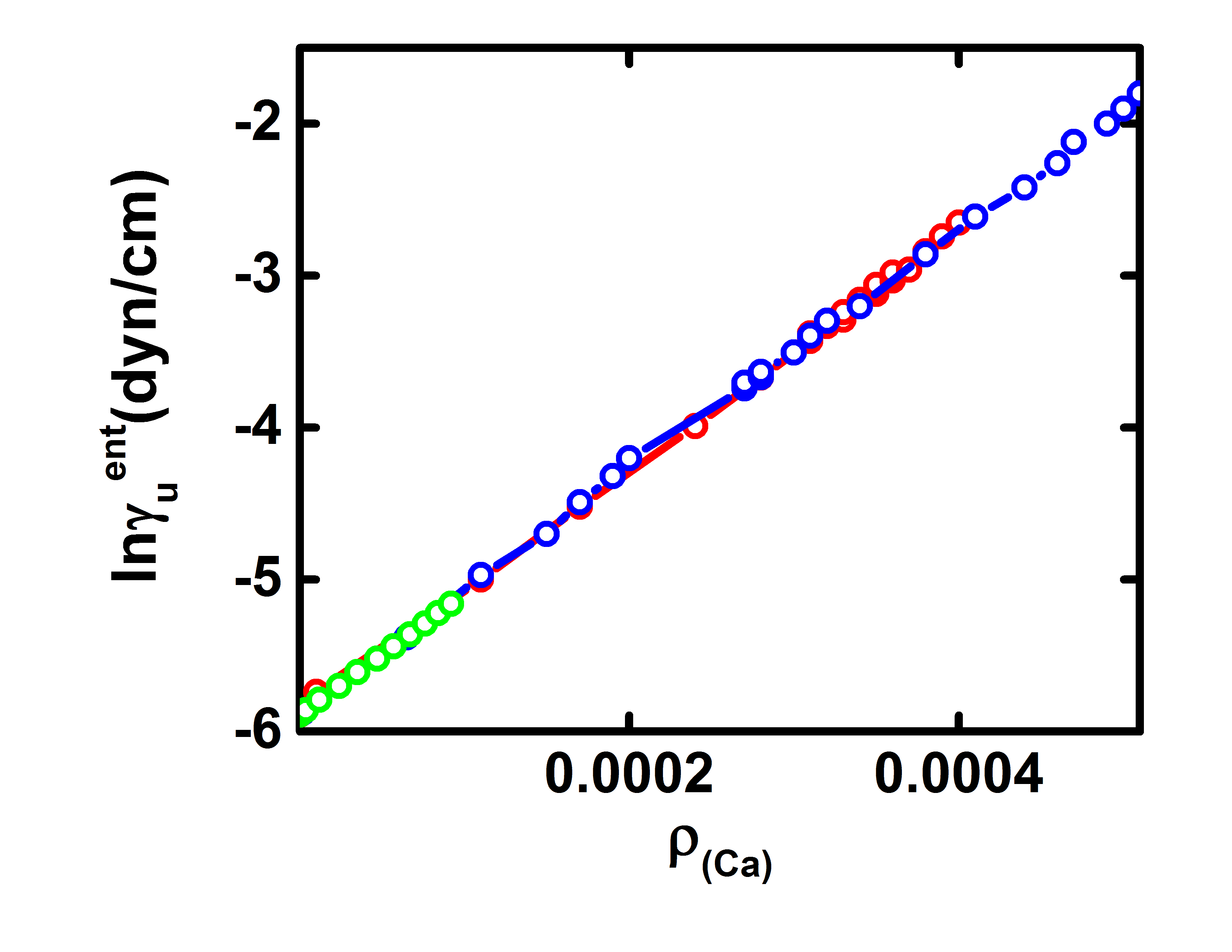}
            \caption{}
        \end{subfigure}
        \hspace{0.12cm}
        \begin{subfigure}[b]{0.48\linewidth}
     \includegraphics[width=\linewidth]{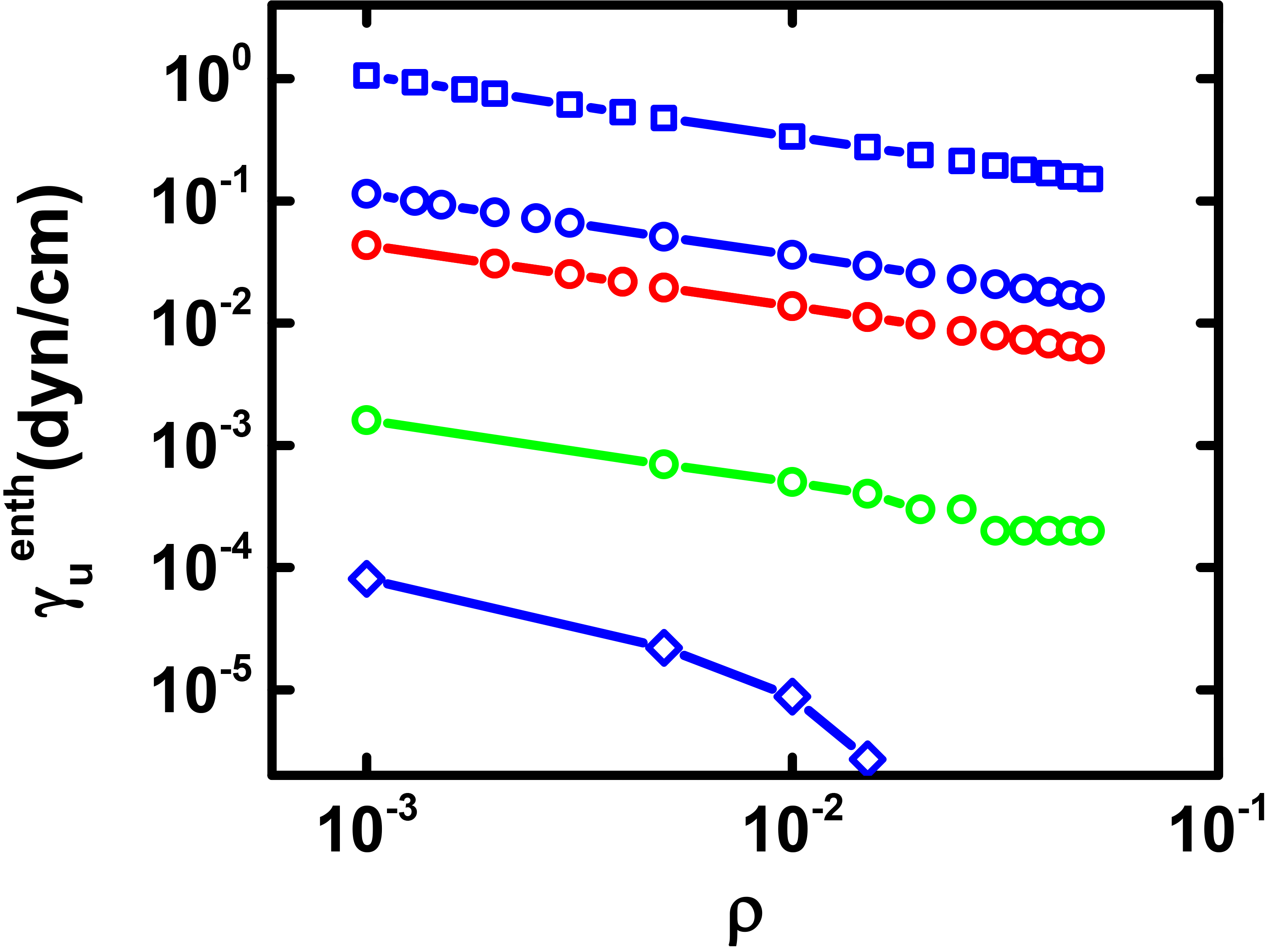}
            \caption{}
        \end{subfigure}
       \hspace{0.12cm}
            \begin{subfigure}[b]{0.48\linewidth}
  \includegraphics[width=\linewidth]{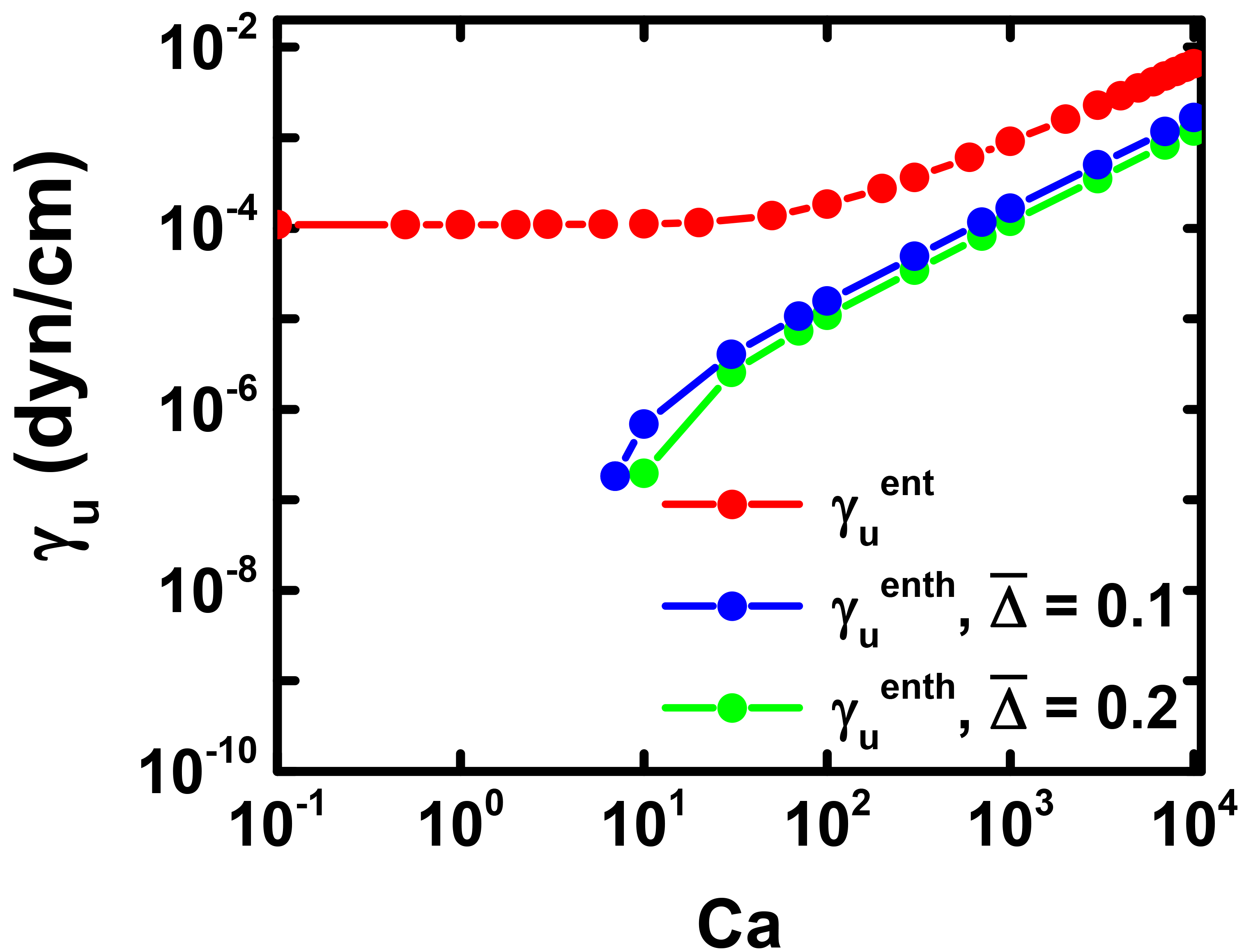}
                \caption{}
            \end{subfigure}    
  \caption{Variation of uniform membrane tension with apparent excess area ($\rho=\Delta/4\pi$) in a pure quadrupole AC electric field by: (a) entropic tension approach for $\bar{\omega}$=0.001 (blue circle),$\bar{\omega}$=1 (red circle), $\bar{\omega}$=1000 (green circle)  (b) enthalpic tension approach at $Ca=9684$ for {$\bar{\omega}$=1 (blue circle),$\bar{\omega}$=0.001 (red circle), $\bar{\omega}$=1000 (green circle)} and at $Ca$=9 (blue diamond), 90000 (blue square) for $\bar{\omega}=1$; (c) Variaton of entropic and enthalpic tension with capillary number for $\bar{\omega}=1$. ($\bar{C}_m$=125, $\bar{G}_m$=0, $\epsilon_r=\mu_r=1, \sigma_r=280,\bar{\gamma}_{init,0}=668, Ca=9684$)} 
        \label{PureTension1}
   \end{figure}
    \begin{figure}[tp]
       \centering
        \begin{subfigure}[b]{0.4\linewidth}
         \includegraphics[width=\linewidth]{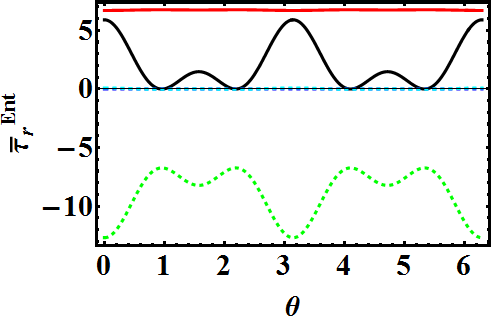}
          \caption{}
        \end{subfigure}
        \hspace{0.12cm}
        \begin{subfigure}[b]{0.49\linewidth}
          \includegraphics[width=\linewidth]{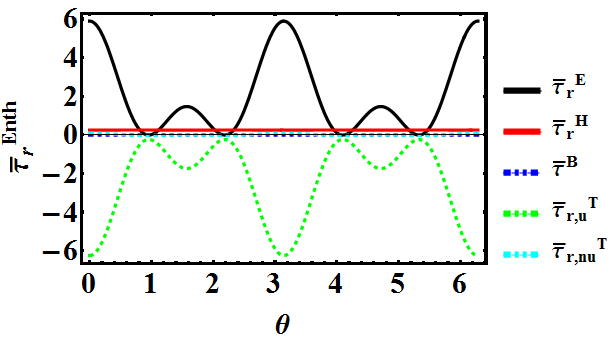}
         \caption{}
       \end{subfigure}
         \begin{subfigure}[b]{0.41\linewidth}
          \includegraphics[width=\linewidth]{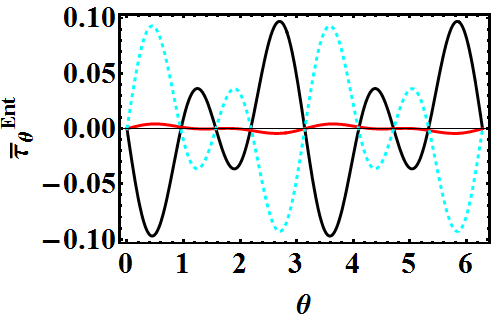}
           \caption{}
         \end{subfigure}
         \hspace{0.12cm}
         \begin{subfigure}[b]{0.52\linewidth}
           \includegraphics[width=\linewidth]{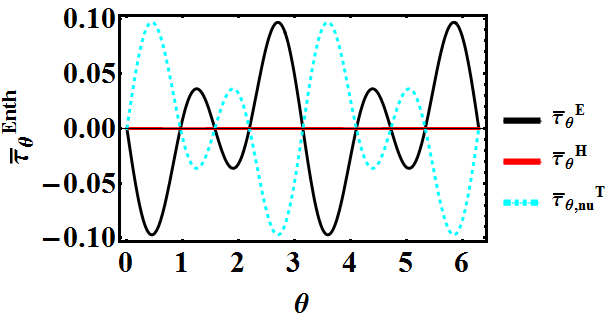}
          \caption{}
        \end{subfigure}
          \begin{subfigure}[b]{0.42\linewidth}
            \includegraphics[width=\linewidth]{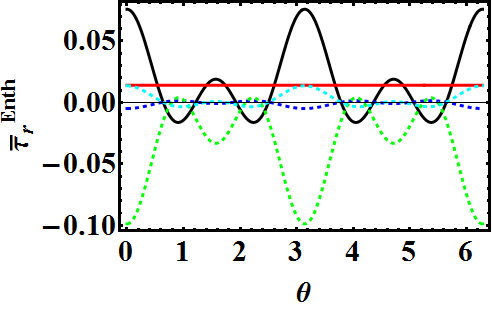}
             \caption{}
           \end{subfigure}
           \hspace{0.12cm}
           \begin{subfigure}[b]{0.507\linewidth}
             \includegraphics[width=\linewidth]{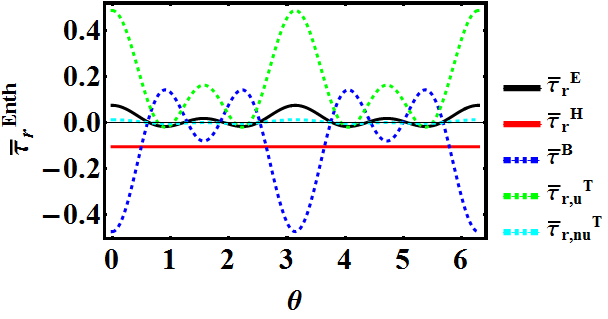}
            \caption{}
          \end{subfigure} 
          \begin{subfigure}[b]{0.4\linewidth}
         \includegraphics[width=\linewidth]{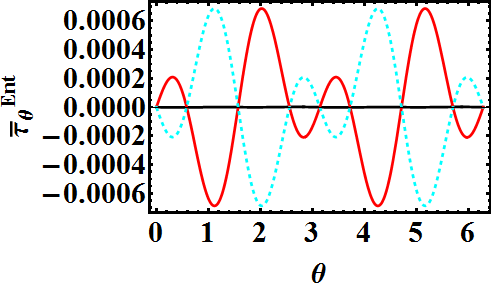}
             \caption{}
           \end{subfigure}
            \hspace{0.12cm}
         \begin{subfigure}[b]{0.52\linewidth}
         \includegraphics[width=\linewidth]{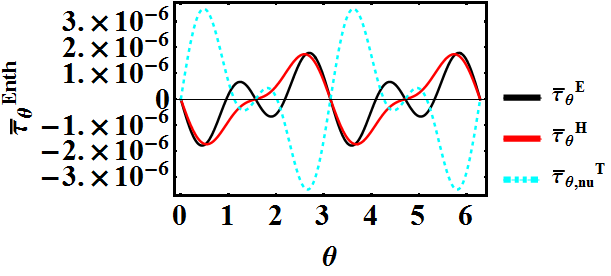}
                  \caption{}
          \end{subfigure}    
        \caption{Relative contribution of all stresses (destabilizing forces: solid, stabilizing forces: dashed) acting on a vesicle from $\theta=0-2\pi$ when $\sigma_r=280$. (a),(c): Entropic approach ($\bar{\omega}=1, Ca=9684$); (b),(d): Enthalpic approach ($\bar{\omega}=1, Ca=9684$); (e) Enthalpic approach ($\bar{\omega}=1000, Ca=9684$); (f) Enthalpic approach ($\bar{\omega}=1000, Ca=100$); (g) Entropic approach ($\bar{\omega}=0.0001, Ca=9684$); (h) Enthalpic approach ($\bar{\omega}=0.0001, Ca=9684$)
    } 
      \label{RelativeStress}
    \end{figure}  
  Figure \ref{RelativeStress} shows the contribution of different stabilizing membrane forces (bending, uniform and nonuniform tension) as well as deforming electric forces and the resulting hydrodynamic forces. The figure \ref{RelativeStress}(a,b) shows that the deforming normal electric stresses are balanced by the uniform tension in both the entropic regimes at all frequencies and enthalpic regime at intermediate and low frequencies (only $\bar{\omega}=1$ shown in figure). At high frequencies, in the enthalpic regime though,  a small but non-trivial contribution of the bending and nonuniform tension generated normal forces is observed. This is really due to the low absolute value of the normal stresses at high frequencies. The contribution of bending forces is found to increase further and equal to that due to the nonuniform tension at small capillary numbers (figure \ref{RelativeStress}(e,f)).
  
  The tangential electric stresses are balanced by the non uniform tension and the tangential hydrodynamic stress in both the entropic and enthalpic regimes at intermediate and high frequencies (only $\bar{\omega}=1$ shown in figure \ref{RelativeStress}(c,d)).  At very low frequencies, the tangential electric stress tends to zero, and while in the entropic regime, the small tangential hydrodynamic stress is balanced by the non uniform tension, in the enthalpic regime, the values of tangential stress is even lower, and a balance of all the three stresses, hydrodynamic, electrical and non uniform tension is observed (figure \ref{RelativeStress}(g,h)).

 \subsubsection{Entropic (Ca dependent) and Enthalpic tension vs Excess area}
   Figure \ref{PureTension1} presents variation of entropic and enthalpic tension with excess area at three different frequencies. Figure \ref{PureTension1}a shows that in the fluctuation dominated regime, the tension varies exponentially with the excess area, in agreement with the prediction of \cite{evan1990}. Figure \ref{PureTension1}b shows that a vesicle in the enthalpic regime, exhibits a decrease in tension  with an increase in the excess area ($\gamma^{enth}\sim \rho^{-2}$ where $\rho=\Delta/(4\pi)$), for the same $Ca$ and different frequencies, as well as for different capillary numbers for a given frequency. The tension increases with capillary number, and decreases with the frequency. The experimental work by \cite{evan1990} indicates that when a vesicle is deflated from a spherical shape due to application of an external force, the maximum tension could not be more than 0.5 $mN/m$ in the low tension regime. In the high tension regime, the  maximum allowable vesicle tension is known to be typically around $10 mN/m$, thereby limiting the capillary number. 
   The entropic and enthalpic tensions as a function of the capillary number are plotted in figure \ref{PureTension1}c. The figure shows that for a given Ca, the excess area dependent enthalpic tension is higher for smaller excess area, but is always lower than the entropic tension. The entropic tension increases weakly with the capillary number in the low capillary number limit,  and shows a $\gamma_{u} \sim Ca$ scaling at around Ca $\sim$ 50. The enthalpic tensions scales as $\gamma_u \sim Ca$. in the high capillary number limit. It is therefore proposed that a transition from the entropic to enthalpic regime can be expected to occur around Ca $\sim$ 100. The dimensional value of tension around this regime is of the order of 0.5 mN/m.
   
   \subsection{Deformation in mixed field}
   
   To generate a mixed field, non-zero values of $E_o$ and $\Lambda_o$ are required yielding a finite value of $\bar{f}$ as discussed earlier for the case of dielectrophoresis. Using similar electrical parameters as for dielectrophoresis, yield a capillary number, $Ca=71033$ and $\bar{f}=0.5$. 
   
\begin{figure}[tp]
            \centering
             \begin{subfigure}[b]{0.46\linewidth}
     \includegraphics[width=\linewidth]{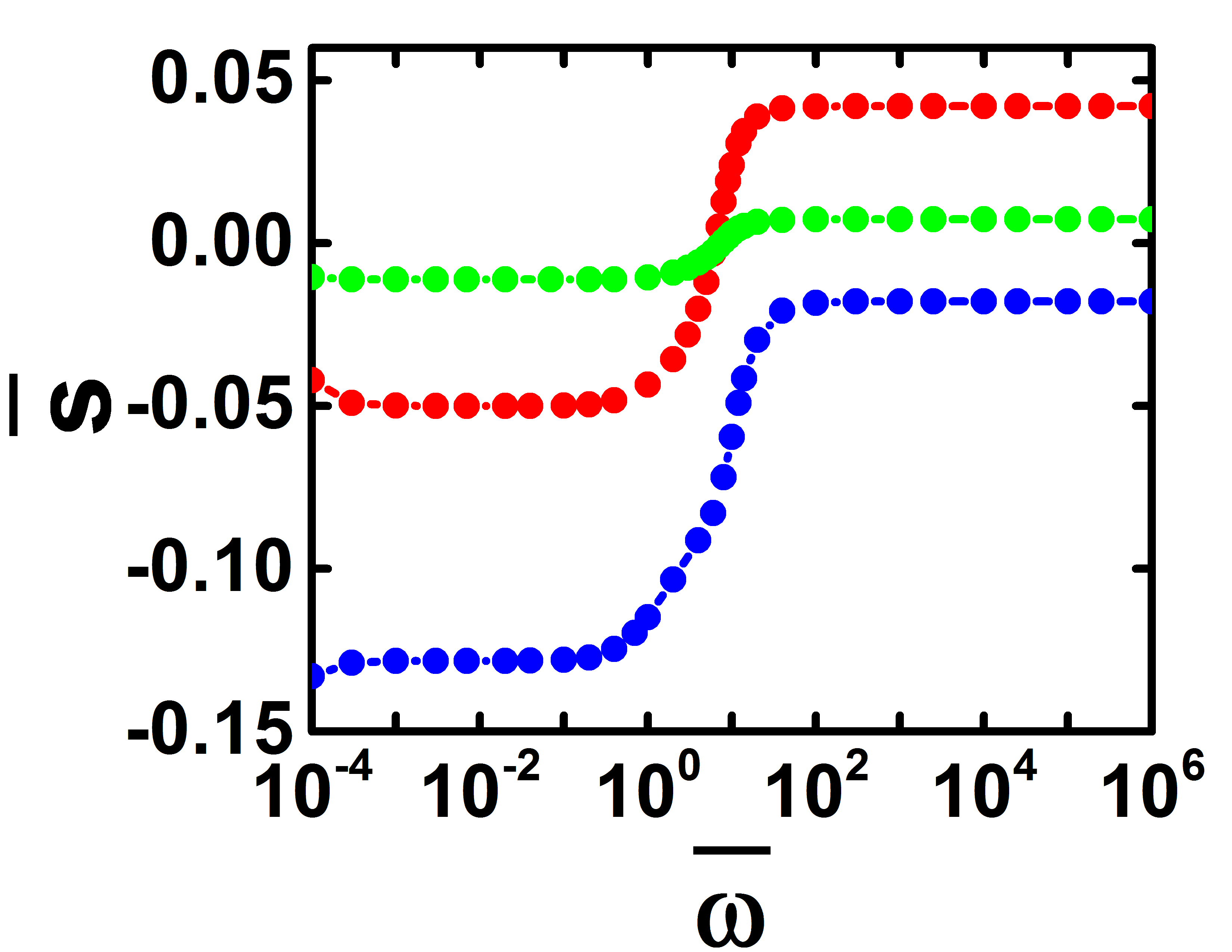}
                  \caption{}
             \end{subfigure} 
             \hspace{0.12cm}
             \begin{subfigure}[b]{0.46\linewidth}
     \includegraphics[width=\linewidth]{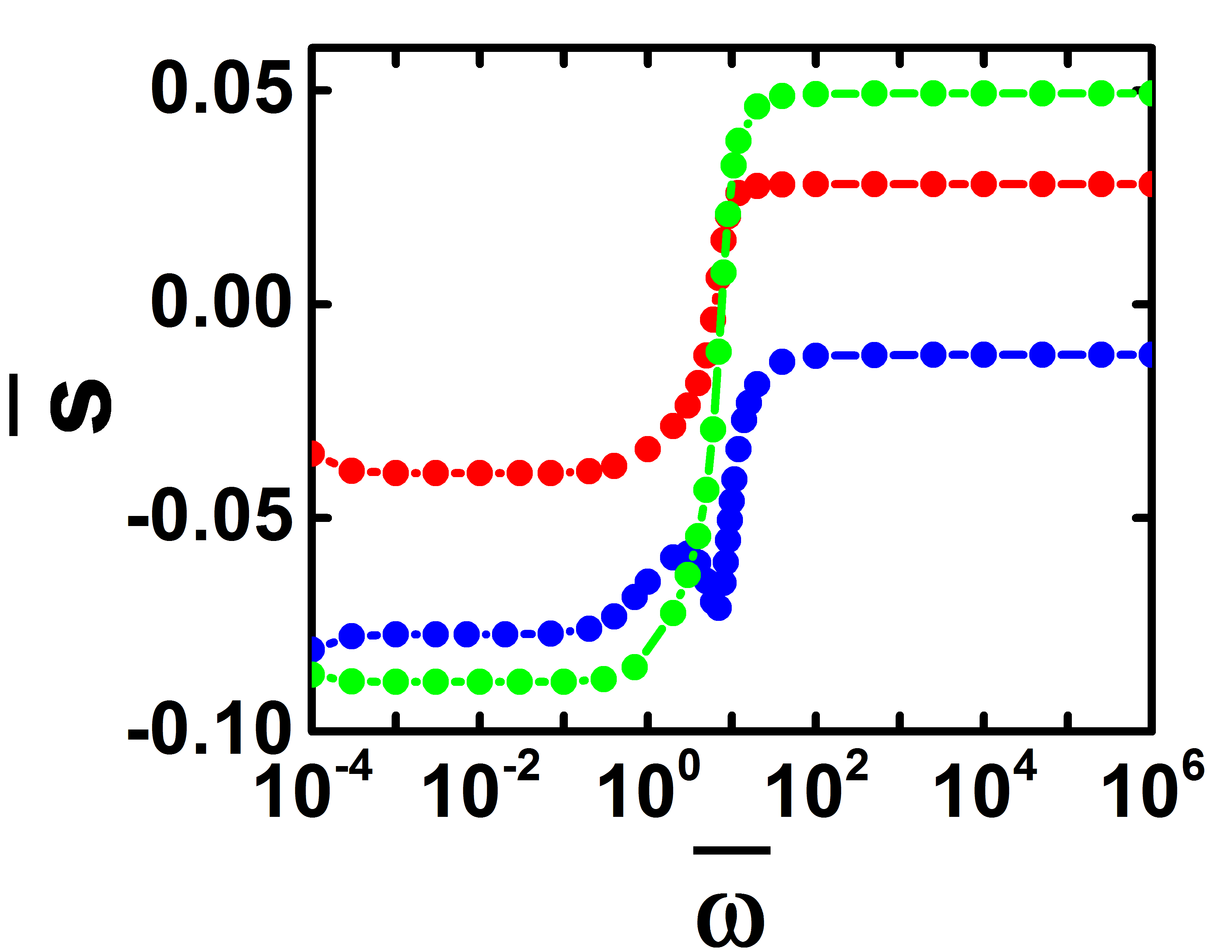}
                  \caption{}
             \end{subfigure}                        
             \begin{subfigure}[b]{0.46\linewidth}
     \includegraphics[width=\linewidth]{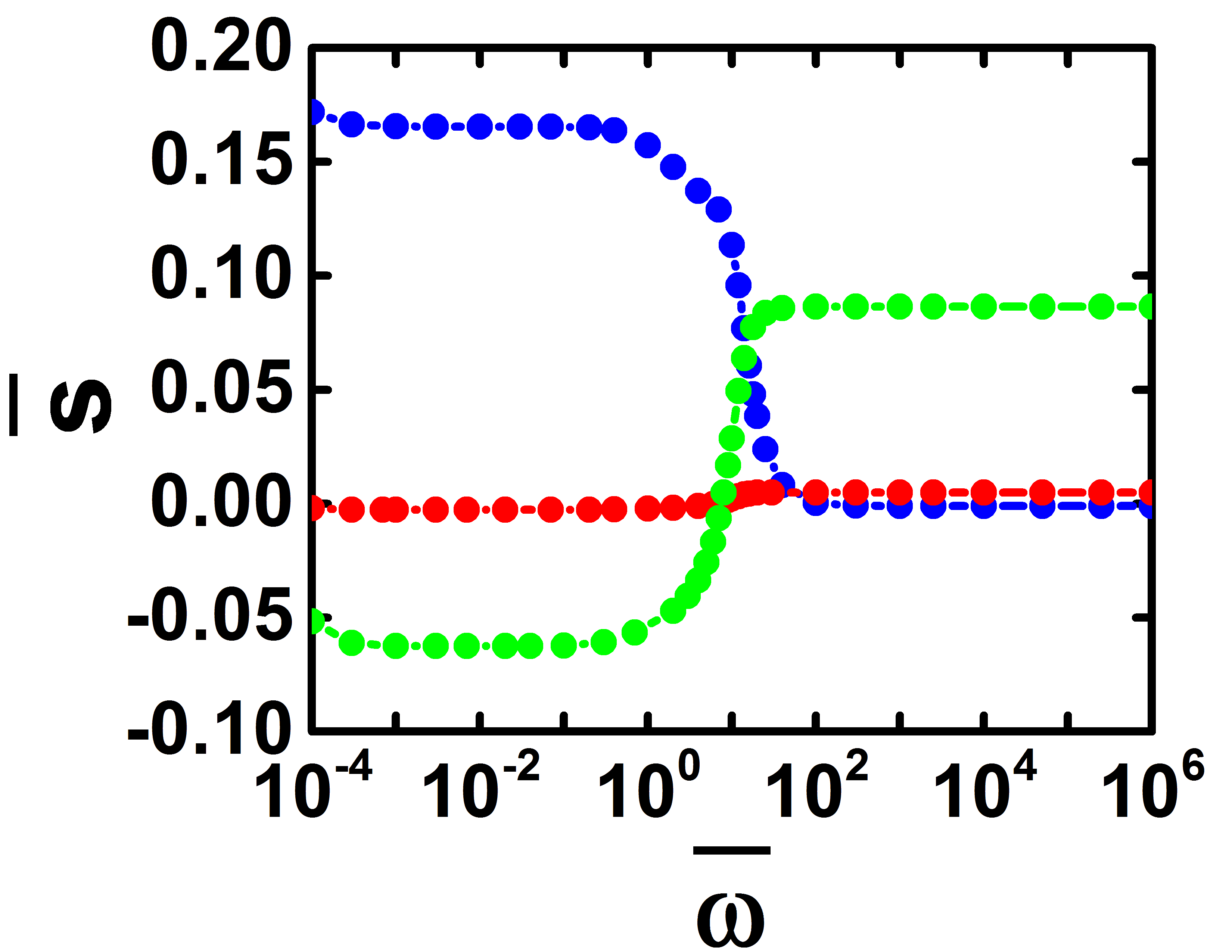}
                 \caption{}
             \end{subfigure}
             \hspace{0.12cm}
             \begin{subfigure}[b]{0.46\linewidth}
     \includegraphics[width=\linewidth]{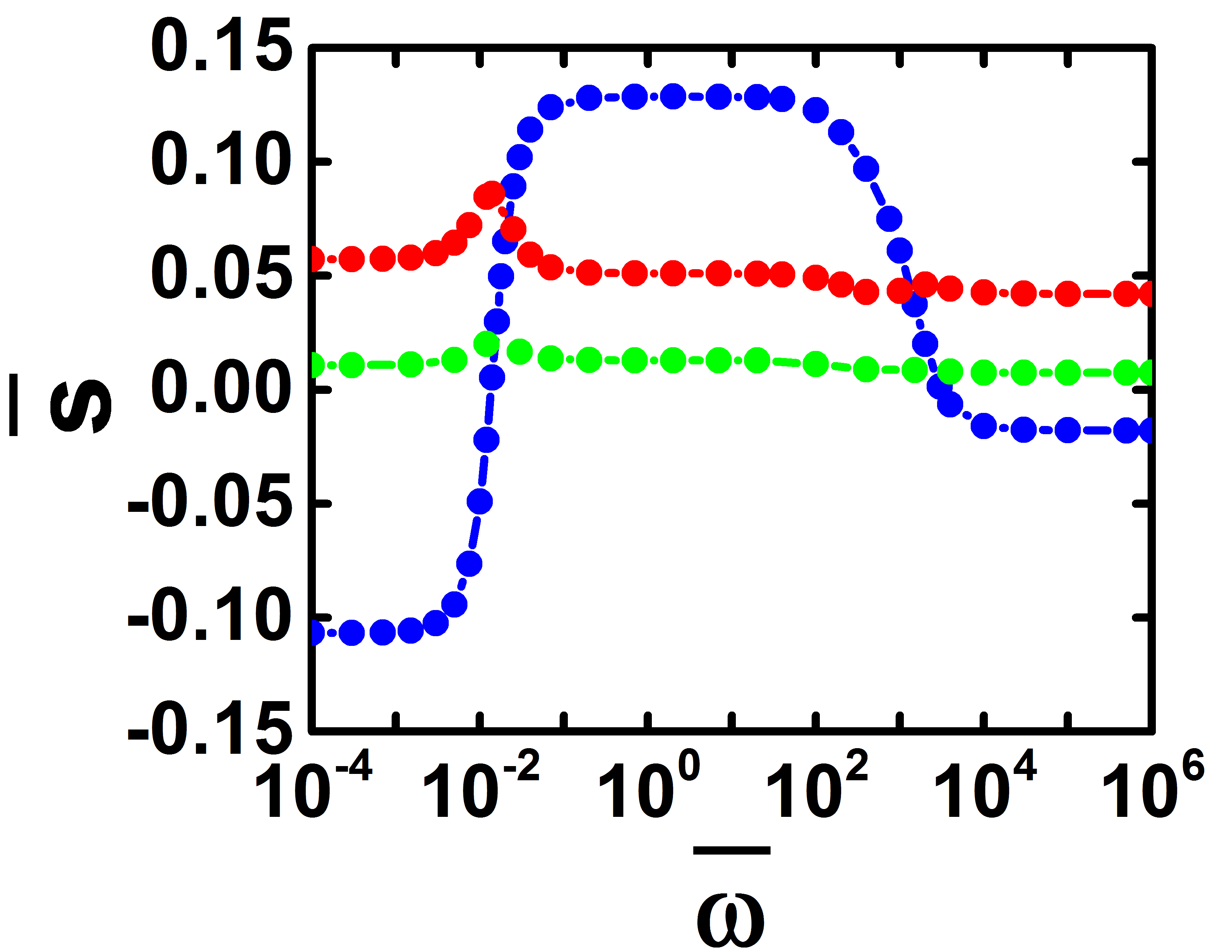}
                  \caption{}
             \end{subfigure}     
     \begin{subfigure}[b]{0.46\linewidth}
     \includegraphics[width=\linewidth]{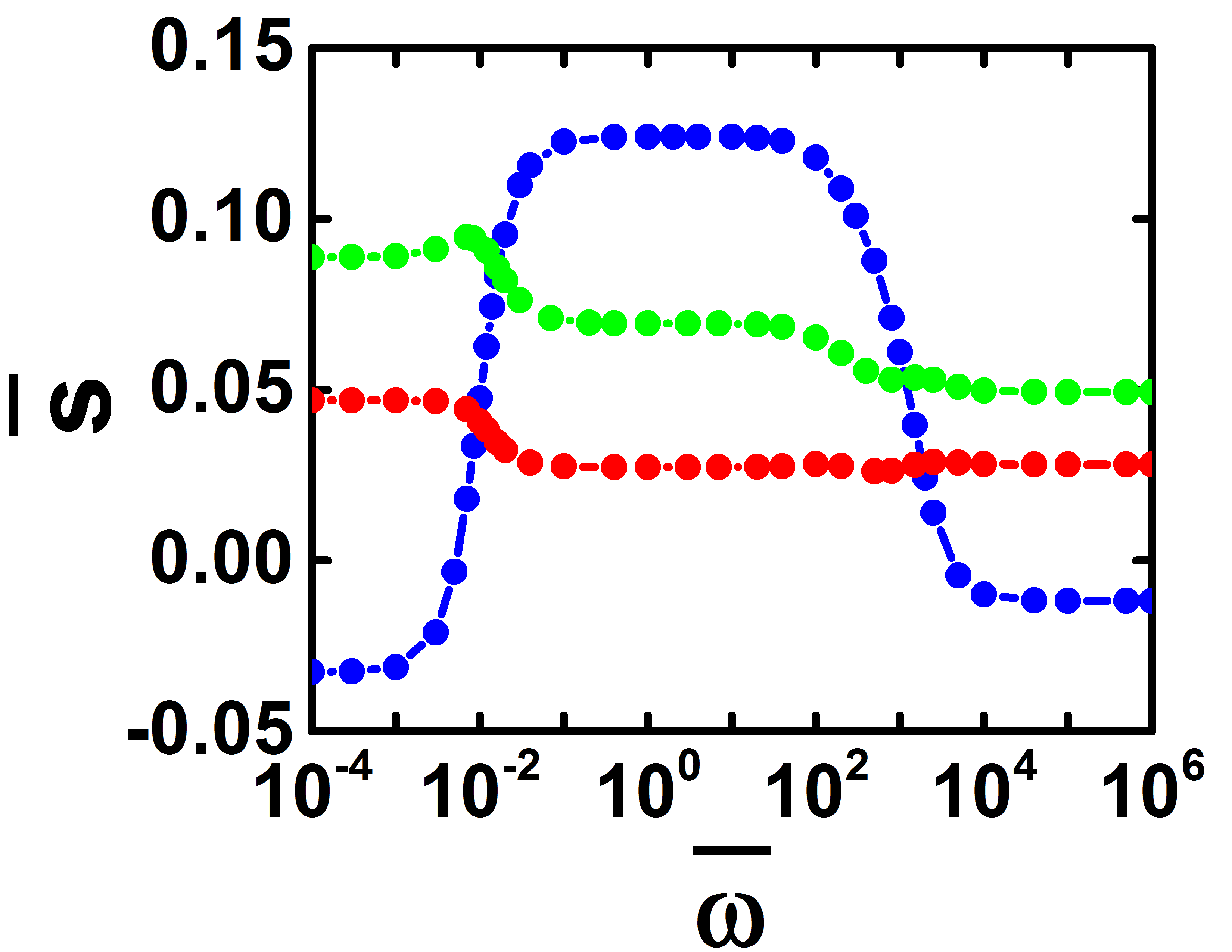}
                 \caption{}
             \end{subfigure}
             \hspace{0.12cm}
             \begin{subfigure}[b]{0.46\linewidth}
     \includegraphics[width=\linewidth]{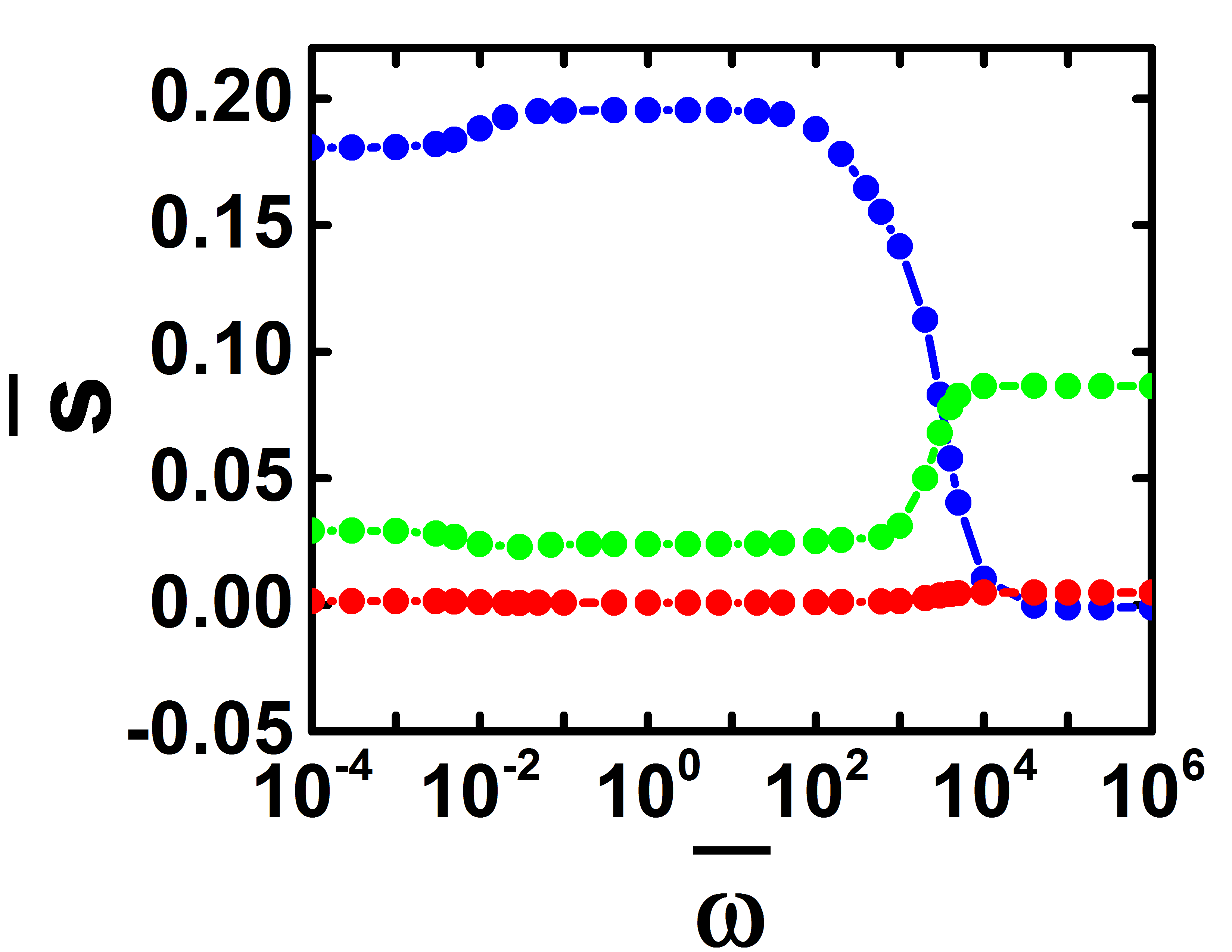}
                 \caption{}
             \end{subfigure}             
    \caption{Variation of amplitudes of shape deformation modes (entropic approach) in mixed AC electric field with frequency: (a)-(c) $\sigma_r$=0.003, and (d)-(f) $\sigma_r$=280 for $\bar{C}_m$=125, $\bar{G}_m$=0, Ca=71033, $\mu_r=\epsilon_r$=1, $\bar{\gamma}_{init,0}=668$. $\bar{f}$ ranges 0.1, 1, 10 left to right. \textcolor{blue}{\textemdash} ($\bar{s}_2$), \textcolor{green}{\textemdash} ($\bar{s}_3$), \textcolor{red}{\textemdash} ($\bar{s}_4$)} 
               \label{MixDefFreqEnt}
        \end{figure}

\begin{figure}[tp]
            \centering
             \begin{subfigure}[b]{0.46\linewidth}
    \includegraphics[width=\linewidth]{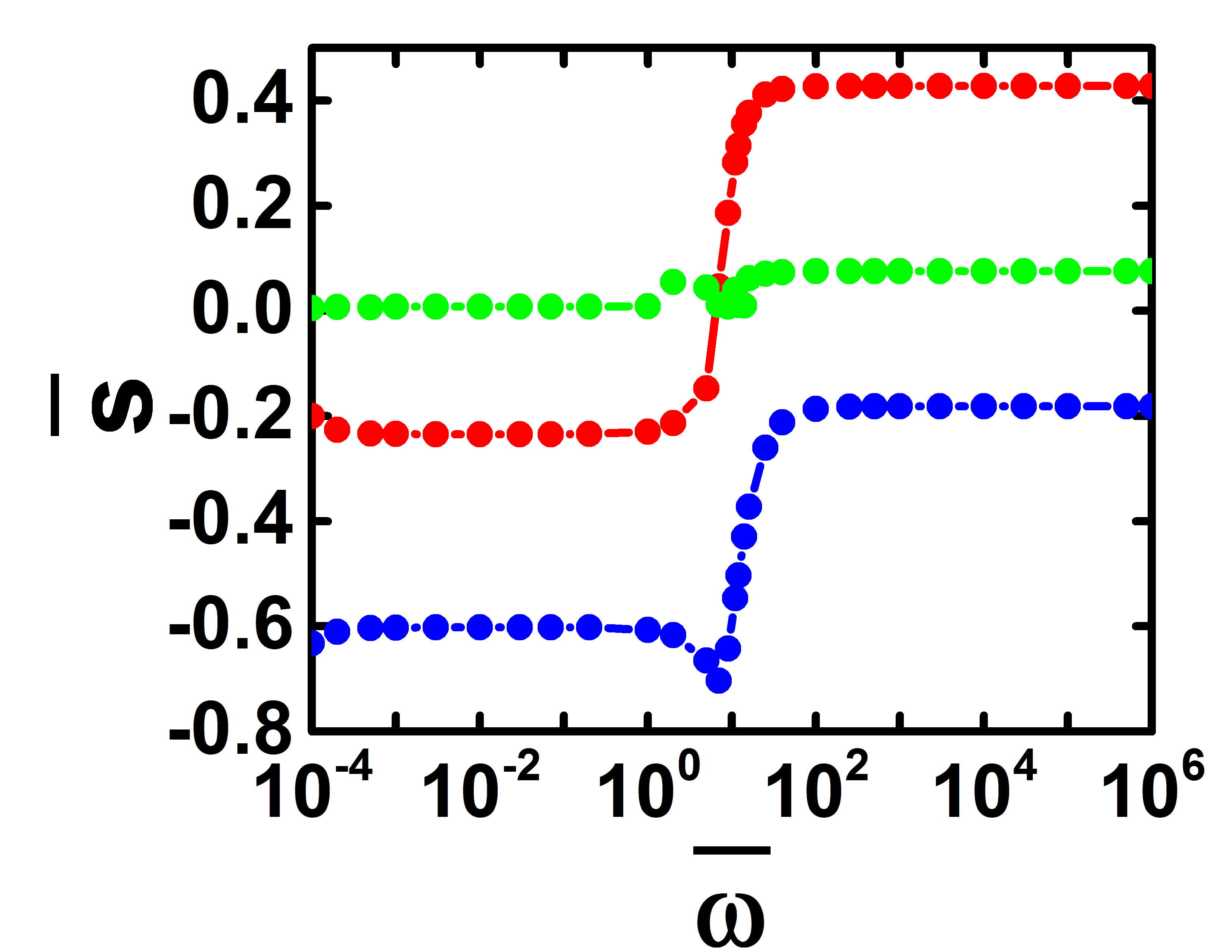}
                  \caption{}
             \end{subfigure} 
             \hspace{0.12cm}
             \begin{subfigure}[b]{0.46\linewidth}
     \includegraphics[width=\linewidth]{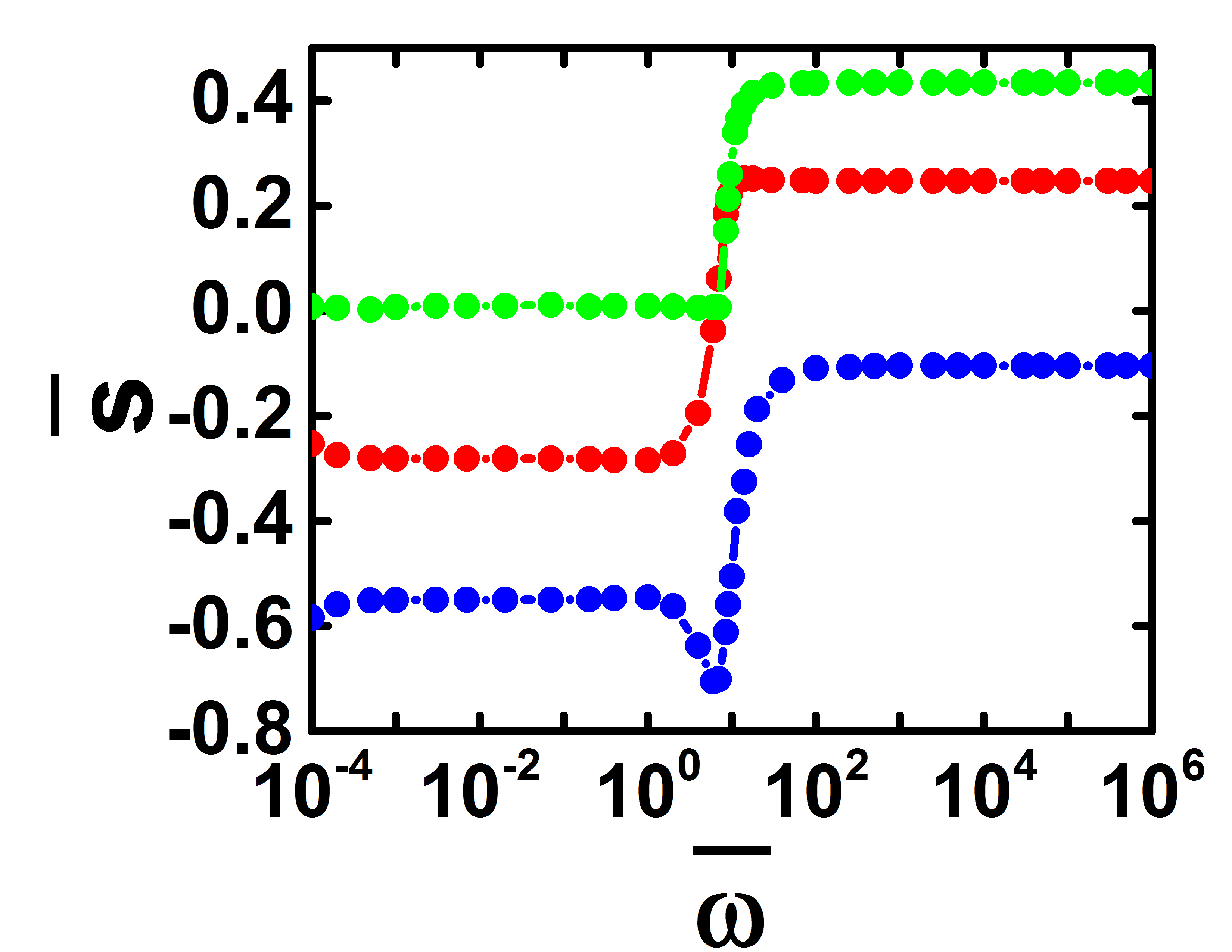}
                  \caption{}
             \end{subfigure}                        
             \begin{subfigure}[b]{0.46\linewidth}
    \includegraphics[width=\linewidth]{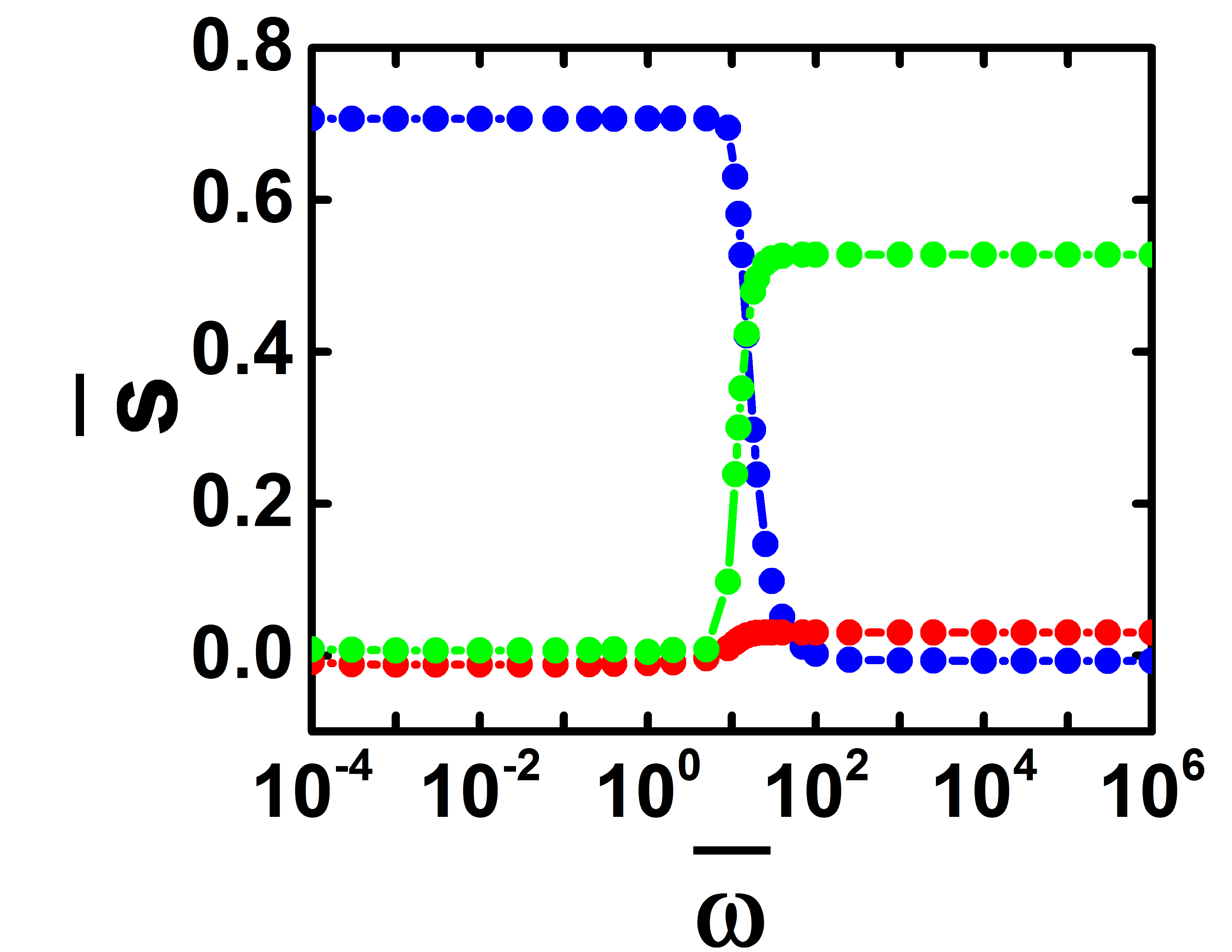}
                 \caption{}
             \end{subfigure}
             \hspace{0.12cm}
             \begin{subfigure}[b]{0.46\linewidth}
     \includegraphics[width=\linewidth]{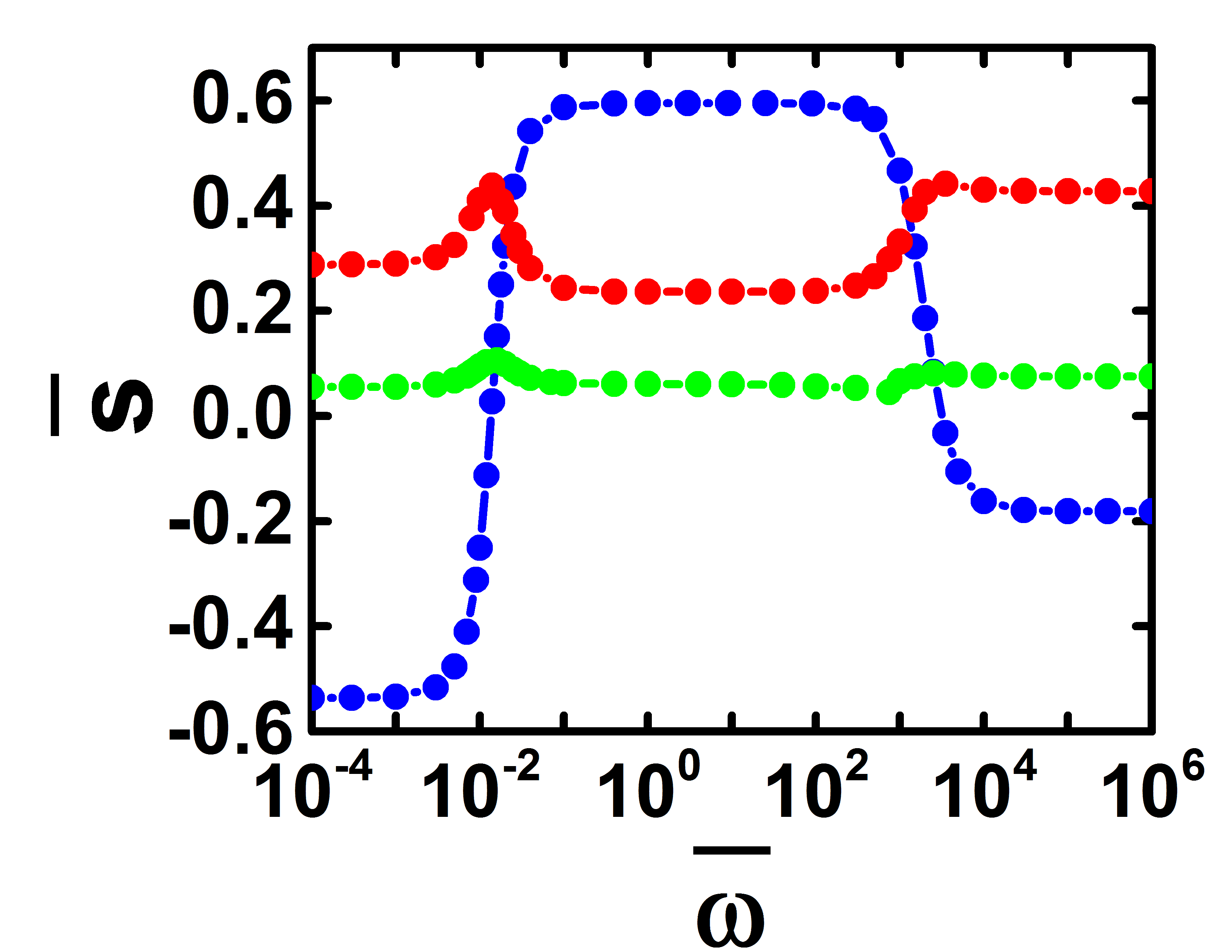}
                  \caption{}
             \end{subfigure}     
     \begin{subfigure}[b]{0.46\linewidth}
      \includegraphics[width=\linewidth]{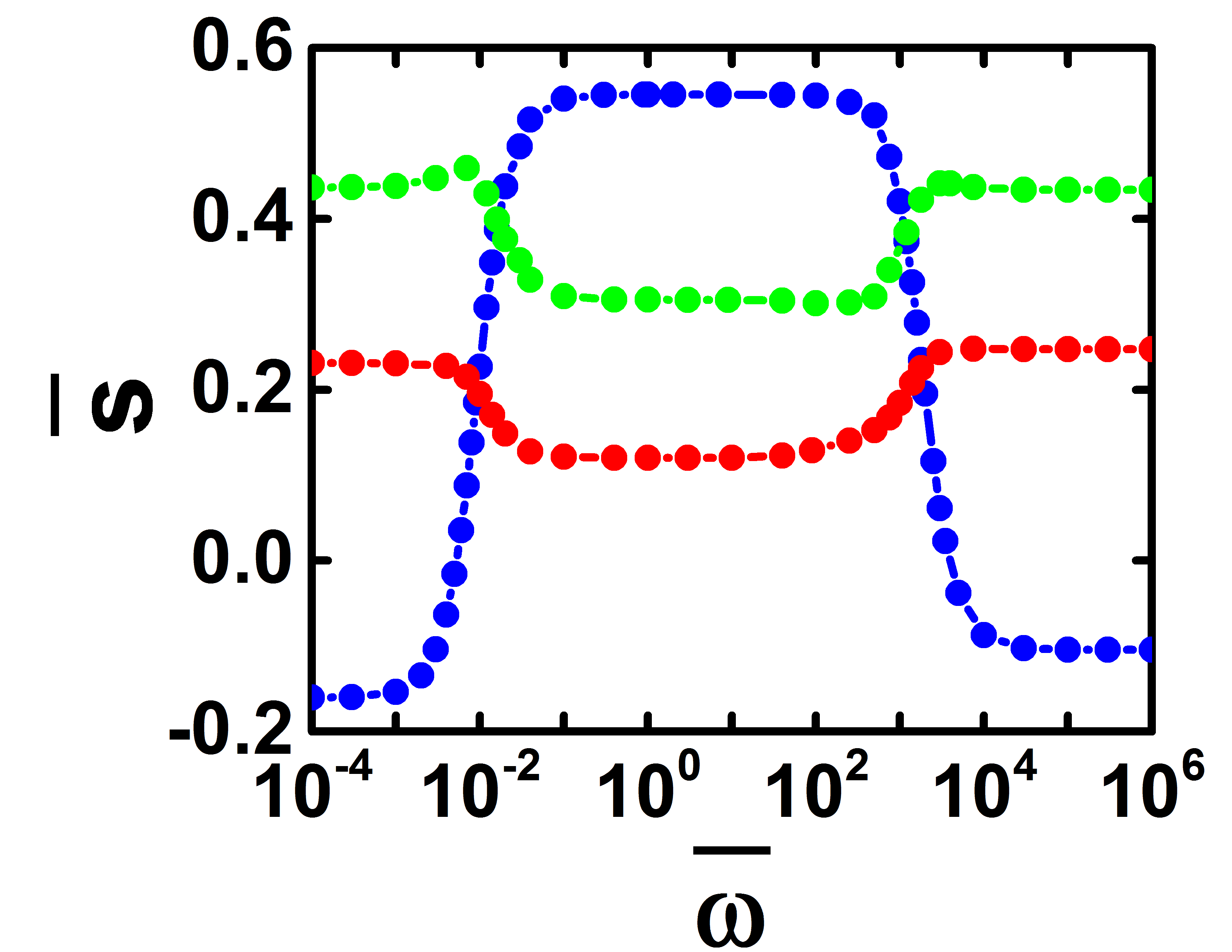}
                 \caption{}
             \end{subfigure}
             \hspace{0.12cm}
             \begin{subfigure}[b]{0.46\linewidth}
    \includegraphics[width=\linewidth]{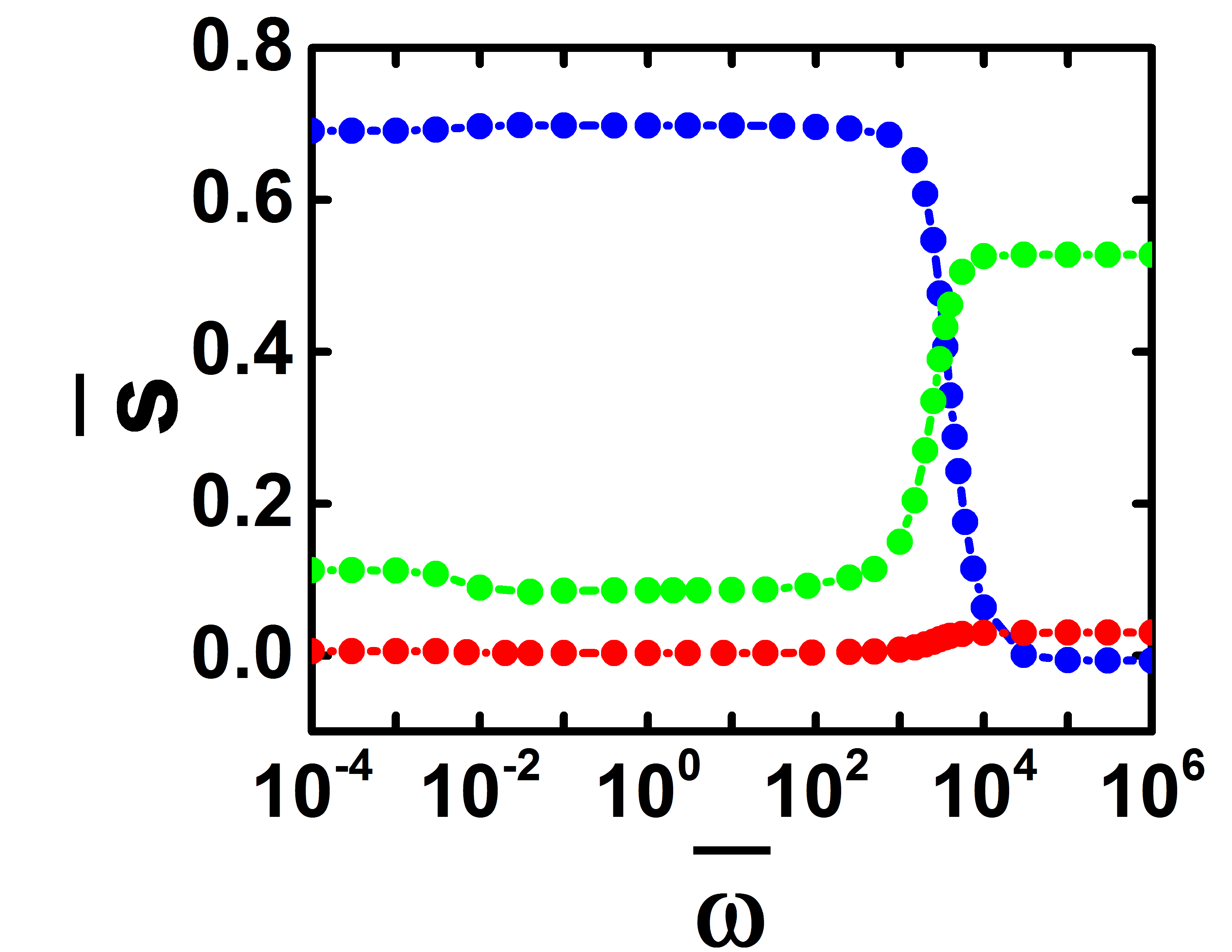}
                 \caption{}
             \end{subfigure}             
    \caption{Variation of the amplitudes of shape deformation modes (enthalpic approach) in mixed AC electric field with frequency: (a)-(c) $\sigma_r$=0.003, and (d)-(f) $\sigma_r$=280 for $\bar{C}_m$=125, $\bar{G}_m$=0, Ca=71033, $\mu_r=\epsilon_r$=1, $\bar{\gamma}_{init,0}=668$, $\bar{\Delta}=0.2$. $\bar{f}$ ranges 0.1, 1, 10 left to right. \textcolor{blue}{\textemdash} ($\bar{s}_2$), \textcolor{green}{\textemdash} ($\bar{s}_3$), \textcolor{red}{\textemdash} ($\bar{s}_4$)} 
     \label{MixDefFreqEnth}
    \end{figure}

\begin{figure}[tp]
         \centering
          \begin{subfigure}[b]{0.488\linewidth}
        \includegraphics[width=\linewidth]{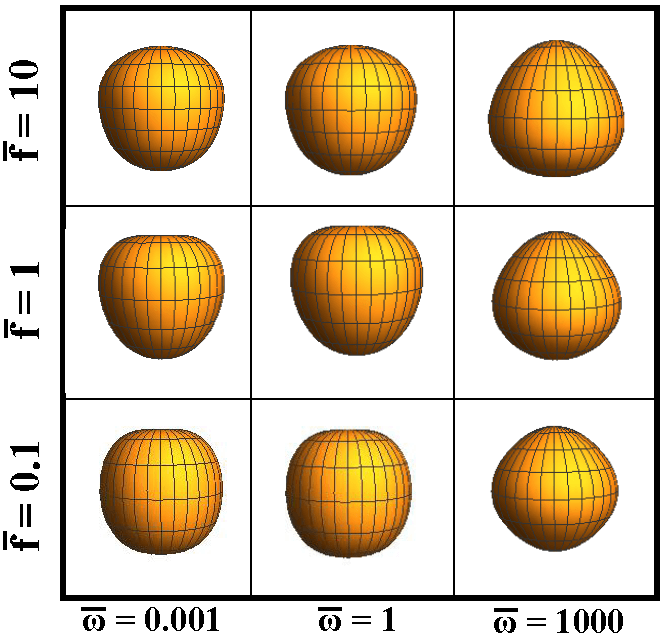}
              \caption{}
          \end{subfigure}
          \hspace{0.12cm}
          \begin{subfigure}[b]{0.45\linewidth}
         \includegraphics[width=\linewidth]{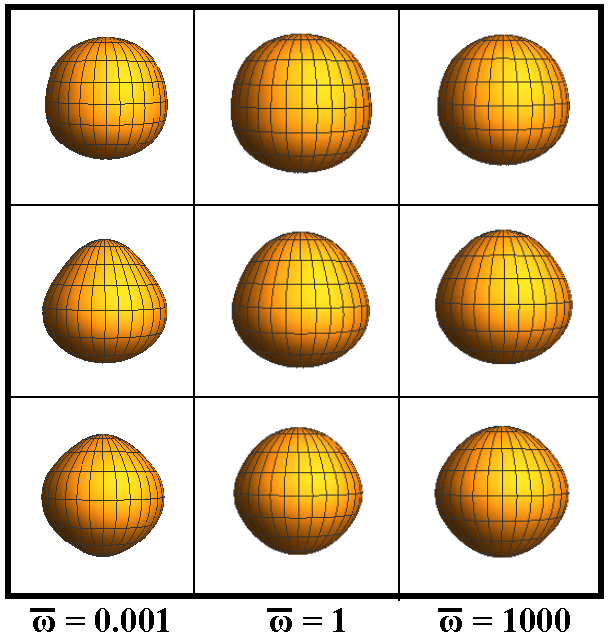}
              \caption{}
          \end{subfigure} 
     \caption{Vesicle shape transition in mixed AC electric field in entropic regime with frequency for (a) $\sigma_r=0.003$, (b) $\sigma_r=280$ at $\bar{f}=0.1, 1.0, 10$ ($\bar{C}_m=125, \bar{G}_m=0, Ca=71033, \bar{\Delta}$=0.2, $\bar{\gamma}_{init,0}$=668)} 
          \label{MixShapeEnt1}
     \end{figure}

   Figures \ref{MixDefFreqEnt} and \ref{MixDefFreqEnth} show the variation of the deformation amplitudes $\bar{s}_2,\bar{s}_3,\bar{s}_4$ with frequency in the entropic  and enthalpic regimes. A clear presence of the $\bar{s}_3$ amplitude is seen indicating that asymmetric shapes are admitted. Depending upon the frequency regime and the conductivity ratio, the three modes have varying magnitudes. The dominance of $P_2$ mode ($\bar{s}_2$) in the deformation increases as the $\bar f$ increases, conforming to the known results of spheroidal deformation in a uniform electric field. Figures \ref{MixShapeEnt1} corroborates these findings, wherein the shapes of the vesicles are shown and asymmetry is seen to increase as $\bar f$ takes intermediate values.  
   
\begin{figure}[tp]
          \centering
           \begin{subfigure}[b]{0.49\linewidth}
      \includegraphics[width=\linewidth]{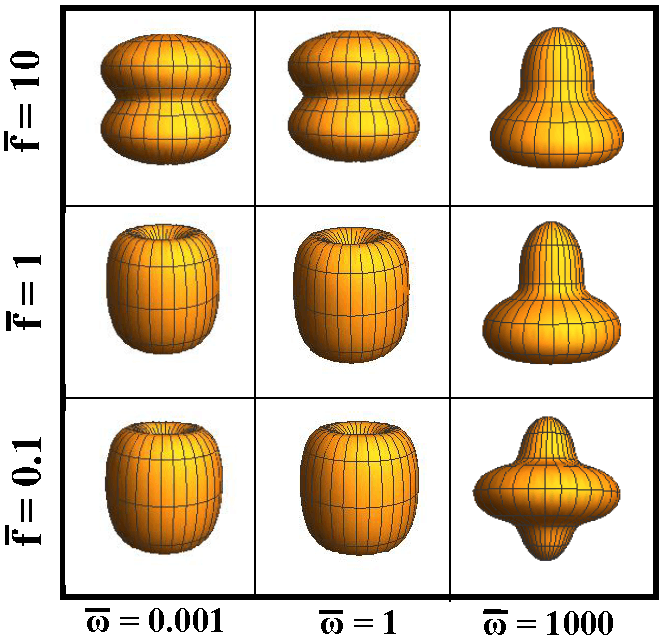}
               \caption{}
           \end{subfigure}
           \hspace{0.12cm}
           \begin{subfigure}[b]{0.45\linewidth}
      \includegraphics[width=\linewidth]{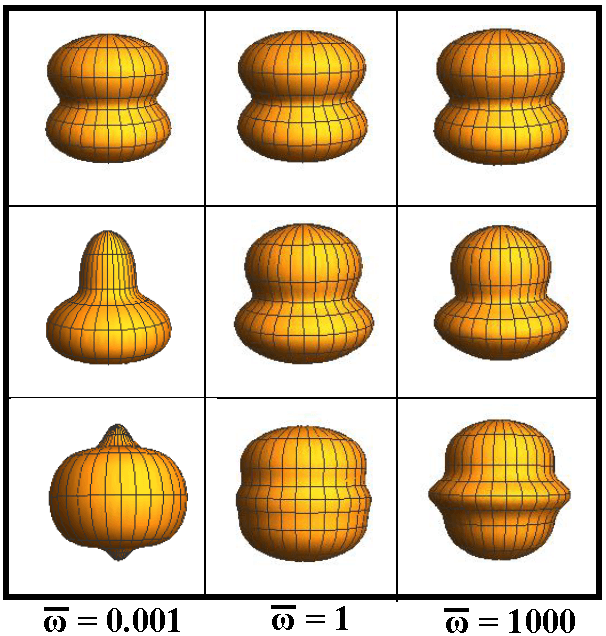}
               \caption{}
           \end{subfigure} 
     \caption{Vesicle shape transition in mixed AC electric field in enthalpic regime with frequency for (a) $\sigma_r=0.003$, (b) $\sigma_r=280$ at $\bar{f}=0.1, 1.0, 10$ ($\bar{C}_m=125, \bar{G}_m=0, Ca=71033, \bar{\Delta}$=0.2, $\bar{\gamma}_{init,0}$=668)} 
                       \label{MixShapeEnth1}
      \end{figure}

   In the enthalpic regime, figure \ref{MixShapeEnth1} shows that  for $\sigma_r<1$ asymmetric modes are stable only at high frequencies. On the contrary, for $\sigma_r>1$, asymmetric shapes are seen only at $\bar f=1$ while near quadrupole and near uniform fields admit symmetric shapes at all frequencies.\\ 


 \subsection{Comments on dielectrophoresis in quadrupole field and the Maxwell stress approach}   
 
 An important question of relevance is the stability (with respect to position) of the vesicle in side the quadrupole field. Vesicles are known to undergo dielectrophoresis in non-uniform electric fields. Typically calculation of dielectrophoretic velocity has two parts, the total electrostatic force acting on the vesicle and the drag force on a moving vesicle. Dielectrophoresis has been studied using a rigorous Maxwell stress approach \cite{sauer1983forces, Rosales2005, Kumar2012, Jarro2007, Kurgan2011, Wang1997} by integrating the electric stress tensor  over the spherical surface of a particle in a leaky media under a slightly nonuniform electric field.  On the other hand the more popular dipole moment method\cite{jones1990, MGreen2003} considers the force exerted by the gradient of the applied field on the polarization vector induced in a spherical particle assuming it is subjected to the electric field at the center of mass. The dipole moment method, although not exact, is more commonly used due to its applicability to arbitrary fields. On the contrary, for composite, concentric spherical systems gets complicated and the Maxwell stress method might be more straightforward, \textit{ass used in this work}.\\
   
     The advantage of Maxwell stress method over the dipole moment method is best demonstrated in the case of  liquid drops, wherein non uniform electric field  has been extensively studied to understand their translation, deformation, levitation, breakup etc \cite{feng1996,kang2003,kang2007,rochish2012,SD2012,SD2013,suman2016,suman2017}. The analysis shows that the tangential electric stresses in a leaky dielectric system (both the drop and the fluid medium in which it is suspended are leaky dielectrics), leads to circulation inside the drop, as well as in the outer fluid medium. This alters the drag on the particle. Therefore to obtain the dielectrophoretic velocity, the problem has to be solved using the Maxwell stress approach. A simplified approach, where the DEP force is calculated from the dipole moment method, and the drag say the Hadamard-Rybczynski  equation, can lead to erroneous results.\\
   
   Similar to a drop interface, the vesicle interface, which is typically a bilayer membrane, is deformable as discussed by \cite{powers2010dynamics} and demonstrated in this work.  There is no apriori reason to not expect this coupling in a vesicle (or a biological cell) as well, since the electric stresses at the interface can in-principle drive fluid motion in the inner and outer side of the vesicle. Thus although the scientific community working on dielectrophoresis of cells and vesicles has been using the net dielectrophoretic force calculated by the dipole moment method, and the net drag as that given by assumption of rigid body hydrodynaimcs, this assumption of a rigid body drag cannot be apriori assumed, and if true, should be shown rigorously. \\
    
   It is therefore important to self consistently solve the electrohydrodynamics problem using the Maxwell-stress and low Re approach. A self consistent calculation, presented in the supplementary material, on the axisymmetric quadrupole electric field yields the following results,
   \begin{enumerate}
   \item The dipole moment method and the Maxwell stress method to describe the dielectrophoresis of vesicles are indeed identical for Quadrupole fields and yield the same dielectrophoretic force.
   \item The correct hydrodynamics in such a case, specifically the drag on a vesicle, is found to obey the drag on a rigid sphere (Stokes drag) thereby {\em vindicating the often used, but not explicitly proved, assumption, typically used in the literature}.
   \end{enumerate}

\section{Concluding remarks}
\label{sec:conclude}
 A systematic analysis of vesicle dielectrophoresis and deformation in non-uniform AC electric field is presented. The deformation of a vesicle in quadrupole field shows a variety of shapes such as cuboid and rhomboid, significantly different that the spheroids seen in uniform field, and these shapes depend upon the regime, entropic or enthalpic, as well as the conductivity ratio and the applied frequency. \\
     
\begin{figure}[tbp]
             \centering
         \begin{subfigure}[b]{0.48\linewidth}
     \includegraphics[width=\linewidth]{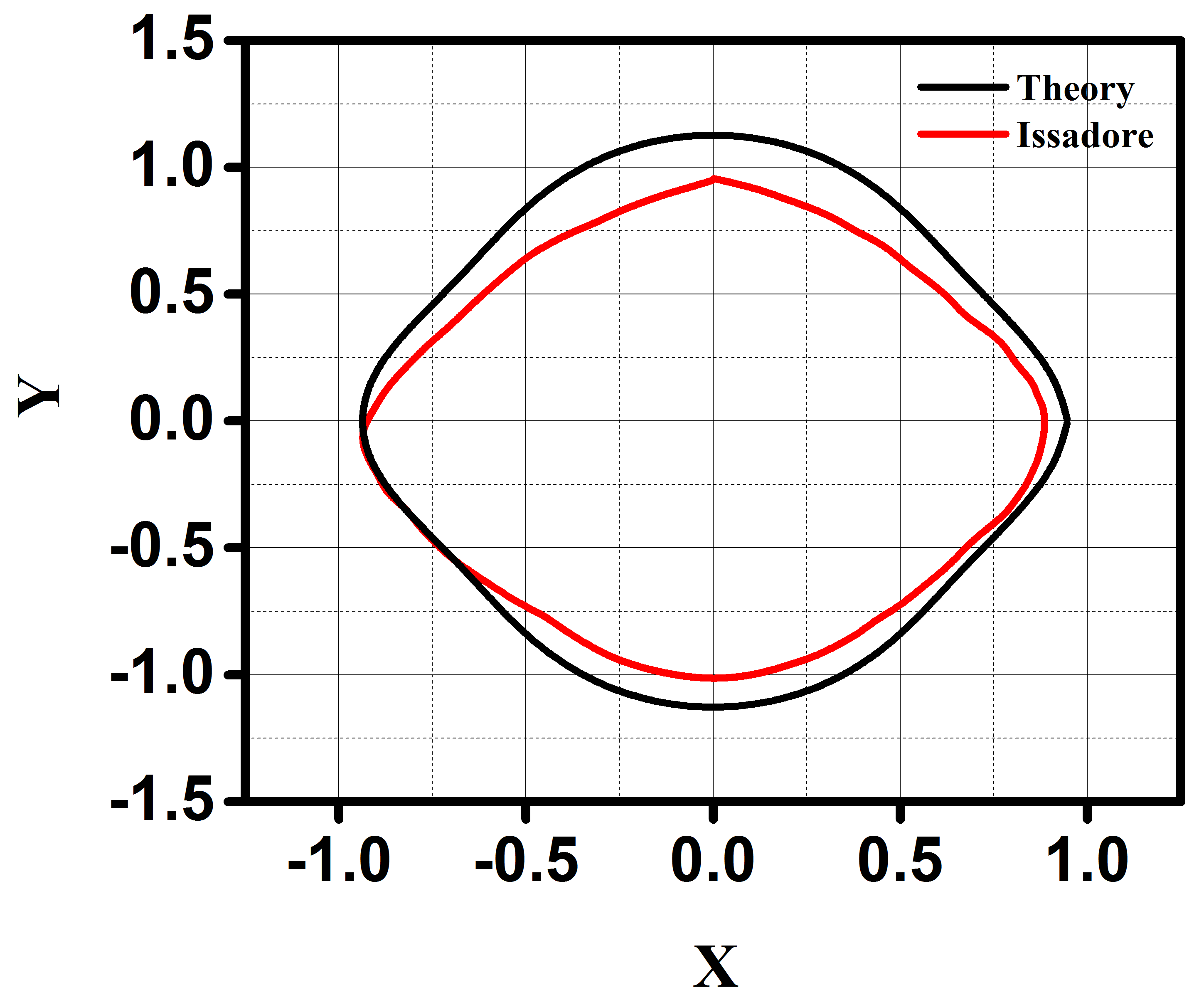}
              \caption{}
          \end{subfigure}
      \hspace{0.12cm}
     \begin{subfigure}[b]{0.48\linewidth}
     \includegraphics[width=\linewidth]{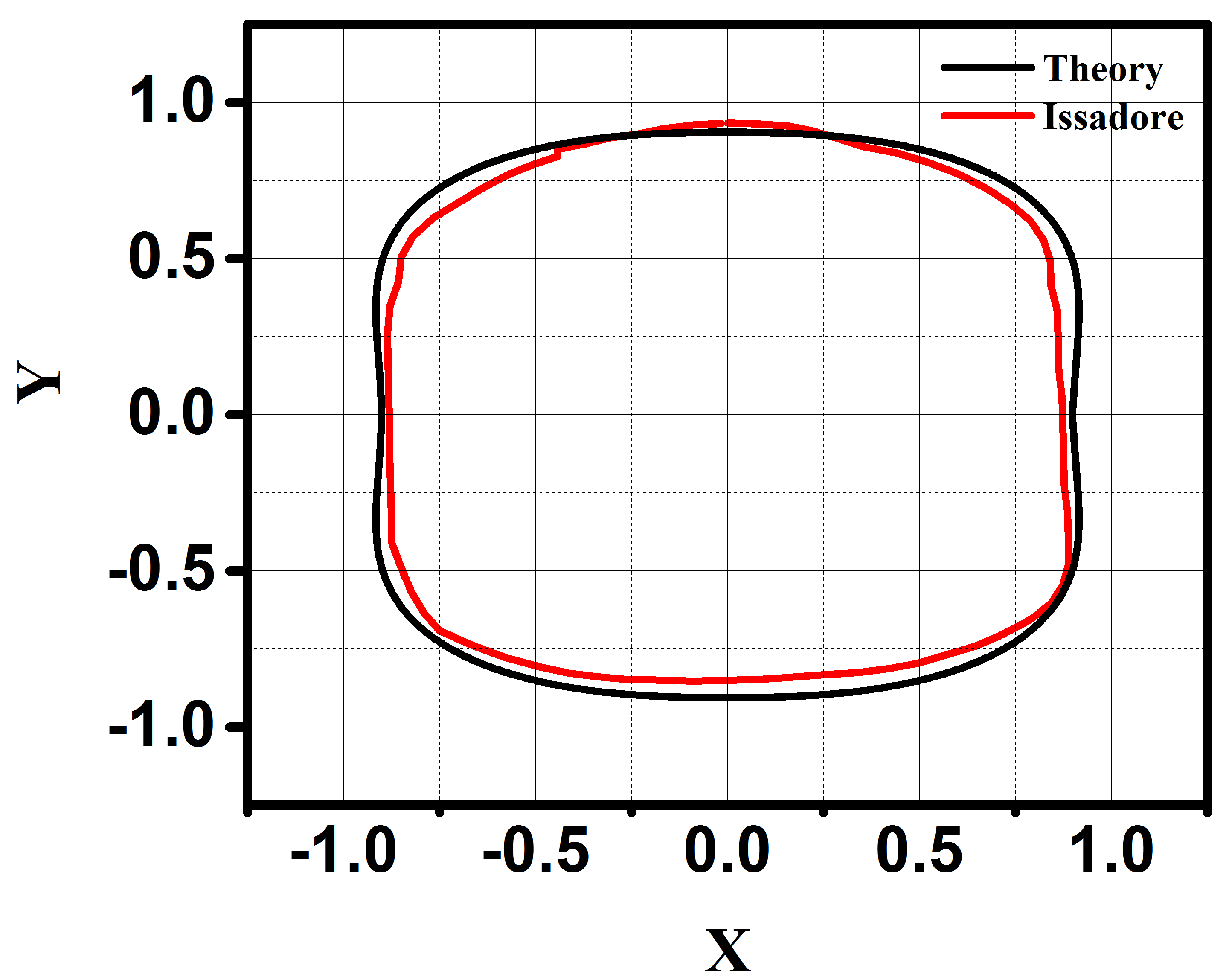}
        \caption{}
      \end{subfigure} 
  \caption{Comparison of prediction of the present theory with that of experiments in  \cite{issadore2010}The dimensional values from \cite{issadore2010} are: $R_o=5 \mu m, \sigma_{in}=0.1 S/m, \sigma_{ex}=0.001 S/m, E_o=10^4 V/m, C_{m}=10^{-2}F/m^2$. The corresponding non-dimensional parameters in our model are: For figure (a): $\sigma_{r}=100, \bar{C}_{m}=C_{m} R_o/\epsilon_{ex}=70, Ca=\epsilon_{ex}R_o^3 E_o^2/\kappa_b=86.08, \bar{\omega}=\omega \epsilon_{ex}/\sigma_{ex}=0.014$, and for figure (b)(on reversing fluid conductivity on each side and increasing frequency): $\sigma_{r}=0.01, \bar{C}_{m}=C_{m} R_o/\epsilon_{m}= 70, Ca=\epsilon_{ex}R_o^3 E_o^2/\kappa_b=86.08, \bar{\omega}=100$  (Here we have assumed $\kappa_b=25 \kappa_B T$ and $\epsilon_{ex}=80 \epsilon_o$, excess area $\bar{\Delta}=0.035$, due to lack of data in \cite{issadore2010})}  
     \label{issadore}
   \end{figure}

   It would be appropriate to compare the experimental results in \cite{issadore2010} and \cite{webb2005} with the analysis conducted in this work while accounting for the planar(quadrupole) and 3D (octupole)  fields in their set-ups respectively as against the axisymmetric quadrupole field in the present case. The squaring of shapes is clearly observed in the experiments of \cite{issadore2010} for planar quadrupole.  Using their experimental parameters it is seen that two shapes from the enthalpic theory presented in this work (figure \ref{issadore}) are identical to their experimental shapes (figures 6(c,e) of \cite{issadore2010}).
 
   Similarly as predicted in this work, intermediate frequency squaring and high frequency near spherical vesicles can be seen in figures 3(b) and 3(c), respectively in \cite{webb2005}.  \\
   
   The electrical parameters as well as the quadrupole electrode design suggested in this work, should allow the method to be used for understanding electrodeformation of vesicles and biological cells in non uniform fields that are more commonly used in experiments and applications. The method shows that in the entropic regime, the vesicles admit higher order shapes, indicating influence of quadrupole field. In the enthalpic regime, the electric field as well as the frequency and the conductivity ratio determines the final shape of the vesicle with a given excess area and high nonlinearity in the shapes is observed.  When a uniform electric field is employed in the enthalpic regime, the shape is prolate or oblate spheroid that satisfies  the excess area constraint, and therefore an interplay of different shape modes cannot be investigated. Thus quadrupole field is the simplest axisymmetric configuration that explores the competition of  $\bar{s}_2$ and $\bar{s}_4$ deformation in determining the final shape. \\
   
   It should be noted that though highly non linear shapes at high capillary numbers are presented in this work, the electrostatics and hydrodynamics are solved on a sphere though. However, it should be mentioned that the membrane deformation leads to non-linear equations due to area incompressibility conditions even at linear order in deformation. Thus although the drag on the non-spherical shapes could be different, this should only lead to slower dynamics, while keeping the shapes similar to what are predicted in the present work. \\
 
  This work  using the more rigorous Maxwell stress tensor method, additionally corroborates that the dipole moment method with Stokes drag for a rigid sphere suffices to estimate the dielectrophoretic velocity of a vesicle in quadrupole field. 

\section*{Acknowledgements}
Authors would like to acknowledge the Department of Science and Technology, India, for financial support.

\bibliographystyle{model1-num-names}
\bibliography{referencesquad}

\begin{thebibliography}{72}
\expandafter\ifx\csname natexlab\endcsname\relax\def\natexlab#1{#1}\fi
\providecommand{\bibinfo}[2]{#2}
\ifx\xfnm\relax \def\xfnm[#1]{\unskip,\space#1}\fi
\bibitem[{Guido et~al.(2009)Guido, Jaeger, and Duschl}]{guido2009}
\bibinfo{author}{I.~Guido}, \bibinfo{author}{M.~S. Jaeger},
  \bibinfo{author}{C.~Duschl},
\newblock \bibinfo{title}{Cell deformation by dielectrophoretic fields},
\newblock \bibinfo{journal}{IFMBE Proceedings} \bibinfo{volume}{25}
  (\bibinfo{year}{2009}) \bibinfo{pages}{21--24}.
\bibitem[{Li et~al.(2011)Li, Ye, and Lam}]{li2011}
\bibinfo{author}{H.~Li}, \bibinfo{author}{T.~Ye}, \bibinfo{author}{K.~Y. Lam},
\newblock \bibinfo{title}{Qualitative and quantitative analysis of dynamic
  deformation of a cell in nonuniform alternating electric field},
\newblock \bibinfo{journal}{Journal of Applied Phys.} \bibinfo{volume}{110}
  (\bibinfo{year}{2011}) \bibinfo{pages}{104701}.
\bibitem[{Weigl et~al.(2003)Weigl, Bardell, and Cabrera}]{weigl2003}
\bibinfo{author}{B.~H. Weigl}, \bibinfo{author}{R.~L. Bardell},
  \bibinfo{author}{C.~R. Cabrera},
\newblock \bibinfo{title}{Lab-on-a-chip for drug development},
\newblock \bibinfo{journal}{Advanced Drug Delivery Reviews}
  \bibinfo{volume}{55} (\bibinfo{year}{2003}) \bibinfo{pages}{349--377}.
\bibitem[{Khoshmanesh et~al.(2011)Khoshmanesh, Kiss, Nahavandi, Evans, Cooper,
  Williams, and Wlodkowic}]{khoshmanesh2011}
\bibinfo{author}{K.~Khoshmanesh}, \bibinfo{author}{N.~Kiss},
  \bibinfo{author}{S.~Nahavandi}, \bibinfo{author}{C.~W. Evans},
  \bibinfo{author}{J.~M. Cooper}, \bibinfo{author}{D.~E. Williams},
  \bibinfo{author}{D.~Wlodkowic},
\newblock \bibinfo{title}{Trapping and imaging of micron-sized embryos using
  dielectrophoresis},
\newblock \bibinfo{journal}{Electrophoresis} \bibinfo{volume}{32}
  (\bibinfo{year}{2011}) \bibinfo{pages}{3129--3132}.
\bibitem[{Jubery et~al.(2014)Jubery, Srivastava, and Dutta}]{jubery2014}
\bibinfo{author}{T.~Z. Jubery}, \bibinfo{author}{S.~K. Srivastava},
  \bibinfo{author}{P.~Dutta},
\newblock \bibinfo{title}{Dielectrophoretic separation of bioparticles in
  microdevices: A review},
\newblock \bibinfo{journal}{Electrophoresis} \bibinfo{volume}{35}
  (\bibinfo{year}{2014}) \bibinfo{pages}{61--713}.
\bibitem[{Dey et~al.(2015)Dey, Shaik, Chakraborty, Ghosal, and
  Chakraborty}]{dey2015}
\bibinfo{author}{R.~Dey}, \bibinfo{author}{V.~A. Shaik},
  \bibinfo{author}{D.~Chakraborty}, \bibinfo{author}{S.~Ghosal},
  \bibinfo{author}{S.~Chakraborty},
\newblock \bibinfo{title}{Ac electric field-induced trapping of microparticles
  in pinched microconfinements},
\newblock \bibinfo{journal}{Langmuir} \bibinfo{volume}{31}
  (\bibinfo{year}{2015}) \bibinfo{pages}{5952--5961}.
\bibitem[{Pohl and Crane(1971)}]{pohl1971}
\bibinfo{author}{H.~A. Pohl}, \bibinfo{author}{J.~S. Crane},
\newblock \bibinfo{title}{Dielectrophoresis of cells},
\newblock \bibinfo{journal}{Biophy. J.} \bibinfo{volume}{11}
  (\bibinfo{year}{1971}) \bibinfo{pages}{711--727}.
\bibitem[{Crane and Pohl(1972)}]{crane1971}
\bibinfo{author}{J.~S. Crane}, \bibinfo{author}{H.~A. Pohl},
\newblock \bibinfo{title}{Theoretical models for cellular dielectrophoresis},
\newblock \bibinfo{journal}{J. Theor. Biol.} \bibinfo{volume}{37}
  (\bibinfo{year}{1972}) \bibinfo{pages}{15--41}.
\bibitem[{Gascoyne et~al.(1997)Gascoyne, Wang, Huang, and
  Becker}]{gascoyne1997}
\bibinfo{author}{P.~R.~C. Gascoyne}, \bibinfo{author}{X.~B. Wang},
  \bibinfo{author}{Y.~Huang}, \bibinfo{author}{F.~F. Becker},
\newblock \bibinfo{title}{Dielectrophoretic separation of cancer cells from
  blood},
\newblock \bibinfo{journal}{IEEE Trans. Ind. Appl.} \bibinfo{volume}{33}
  (\bibinfo{year}{1997}) \bibinfo{pages}{670--678}.
\bibitem[{Leonard et~al.(2008)Leonard, Rutan, Reeves, Pate, Walton, and
  Thompson}]{leonard2008}
\bibinfo{author}{K.~Leonard}, \bibinfo{author}{E.~Rutan},
  \bibinfo{author}{S.~Reeves}, \bibinfo{author}{A.~Pate},
  \bibinfo{author}{M.~Walton}, \bibinfo{author}{S.~Thompson},
\newblock \bibinfo{title}{Dielectrophoretic characterization of red blood
  cells},
\newblock \bibinfo{journal}{AIChE Annual meeting}  (\bibinfo{year}{2008}).
\bibitem[{Nakano et~al.(2016)Nakano, Ding, and Suehiro}]{nakano2016}
\bibinfo{author}{M.~Nakano}, \bibinfo{author}{Z.~Ding},
  \bibinfo{author}{J.~Suehiro},
\newblock \bibinfo{title}{Dielectrophoresis and dielectrophoretic impedance
  detection of adenovirus and rotavirus},
\newblock \bibinfo{journal}{Japanese Journal of Applied Physics}
  \bibinfo{volume}{55} (\bibinfo{year}{2016}) \bibinfo{pages}{017001}.
\bibitem[{Chou et~al.(2002)Chou, Tegenfeldt, Bakajin, Chan, and Cox}]{chou2002}
\bibinfo{author}{C.~F. Chou}, \bibinfo{author}{J.~O. Tegenfeldt},
  \bibinfo{author}{O.~Bakajin}, \bibinfo{author}{S.~S. Chan},
  \bibinfo{author}{E.~C. Cox},
\newblock \bibinfo{title}{Electrodeless dielectrophoresis of single and double
  stranded dna},
\newblock \bibinfo{journal}{Biophy. J.} \bibinfo{volume}{83}
  (\bibinfo{year}{2002}) \bibinfo{pages}{2170--2179}.
\bibitem[{Tuukkanen et~al.(2005)Tuukkanen, Kuzyk, Toppari, Hytonen, Ihalainen,
  and Torma}]{tuukkanen2005}
\bibinfo{author}{S.~Tuukkanen}, \bibinfo{author}{A.~Kuzyk},
  \bibinfo{author}{J.~J. Toppari}, \bibinfo{author}{V.~P. Hytonen},
  \bibinfo{author}{T.~Ihalainen}, \bibinfo{author}{P.~Torma},
\newblock \bibinfo{title}{Dielectrophorisis of nanoscale double-stranded dna
  and humidity effects on its electrical conductivity},
\newblock \bibinfo{journal}{Appl. Phys. Lett.} \bibinfo{volume}{87}
  (\bibinfo{year}{2005}) \bibinfo{pages}{183102}.
\bibitem[{Nakano et~al.(2011)Nakano, Chao, Camacho-Alanis, and
  Ros}]{nakano2011}
\bibinfo{author}{A.~Nakano}, \bibinfo{author}{T.~C. Chao},
  \bibinfo{author}{F.~Camacho-Alanis}, \bibinfo{author}{A.~Ros},
\newblock \bibinfo{title}{Immunoglobulin g and bovin serum albumin streaming
  dielectrophoresis in a microfluidic device},
\newblock \bibinfo{journal}{Electrophoresis} \bibinfo{volume}{32}
  (\bibinfo{year}{2011}) \bibinfo{pages}{2314--2322}.
\bibitem[{Gagnon(2011)}]{gagnon2011}
\bibinfo{author}{Z.~R. Gagnon},
\newblock \bibinfo{title}{Cellular dielectrophoresis: Application to the
  characterization, manipulation, separation and patterning of cells},
\newblock \bibinfo{journal}{Electrophoresis} \bibinfo{volume}{32}
  (\bibinfo{year}{2011}) \bibinfo{pages}{2466--2487}.
\bibitem[{Cemazar et~al.(2013)Cemazar, Miklavcic, and Kotnik}]{jaka2013}
\bibinfo{author}{J.~Cemazar}, \bibinfo{author}{D.~Miklavcic},
  \bibinfo{author}{T.~Kotnik},
\newblock \bibinfo{title}{Microfluidic devices for manipulation, modification
  and characterization of biological cells in electric fields- a review},
\newblock \bibinfo{journal}{Journal of Microelectronics, Electronic Components
  and Materials} \bibinfo{volume}{43} (\bibinfo{year}{2013})
  \bibinfo{pages}{143--161}.
\bibitem[{Kang et~al.(2008)Kang, Li, and Kalams}]{kang2008}
\bibinfo{author}{Y.~Kang}, \bibinfo{author}{D.~Li}, \bibinfo{author}{S.~A.
  Kalams},
\newblock \bibinfo{title}{Dc dielectrophoretic separation of biological cells
  by size},
\newblock \bibinfo{journal}{Biomed Microdevices} \bibinfo{volume}{10}
  (\bibinfo{year}{2008}) \bibinfo{pages}{243--249}.
\bibitem[{Meighan et~al.(2009)Meighan, Staton, and Hayes}]{meighan2009}
\bibinfo{author}{M.~M. Meighan}, \bibinfo{author}{S.~J.~R. Staton},
  \bibinfo{author}{M.~A. Hayes},
\newblock \bibinfo{title}{Bioanalytical separation using electric field
  gradient techniques},
\newblock \bibinfo{journal}{Electrophoresis} \bibinfo{volume}{30}
  (\bibinfo{year}{2009}) \bibinfo{pages}{852--865}.
\bibitem[{Lewpiriyawong et~al.(2011)Lewpiriyawong, Kandaswamy, Yang, Ivanov,
  and Stocker}]{nuttawut2011}
\bibinfo{author}{N.~Lewpiriyawong}, \bibinfo{author}{K.~Kandaswamy},
  \bibinfo{author}{C.~Yang}, \bibinfo{author}{V.~Ivanov},
  \bibinfo{author}{R.~Stocker},
\newblock \bibinfo{title}{Microfluidic characterization and continuous
  separation of cells and particles using conducting poly(dimethyl siloxane)
  electrode induced alternating current-dielectrophoresis},
\newblock \bibinfo{journal}{Analytical chemistry} \bibinfo{volume}{83}
  (\bibinfo{year}{2011}) \bibinfo{pages}{9579--9585}.
\bibitem[{Fiedler et~al.(1998)Fiedler, Shirley, Schnelle, and
  Fuhr}]{stefan1998}
\bibinfo{author}{S.~Fiedler}, \bibinfo{author}{S.~G. Shirley},
  \bibinfo{author}{T.~Schnelle}, \bibinfo{author}{G.~Fuhr},
\newblock \bibinfo{title}{Dielectrophoretic sorting of particles and cells in a
  microsystem},
\newblock \bibinfo{journal}{Anal. Chem.} \bibinfo{volume}{70}
  (\bibinfo{year}{1998}) \bibinfo{pages}{1909--1915}.
\bibitem[{Taff and Voldman(2005)}]{brian2005}
\bibinfo{author}{B.~M. Taff}, \bibinfo{author}{J.~Voldman},
\newblock \bibinfo{title}{A scalable addressable positive dielectrophoretic
  cell-sorting array},
\newblock \bibinfo{journal}{Anal. Chem.} \bibinfo{volume}{77}
  (\bibinfo{year}{2005}) \bibinfo{pages}{7978--7983}.
\bibitem[{Braschler et~al.(2008)Braschler, Demierre, Nascimento, Silva, Oliva,
  and Renaud}]{thomas2007}
\bibinfo{author}{T.~Braschler}, \bibinfo{author}{N.~Demierre},
  \bibinfo{author}{E.~Nascimento}, \bibinfo{author}{T.~Silva},
  \bibinfo{author}{A.~G. Oliva}, \bibinfo{author}{P.~Renaud},
\newblock \bibinfo{title}{Continuous separation of cells by balanced
  dielectrophoretic force at multiple frequencies},
\newblock \bibinfo{journal}{Lab Chip} \bibinfo{volume}{8}
  (\bibinfo{year}{2008}) \bibinfo{pages}{280--286}.
\bibitem[{Reichle et~al.(1999)Reichle, Muller, Schnelle, and
  Fuhr}]{reichle1999}
\bibinfo{author}{C.~Reichle}, \bibinfo{author}{T.~Muller},
  \bibinfo{author}{T.~Schnelle}, \bibinfo{author}{G.~Fuhr},
\newblock \bibinfo{title}{Electrorotation in octopole micro cages},
\newblock \bibinfo{journal}{J. Phys. D: Appl. Phys.} \bibinfo{volume}{32}
  (\bibinfo{year}{1999}) \bibinfo{pages}{2128--2135}.
\bibitem[{Han et~al.(2013)Han, Joo, and Han}]{han2013}
\bibinfo{author}{S.~Han}, \bibinfo{author}{Y.~D. Joo}, \bibinfo{author}{K.~H.
  Han},
\newblock \bibinfo{title}{An electrorotation technique for measuring the
  dielectric properties of cells with simultaneous use of negative quadrupolar
  dielectrophoresis and electrorotation},
\newblock \bibinfo{journal}{Analyst} \bibinfo{volume}{138}
  (\bibinfo{year}{2013}) \bibinfo{pages}{1529}.
\bibitem[{Zimmermann(1982)}]{zimmermann1982}
\bibinfo{author}{U.~Zimmermann},
\newblock \bibinfo{title}{Electric field mediated fusion and related electrical
  phenomena},
\newblock \bibinfo{journal}{Biochimica et Biophysica Acta}
  \bibinfo{volume}{694} (\bibinfo{year}{1982}) \bibinfo{pages}{227--277}.
\bibitem[{Cavallaro et~al.(2012)Cavallaro, Capaccioli, and
  Carloni}]{cavallaro2012}
\bibinfo{author}{D.~Cavallaro}, \bibinfo{author}{S.~Capaccioli},
  \bibinfo{author}{V.~Carloni},
\newblock \bibinfo{title}{Targetting mechanism of cell fusion as a novel
  approach to abrogate multi-drug resistance of metastatic colon cancer},
\newblock \bibinfo{journal}{European Journal of Cancer} \bibinfo{volume}{48}
  (\bibinfo{year}{2012}).
\bibitem[{Yang et~al.(2012)Yang, Li, Weisal, Liu, and Li}]{yang2012}
\bibinfo{author}{W.~J. Yang}, \bibinfo{author}{S.~H. Li},
  \bibinfo{author}{R.~D. Weisal}, \bibinfo{author}{S.~M. Liu},
  \bibinfo{author}{R.~K. Li},
\newblock \bibinfo{title}{Cell fusion contributes to the rescue of apoptotic
  cardiomyocytes by bone marrow cells},
\newblock \bibinfo{journal}{J. Cell. Mol. Med.} \bibinfo{volume}{16}
  (\bibinfo{year}{2012}) \bibinfo{pages}{3085--3095}.
\bibitem[{Ye et~al.(2011)Ye, Li, and Lam}]{yeli2011}
\bibinfo{author}{T.~Ye}, \bibinfo{author}{H.~Li}, \bibinfo{author}{K.~Y. Lam},
\newblock \bibinfo{title}{Motion, deformation and aggregation pf two cells in a
  microchannel by dielectrophoresis},
\newblock \bibinfo{journal}{Electrophoresis} \bibinfo{volume}{32}
  (\bibinfo{year}{2011}) \bibinfo{pages}{3147--3158}.
\bibitem[{Patel and Markx(2008)}]{patel2008}
\bibinfo{author}{P.~Patel}, \bibinfo{author}{G.~H. Markx},
\newblock \bibinfo{title}{Dielectric measurement of cell death},
\newblock \bibinfo{journal}{Enzyme and Microbial Technology}
  \bibinfo{volume}{43} (\bibinfo{year}{2008}) \bibinfo{pages}{463--470}.
\bibitem[{Alshareef et~al.(2013)Alshareef, Metrakos, Perez, Azer, Yang, Yang,
  and Wang}]{alshareef2013}
\bibinfo{author}{M.~Alshareef}, \bibinfo{author}{N.~Metrakos},
  \bibinfo{author}{E.~J. Perez}, \bibinfo{author}{F.~Azer},
  \bibinfo{author}{F.~Yang}, \bibinfo{author}{X.~Yang},
  \bibinfo{author}{G.~Wang},
\newblock \bibinfo{title}{Separation of tumor cells with dielectrophoresis
  based microfluidic chip},
\newblock \bibinfo{journal}{Biomicrofluidics} \bibinfo{volume}{7}
  (\bibinfo{year}{2013}) \bibinfo{pages}{011803}.
\bibitem[{Gascoyne and Shim(2014)}]{peter2014}
\bibinfo{author}{P.~R.~C. Gascoyne}, \bibinfo{author}{S.~Shim},
\newblock \bibinfo{title}{Isolation of circulating tumor cells by
  dielectrophoresis},
\newblock \bibinfo{journal}{Cancer} \bibinfo{volume}{6} (\bibinfo{year}{2014})
  \bibinfo{pages}{545--579}.
\bibitem[{Sonnenberg et~al.(2014)Sonnenberg, Marciniak, Skowronski,
  Manouchehri, Rassenti, Ghia, Widhopf~II, Kipps, and Heller}]{sonnenberg2014}
\bibinfo{author}{A.~Sonnenberg}, \bibinfo{author}{J.~Y. Marciniak},
  \bibinfo{author}{E.~A. Skowronski}, \bibinfo{author}{S.~Manouchehri},
  \bibinfo{author}{L.~Rassenti}, \bibinfo{author}{E.~M. Ghia},
  \bibinfo{author}{G.~F. Widhopf~II}, \bibinfo{author}{T.~J. Kipps},
  \bibinfo{author}{M.~J. Heller},
\newblock \bibinfo{title}{Dielectrophoretic isolation and detection of
  cance-related circulating cell-free dna biomarkers from blood and plasma},
\newblock \bibinfo{journal}{Electrophoresis} \bibinfo{volume}{00}
  (\bibinfo{year}{2014}) \bibinfo{pages}{1--9}.
\bibitem[{Jang et~al.(2009)Jang, Huang, and Lan}]{jang2009}
\bibinfo{author}{L.~S. Jang}, \bibinfo{author}{P.~H. Huang},
  \bibinfo{author}{K.~C. Lan},
\newblock \bibinfo{title}{Single-cell trapping utilizing negative
  dielectrophoretic quadrupole and microwell electrodes},
\newblock \bibinfo{journal}{Biosensors and Bioelectronics} \bibinfo{volume}{24}
  (\bibinfo{year}{2009}) \bibinfo{pages}{3637--3644}.
\bibitem[{Wang et~al.(2013)Wang, Lan, Chen, Wang, and Jang}]{wang2013}
\bibinfo{author}{C.~C. Wang}, \bibinfo{author}{K.~C. Lan},
  \bibinfo{author}{M.~K. Chen}, \bibinfo{author}{M.~H. Wang},
  \bibinfo{author}{L.~S. Jang},
\newblock \bibinfo{title}{Adjustable trapping position for single cells using
  voltage phase-controlled method},
\newblock \bibinfo{journal}{Biosensors and Bioelectronics} \bibinfo{volume}{49}
  (\bibinfo{year}{2013}) \bibinfo{pages}{297--304}.
\bibitem[{Huang et~al.(2014)Huang, Liu, Loo, Stakenborg, and Lagae}]{huang2014}
\bibinfo{author}{C.~Huang}, \bibinfo{author}{C.~Liu}, \bibinfo{author}{J.~Loo},
  \bibinfo{author}{T.~Stakenborg}, \bibinfo{author}{L.~Lagae},
\newblock \bibinfo{title}{Single cell vibility observation in cell
  dielectrophoretic trapping on a microphip},
\newblock \bibinfo{journal}{Applied Phys. Lett.} \bibinfo{volume}{104}
  (\bibinfo{year}{2014}) \bibinfo{pages}{013703}.
\bibitem[{Shafiee et~al.(2010)Shafiee, Caldwell, and Davalos}]{hadi2010}
\bibinfo{author}{H.~Shafiee}, \bibinfo{author}{L.~Caldwell},
  \bibinfo{author}{R.~V. Davalos},
\newblock \bibinfo{title}{A microfluidic system for biological particle
  enrichment using contactless dielectrophoresis},
\newblock \bibinfo{journal}{Journal of the Association for Laboratory
  Automation} \bibinfo{volume}{15} (\bibinfo{year}{2010})
  \bibinfo{pages}{224--232}.
\bibitem[{Stoicheva and Hui(1994)}]{stoicheva1994}
\bibinfo{author}{N.~G. Stoicheva}, \bibinfo{author}{S.~W. Hui},
\newblock \bibinfo{title}{Dielectrophoresis of cell-size liposomes},
\newblock \bibinfo{journal}{Biochimica et Biophysica Acta}
  \bibinfo{volume}{1195} (\bibinfo{year}{1994}) \bibinfo{pages}{39--44}.
\bibitem[{Hadady et~al.(2015)Hadady, Montiel, Wetta, and Geiger}]{hadady2015}
\bibinfo{author}{H.~Hadady}, \bibinfo{author}{C.~Montiel},
  \bibinfo{author}{D.~Wetta}, \bibinfo{author}{E.~J. Geiger},
\newblock \bibinfo{title}{Liposome as a model for the study of high frequency
  dielectrophoresis},
\newblock \bibinfo{journal}{Electrophoresis} \bibinfo{volume}{36}
  (\bibinfo{year}{2015}) \bibinfo{pages}{1423--1428}.
\bibitem[{Froude and Zhu(2009)}]{victoria2009}
\bibinfo{author}{V.~E. Froude}, \bibinfo{author}{Y.~Zhu},
\newblock \bibinfo{title}{Dielectrophoresis of functionalized lipid unilamellar
  vesicles (liposomes) with contrasting surface constructs},
\newblock \bibinfo{journal}{J. Phys. Chem. B} \bibinfo{volume}{36}
  (\bibinfo{year}{2009}) \bibinfo{pages}{1552--1558}.
\bibitem[{Kodama et~al.(2013)Kodama, Osaki, Kawano, Kamiya, Miki, and
  Takeuchi}]{kodama2013}
\bibinfo{author}{T.~Kodama}, \bibinfo{author}{T.~Osaki},
  \bibinfo{author}{R.~Kawano}, \bibinfo{author}{K.~Kamiya},
  \bibinfo{author}{N.~Miki}, \bibinfo{author}{S.~Takeuchi},
\newblock \bibinfo{title}{Contactless catch-and-release system for giant
  liposomes based on negative dielectrophoresis},
\newblock \bibinfo{journal}{IEEE 26th International Conference on MEMS}
  (\bibinfo{year}{2013}) \bibinfo{pages}{1169--1170}.
\bibitem[{Kaler and Jones(1990)}]{jones1990}
\bibinfo{author}{K.~V. I.~S. Kaler}, \bibinfo{author}{T.~B. Jones},
\newblock \bibinfo{title}{Dielectrophoretic spectra of single cells determined
  by feedback-controlled levitation},
\newblock \bibinfo{journal}{Biophys. J.} \bibinfo{volume}{57}
  (\bibinfo{year}{1990}) \bibinfo{pages}{173--182}.
\bibitem[{Korlach et~al.(2005)Korlach, Reichle, Muller, Schnelle, and
  Webb}]{webb2005}
\bibinfo{author}{J.~Korlach}, \bibinfo{author}{C.~Reichle},
  \bibinfo{author}{T.~Muller}, \bibinfo{author}{T.~Schnelle},
  \bibinfo{author}{W.~W. Webb},
\newblock \bibinfo{title}{Trapping, deformation, and rotation of giant
  unilammelar vesicles in octode dielectrophoretic field cages},
\newblock \bibinfo{journal}{Biophys. J.} \bibinfo{volume}{8}
  (\bibinfo{year}{2005}) \bibinfo{pages}{554--562}.
\bibitem[{Desai et~al.(2009)Desai, Vahey, and Voldman}]{desai2009}
\bibinfo{author}{S.~P. Desai}, \bibinfo{author}{M.~D. Vahey},
  \bibinfo{author}{J.~Voldman},
\newblock \bibinfo{title}{Electrically addressable vesicles- tools for
  dielectrophoresis metrology},
\newblock \bibinfo{journal}{Langmuir} \bibinfo{volume}{25}
  (\bibinfo{year}{2009}) \bibinfo{pages}{3867--3875}.
\bibitem[{Vlahovska et~al.(2009)Vlahovska, Gracia, Espinoza, and
  Dimova}]{Vla2009}
\bibinfo{author}{P.~M. Vlahovska}, \bibinfo{author}{R.~S. Gracia},
  \bibinfo{author}{S.~A. Espinoza}, \bibinfo{author}{R.~Dimova},
\newblock \bibinfo{title}{Electrohydrodynamic model of vesicle deformation in
  alternating electric fields},
\newblock \bibinfo{journal}{Biophysical Journal} \bibinfo{volume}{96}
  (\bibinfo{year}{2009}) \bibinfo{pages}{4789--4803}.
\bibitem[{Yamamoto et~al.(2010)Yamamoto, Espinoza, Dimova, and
  Lipowsky}]{Dimova2010}
\bibinfo{author}{T.~Yamamoto}, \bibinfo{author}{S.~A. Espinoza},
  \bibinfo{author}{R.~Dimova}, \bibinfo{author}{R.~Lipowsky},
\newblock \bibinfo{title}{Stability of spherical vesicle in electric fields},
\newblock \bibinfo{journal}{Langmuir} \bibinfo{volume}{26}
  (\bibinfo{year}{2010}) \bibinfo{pages}{12390--12407}.
\bibitem[{Antonova et~al.(2010)Antonova, Vitkova, and Mitov}]{Antonova10}
\bibinfo{author}{K.~Antonova}, \bibinfo{author}{V.~Vitkova},
  \bibinfo{author}{M.~D. Mitov},
\newblock \bibinfo{title}{Deformation of giant vesicles in ac electric
  fields-dependence of the prolate-to-oblate transition frequency on vesicle
  radius},
\newblock \bibinfo{journal}{EPL} \bibinfo{volume}{89} (\bibinfo{year}{2010})
  \bibinfo{pages}{38004}.
\bibitem[{Peterlin(2010)}]{Peterlin10}
\bibinfo{author}{P.~Peterlin},
\newblock \bibinfo{title}{Frequency dependent electrodeformation of giant
  phospholipid vesicles in ac electric field},
\newblock \bibinfo{journal}{J Biol Phys} \bibinfo{volume}{36}
  (\bibinfo{year}{2010}) \bibinfo{pages}{339--354}.
\bibitem[{Salipante and Vlahovska(2014)}]{Vla2014}
\bibinfo{author}{P.~F. Salipante}, \bibinfo{author}{P.~M. Vlahovska},
\newblock \bibinfo{title}{Vesicle deformation in dc pulses},
\newblock \bibinfo{journal}{Soft Matter} \bibinfo{volume}{10}
  (\bibinfo{year}{2014}) \bibinfo{pages}{3386--3393}.
\bibitem[{McConnell et~al.(2013)McConnell, Miksis, and Vlahovska}]{Mc2013}
\bibinfo{author}{L.~C. McConnell}, \bibinfo{author}{M.~J. Miksis},
  \bibinfo{author}{P.~M. Vlahovska},
\newblock \bibinfo{title}{Vesicle electrohydrodynamics in dc electric fields},
\newblock \bibinfo{journal}{IMA Journal of Applied Mathematics}
  \bibinfo{volume}{78} (\bibinfo{year}{2013}) \bibinfo{pages}{797--817}.
\bibitem[{C. et~al.(2015)C., Vlahovska, and Miksis}]{Vlah2015}
\bibinfo{author}{M.~L. C.}, \bibinfo{author}{P.~M. Vlahovska},
  \bibinfo{author}{M.~J. Miksis},
\newblock \bibinfo{title}{Vesicle dynamics in uniform electric fields:squaring
  and breathing},
\newblock \bibinfo{journal}{Soft Matter} \bibinfo{volume}{11}
  (\bibinfo{year}{2015}) \bibinfo{pages}{4840--4846}.
\bibitem[{Dimova et~al.(2009)Dimova, Bezlyepkina, Jord, Knorr, Riske, Staykova,
  Vlahovska, Yamamoto, Yang, and Lipowsky}]{RDimova2009}
\bibinfo{author}{R.~Dimova}, \bibinfo{author}{N.~Bezlyepkina},
  \bibinfo{author}{M.~D. Jord}, \bibinfo{author}{R.~L. Knorr},
  \bibinfo{author}{K.~A. Riske}, \bibinfo{author}{M.~Staykova},
  \bibinfo{author}{P.~M. Vlahovska}, \bibinfo{author}{T.~Yamamoto},
  \bibinfo{author}{P.~Yang}, \bibinfo{author}{R.~Lipowsky},
\newblock \bibinfo{title}{Vesicle in electric fields: Some novel aspects of
  membrane behaviur},
\newblock \bibinfo{journal}{Soft Matter} \bibinfo{volume}{5}
  (\bibinfo{year}{2009}) \bibinfo{pages}{3201--3212}.
\bibitem[{Issadore et~al.(2010)Issadore, Franke, Brown, and
  Westervelt}]{issadore2010}
\bibinfo{author}{D.~Issadore}, \bibinfo{author}{T.~Franke},
  \bibinfo{author}{K.~A. Brown}, \bibinfo{author}{R.~M. Westervelt},
\newblock \bibinfo{title}{A microfluidic microprocessor: controlling biomimetic
  containers and cells using hybrid integrated circuit/microfluidic chips},
\newblock \bibinfo{journal}{Lab on a Chip} \bibinfo{volume}{10}
  (\bibinfo{year}{2010}) \bibinfo{pages}{2937--2943}.
\bibitem[{Sinha and Thaokar(2017)}]{PS2017}
\bibinfo{author}{P.~S. Sinha}, \bibinfo{author}{R.~M. Thaokar},
\newblock \bibinfo{title}{Electrohydrodynamics of a compound vesicle under an
  ac electric field},
\newblock \bibinfo{journal}{Journal of Phys.: Condensed Matter}
  \bibinfo{volume}{29} (\bibinfo{year}{2017}) \bibinfo{pages}{275101}.
\bibitem[{Vlahovska(2007)}]{shear2007}
\bibinfo{author}{R.~S. Vlahovska, P. M.and~Gracia},
\newblock \bibinfo{title}{Dynamics of a viscous vesicle in linear flows},
\newblock \bibinfo{journal}{Physical Review E} \bibinfo{volume}{75}
  (\bibinfo{year}{2007}) \bibinfo{pages}{016313}.
\bibitem[{Evans and Rawicz(1990)}]{evan1990}
\bibinfo{author}{E.~Evans}, \bibinfo{author}{W.~Rawicz},
\newblock \bibinfo{title}{Entropy-driven tension and bending elasticity in
  condensed-fluid membrane},
\newblock \bibinfo{journal}{Phys. Rev. Lett.} \bibinfo{volume}{64}
  (\bibinfo{year}{1990}) \bibinfo{pages}{2094--2097}.
\bibitem[{Thaokar(2016)}]{RT2016PRE}
\bibinfo{author}{R.~M. Thaokar},
\newblock \bibinfo{title}{Time-dependent electrohydrodynamics of a compressible
  viscoelastic capsule in the small-deformation limit},
\newblock \bibinfo{journal}{Physical Review E} \bibinfo{volume}{94}
  (\bibinfo{year}{2016}) \bibinfo{pages}{042607}.
\bibitem[{Deshmukh and Thaokar(2012)}]{SD2012}
\bibinfo{author}{S.~D. Deshmukh}, \bibinfo{author}{R.~M. Thaokar},
\newblock \bibinfo{title}{Deformation, breakup and motion of a perfect
  dielectric drop in a quadrupole electric field},
\newblock \bibinfo{journal}{Phys. of Fluids} \bibinfo{volume}{24}
  (\bibinfo{year}{2012}) \bibinfo{pages}{032105}.
\bibitem[{Sauer(1983)}]{sauer1983forces}
\bibinfo{author}{F.~A. Sauer},
\newblock \bibinfo{title}{Forces on suspended particles in the electromagnetic
  field},
\newblock in: \bibinfo{booktitle}{Coherent excitations in biological systems},
  \bibinfo{publisher}{Springer}, \bibinfo{year}{1983}, pp.
  \bibinfo{pages}{134--144}.
\bibitem[{Rosales and Lim(2005)}]{Rosales2005}
\bibinfo{author}{C.~Rosales}, \bibinfo{author}{K.~M. Lim},
\newblock \bibinfo{title}{Numerical comparison between maxwell stress method
  and equivalent multipole approach for calculation of the dielectrophoretic
  force in single-cell traps},
\newblock \bibinfo{journal}{Electrophoresis} \bibinfo{volume}{26}
  (\bibinfo{year}{2005}) \bibinfo{pages}{2057--2065}.
\bibitem[{Kumar and Hesketh(2012)}]{Kumar2012}
\bibinfo{author}{S.~Kumar}, \bibinfo{author}{P.~Hesketh},
\newblock \bibinfo{title}{Interpretation of ac dielectrophoretic behavior of
  tin oxide nanobelts using maxwell stress tensor approach modeling},
\newblock \bibinfo{journal}{Sensors and Actuators B} \bibinfo{volume}{161}
  (\bibinfo{year}{2012}) \bibinfo{pages}{1198--1208}.
\bibitem[{Jarro et~al.(2007)Jarro, Paul, Thomas, Crowe, Sawyer, Rose, and
  Shakesheff}]{Jarro2007}
\bibinfo{author}{A.~A. Jarro}, \bibinfo{author}{J.~Paul},
  \bibinfo{author}{D.~W.~P. Thomas}, \bibinfo{author}{J.~Crowe},
  \bibinfo{author}{N.~Sawyer}, \bibinfo{author}{F.~R.~A. Rose},
  \bibinfo{author}{K.~M. Shakesheff},
\newblock \bibinfo{title}{Direct calculation of maxwell stress tensor for
  accurate trajectory prediction during dep for 2d and 3d structures},
\newblock \bibinfo{journal}{Journal of Physics D: Applied Physics}
  \bibinfo{volume}{40} (\bibinfo{year}{2007}) \bibinfo{pages}{71--77}.
\bibitem[{Kurgan(2011)}]{Kurgan2011}
\bibinfo{author}{E.~Kurgan},
\newblock \bibinfo{title}{Comparision of different force calculation methods in
  dc dielectrophoresis},
\newblock \bibinfo{journal}{Electr. Rev.} \bibinfo{volume}{88}
  (\bibinfo{year}{2011}) \bibinfo{pages}{0033--2097}.
\bibitem[{Wang and Gascoyne(1997)}]{Wang1997}
\bibinfo{author}{X.~B. Wang}, \bibinfo{author}{P.~R.~C. Gascoyne},
\newblock \bibinfo{title}{General expressions for dielectrophoretic force and
  electrorotational torque derived using the maxwell stress tensor method.},
\newblock \bibinfo{journal}{J. Electrostat.} \bibinfo{volume}{39}
  (\bibinfo{year}{1997}) \bibinfo{pages}{277–95}.
\bibitem[{Morgan and Green(2003)}]{MGreen2003}
\bibinfo{author}{H.~Morgan}, \bibinfo{author}{N.~G. Green},
\newblock \bibinfo{title}{Ac electrokinetics: Colloids and nanoparticles},
\newblock \bibinfo{publisher}{Research Studies Press}, \bibinfo{year}{2003}.
\bibitem[{Feng(1996)}]{feng1996}
\bibinfo{author}{J.~Q. Feng},
\newblock \bibinfo{title}{Dielectrophoresis of a deformable fluid particle in a
  nonuniform electric field},
\newblock \bibinfo{journal}{Phys. Rev. E} \bibinfo{volume}{54}
  (\bibinfo{year}{1996}) \bibinfo{pages}{4438--4441}.
\bibitem[{Im and Kang(2003)}]{kang2003}
\bibinfo{author}{D.~J. Im}, \bibinfo{author}{I.~S. Kang},
\newblock \bibinfo{title}{Electrohydrodynamics of a drop under nonaxisymmetric
  electric fields},
\newblock \bibinfo{journal}{Journal of Colloid and Interface Science}
  \bibinfo{volume}{266} (\bibinfo{year}{2003}) \bibinfo{pages}{127--140}.
\bibitem[{Kim et~al.(2007)Kim, Im, and Kang}]{kang2007}
\bibinfo{author}{J.~G. Kim}, \bibinfo{author}{D.~J. Im}, \bibinfo{author}{I.~S.
  Kang},
\newblock \bibinfo{title}{Deformation and motion of a charged conducting drop
  in a dielectric liquid under a nonaxisymmetric electric fields},
\newblock \bibinfo{journal}{Journal of Colloid and Interface Science}
  \bibinfo{volume}{310} (\bibinfo{year}{2007}) \bibinfo{pages}{599--606}.
\bibitem[{Thaokar(2012)}]{rochish2012}
\bibinfo{author}{R.~M. Thaokar},
\newblock \bibinfo{title}{Dielectric and deformation of a liquid drop in a
  nonuniform, axisymmetric ac electric fields},
\newblock \bibinfo{journal}{Eur. Phys. J. E} \bibinfo{volume}{35}
  (\bibinfo{year}{2012}) \bibinfo{pages}{1--15}.
\bibitem[{Deshmukh and Thaokar(2013)}]{SD2013}
\bibinfo{author}{S.~D. Deshmukh}, \bibinfo{author}{R.~M. Thaokar},
\newblock \bibinfo{title}{Deformation and breakup of a leaky dielectric drop in
  a quadrupole electric field},
\newblock \bibinfo{journal}{J. Fluid Mech.} \bibinfo{volume}{731}
  (\bibinfo{year}{2013}) \bibinfo{pages}{713--733}.
\bibitem[{Mandal et~al.(2016)Mandal, Bandopadhyay, and Chakraborty}]{suman2016}
\bibinfo{author}{S.~Mandal}, \bibinfo{author}{A.~Bandopadhyay},
  \bibinfo{author}{S.~Chakraborty},
\newblock \bibinfo{title}{Surface charge convection and shape deformation on
  the dielectrophoretic motion of a liquid drop},
\newblock \bibinfo{journal}{Phys. Rev. E} \bibinfo{volume}{93}
  (\bibinfo{year}{2016}) \bibinfo{pages}{043127}.
\bibitem[{Mandal et~al.(2017)Mandal, Bandopadhyay, and Chakraborty}]{suman2017}
\bibinfo{author}{S.~Mandal}, \bibinfo{author}{A.~Bandopadhyay},
  \bibinfo{author}{S.~Chakraborty},
\newblock \bibinfo{title}{The effect of surface charge convection and shape
  deformation on the settling velocity of drops in nonuniform electric field},
\newblock \bibinfo{journal}{Phys. of Fluids} \bibinfo{volume}{29}
  (\bibinfo{year}{2017}) \bibinfo{pages}{012101}.
\bibitem[{Powers(2010)}]{powers2010dynamics}
\bibinfo{author}{T.~R. Powers},
\newblock \bibinfo{title}{Dynamics of filaments and membranes in a viscous
  fluid},
\newblock \bibinfo{journal}{Reviews of Modern Physics} \bibinfo{volume}{82}
  (\bibinfo{year}{2010}) \bibinfo{pages}{1607}.

\end{thebibliography}

\appendix
\section*{\LARGE Appendix}

\section{Electric potential coefficients}
\begin{align}
\ A_1=&(E_o R_o^3 (-\sigma_r - 
     I (\bar{C}_m + \epsilon_r + \sigma_r - 
        \bar{C}_m \sigma_r) \bar{\omega} + (\bar{C}_m + \epsilon_r - 
        \bar{C}_m \epsilon_r) \bar{\omega}^2 + 
     \bar{G}_m (-1 + \sigma_r  + 
        I (-1& \nonumber \\
 & + \epsilon_r) \bar{\omega})))/(2 \bar{G}_m + 2 \sigma_r + \bar{G}_m \sigma_r + 
   I (\bar{G}_m (2 + \epsilon_r) + \bar{C}_m (2 + \sigma_r) + 2 (\epsilon_r + \sigma_r)) \bar{\omega} - (2 \epsilon_r & \nonumber \\
    &+ \bar{C}_m (2 + \epsilon_r)) \bar{\omega}^2) &\\
\ A_2=&(2 R_o^5 \Lambda_o (-\bar{G}_m - 2 \sigma_r + \bar{G}_m \sigma_r + 
     I (\bar{G}_m (-1 + \epsilon_r) + \bar{C}_m (-1 + \sigma_r) - 
        2 (\epsilon_r + \sigma_r)) \bar{\omega} & \nonumber \\
          &+ (\bar{C}_m + 
        2 \epsilon_r - \bar{C}_m \epsilon_r) \bar{\omega}^2))/(3 \bar{G}_m + 
   6 \sigma_r + 2 \bar{G}_m \sigma_r + 
   I (\bar{G}_m (3 + 2 \epsilon_r) + 6 (\epsilon_r + \sigma_r) & \nonumber \\
             &+ 
      \bar{C}_m (3 + 2 \sigma_r)) \bar{\omega} - (3 \bar{C}_m + 
      2 (3 + \bar{C}_m) \epsilon_r) \bar{\omega}^2) &\\
\ B_1=&-((3 E_o (1 + I \bar{\omega}) (\bar{G}_m + I \bar{C}_m \bar{\omega})))/((2 \bar{G}_m + 2 \sigma_r + \bar{G}_m \sigma_r + 
  I (\bar{G}_m (2 + \epsilon_r) + \bar{C}_m (2 + \sigma_r) & \nonumber \\
 &+ 2 (\epsilon_r + \sigma_r)) \bar{\omega} - (2 \epsilon_r + 
     \bar{C}_m (2 + \epsilon_r)) \bar{\omega}^2)) &\\
\ B_2=&((5 \Lambda_o (1 + I \bar{\omega}) (\bar{G}_m + I \bar{C}_m \bar{\omega})))/((-3 \bar{G}_m - 6 \sigma_r - 2 \bar{G}_m \sigma_r - 
  I (\bar{G}_m (3 + 2 \epsilon_r) + 6 (\epsilon_r + \sigma_r) & \nonumber \\
   &+ 
     \bar{C}_m (3 + 2 \sigma_r)) \bar{\omega} + (3 \bar{C}_m + 
     2 (3 + \bar{C}_m) \epsilon_r) \bar{\omega}^2)) &
\end{align}

\section{Transmembrane potential}
Orthogonality of Legendre polynomial 
\begin{align}
\int_0^\pi P_n (\cos\theta) P_m (\cos\theta) \sin\theta d\theta=\frac{1}{2n+1} \delta_{mn}
\end{align}
Solution for transmembrane potential in non-dimensional form 
\begin{equation}
 \bar{V}_{m1}=-\frac{3(-I + \bar{\omega})(-I \sigma_r + \epsilon_r \bar{\omega})}{2 \bar{G}_m + 2 \sigma_r + \bar{G}_m \sigma_r + 
 I (\bar{G}_m (2 + \epsilon_r) + \bar{C}_m (2 + \sigma_r) + 
    2 (\epsilon_r + \sigma_r)) \bar{\omega} - (2 \epsilon_r + 
    \bar{C}_m (2 + \epsilon_r)) \bar{\omega}^2} 
\label{Vm1}
\end{equation}
 \begin{equation}
 \bar{V}_{m2}=\frac{10(-I + \bar{\omega})(-I \sigma_r + \
\epsilon_r \bar{\omega})}{-3 \bar{G}_m - 6 \sigma_r - 2 \bar{G}_m \sigma_r - 
 I (\bar{G}_m (3 + 2 \epsilon_r) + 6 (\epsilon_r + \sigma_r) + 
    \bar{C}_m (3 + 2 \sigma_r)) \bar{\omega} + (3 \bar{C}_m + 
    2 (3 + \bar{C}_m) \epsilon_r) \bar{\omega}^2}
  \label{Vm2}  
\end{equation}
\section{Electric Maxwell's Stress}
\begin{align}
&\tau_{r}^E=X_0 P_0(\cos\theta) + X_1 P_1(\cos\theta) + X_2 P_2(\cos\theta) + X_3 P_3(\cos\theta) + X_4 P_4(\cos\theta) & \\
 &\tau_{\theta}^e= -(3(Y_1 \cos\theta - 
5(Y_0 + Y_2 \cos 2\theta + 
Y_3 \cos 3\theta)) \sin\theta)/Y&
\end{align}
here, $X_0, X_1, X_2, X_3, X_4$ and $Y_0, Y_1, Y_2, Y_3, Y$ are coefficients of normal and tangential electric stress, respectively. These are lengthy expressions and so not provided in the appendix. $\bar{X}_0, \bar{X}_1, \bar{X}_2, \bar{X}_3, \bar{X}_4, \bar{Y}_0, \bar{Y}_1,\bar{Y}_2, \bar{Y}_3, \bar{Y}$ represents their non-dimensional form.
\section{Gegenbauer's functions}
\begin{align}
&\ G_2=\frac{1}{2}(1-\cos^2\theta) &\\
&\ G_3=\frac{1}{2}(1-\cos^2\theta) \cos\theta &\\
&\ G_4=\frac{1}{8}(1-\cos^2\theta) (5\cos^2\theta-1) &\\
&\ G_5=\frac{1}{8}(1-\cos^2\theta)(7\cos^2\theta-3)\cos\theta &
\end{align}
\section{Enthalpic uniform membrane tension}
The enthalpic tension is given by
 \begin{align}
 \gamma_u^{enth}=- \frac{\zeta_1+\zeta_2+\zeta_3}{\zeta_4}
 \end{align}
 where
\begin{align}
&\zeta_1=\frac{350 s_4 (-7 R_o^4 X_4 Y+756 s_4 Y \kappa_b+48 R_o^5 Y_3 \epsilon_{ex} \Lambda_o)}{20 \mu_{ex} + 19 \mu_{in}} &\\
& \zeta_2=\frac{297 s_2 (7 R_o^4 (X_2 Y + 2 Y_1 \epsilon_{ex}) -168 s_2 Y \kappa_b +90 R_o^5 Y_3 \epsilon_e \Lambda_o)}{32 \mu_{ex} + 23 \mu_{in}}&\\
& \zeta_3=\frac{7425 s_3 (-R_o^4 X_3 Y + 60 s_3 Y \kappa_b + 
   8 R_o^5 Y_2 \epsilon_{ex} \Lambda_o)}{85 \mu_{ex} + 76 \mu_{in}}& \\
& \zeta_4=18 R_o^2 Y \left(\frac{2450 s_4^2}{20 \mu_{ex} + 19 \mu_{in}}+\frac{462 s_2^2}{32 \mu_{ex} + 23 \mu_{in}}+\frac{4125 s_3^2}{85 \mu_{ex} + 76 \mu_{in}} \right) &   
\end{align}
Enthalpic membrane tension in non-dimensional form  
\begin{align}
 \bar{\gamma}_u^{enth} = &-((350 \bar{s}_4 (48 Ca \bar{Y}_3 + 756 \bar{s}_4 \bar{Y} - 7 Ca \bar{X}_4 \bar{Y}))/(20 + 19 \mu_r) -
    297 \bar{s}_2 (-168 \bar{s}_2 \bar{Y}+ Ca (14 \bar{Y}_{1a}& \nonumber \\
     &+ 90 \bar{Y}_3+7 \bar{X}_{2a} \bar{Y} + 7 \bar{f}^2 (2 \bar{Y}_{1b}+\bar{X}_{2b} \bar{Y}))))/(32 + 23 \mu_r) + (7425 \bar{s}_3 (60 \bar{s}_3 \bar{Y} + Ca \bar{f} (8 \bar{Y}_2 & \nonumber \\
      &- \bar{X}_3 \bar{Y})))/(
    85 + 76 \mu_r))/(18 \bar{Y} ((2450 \bar{s}_4^2)/(20 + 19 \mu_r) + (462 \bar{s}_2^2)/(32 + 23 \mu_r) & \nonumber \\
     &+ (4125 \bar{s}_3^2)/(85 + 76 \mu_r)))&
\end{align}
here all $\bar{X}'s$ and $\bar{Y}'s$ are part of electric normal and tangential stress, respectively.
\section{Non-uniform membrane tension}

\begin{align}
\ \gamma_{nu1}=&-((-64 (3 + 4 Bq) R_o^4 X_1 Y + 63 (-3 + 8 Bq) R_o^2 s_5 Y \gamma_u + 378 (-3 + 8 Bq) s_5 Y \kappa_b & \nonumber \\ &+ 192 (-3 + 8 Bq) R_o^5 (5 Y_0 - 3 Y_2) \epsilon_{ex} \Lambda_o))/(576 R_o^3 Y)&\\
\ \gamma_{nu2}=&(135 R_o^5 Y_3 \epsilon_{ex} \Lambda_o (8 (-1 + Bq) \mu_{ex} - 
 7 \mu_{in}) - 28 R_o^2 s_2 Y \gamma_u (4 (1 + 3 Bq) \mu_{ex} + \mu_{in})& \nonumber \\ & - 
  168 s_2 Y \kappa_b (4 (1 + 3 Bq) \mu_{ex} + \mu_{in}) + 7 R_o^4 (3 Y_1 \epsilon_{ex} (8 (-1 + Bq) \mu_{ex} - 7 \mu_{in}) & \nonumber \\ &
  + X_2 Y (4 \mu_{ex} + 12 Bq \mu_{ex} + \mu_{in})))/(7 R_o^3 Y (32 \mu_{ex} + 23 \mu_{in}))&\\
\ \gamma_{nu3}=&-(96 R_o^5 Y_2 \epsilon_{ex} \Lambda_o ((-25 + 16 Bq) \mu_{ex} - 24 \mu_{in}) - 8 R_o^4 X_3 Y ((5 + 24 Bq) \mu_{ex} + 2 \mu_{in})& \nonumber \\ & + R_o^2 Y \gamma_u (147 s_5 (-25 \mu_{ex} + 16 Bq \mu_{ex} - 24 \mu_{in}) + 80 s_3 (5 \mu_{ex} + 24 Bq \mu_{ex} + 2 \mu_{in}))& \nonumber \\ & + 6 Y \kappa_b (147 s_5 (-25 \mu_{ex} + 16 Bq \mu_{ex} - 24 \mu_{in}) + 80 s_3 (5 \mu_{ex} + 24 Bq \mu_{ex} + 2 \mu_{in})))/& \nonumber \\ &(8 R_o^3 Y (85 \mu_{ex} + 76 \mu_{in}))&\\
\ \gamma_{nu4}=&(7 R_o^4 X_4 Y ((6 + 40 Bq) \mu_{ex} + 3 \mu_{in}) - 126 R_o^2 s_4 Y \gamma_u ((6 + 40 Bq) \mu_{ex} + 3 \mu_{in})& \nonumber \\ & - 756 s_4 Y \kappa_b ((6 + 40 Bq) \mu_{ex} + 3 \mu_{in}) + 24 R_o^5 Y_3 \epsilon_{ex} \Lambda_o ((168 - 80 Bq) \mu_{ex} + 165 \mu_{in}))/& \nonumber \\ &(63 R_o^3 Y (20 \mu_{ex} + 19 \mu_{in}))&
\end{align}
\section{Hydrodynamic coefficients with $Bq=0$}
\begin{align}
\ p_0 = &-X_0 + (2 \gamma_u)/R_o&\\
\ C_{1i}=&(-2 R_o (C_{9i} R_o + X_1) + 4 \gamma_{nu1} + 
 3 C_{2i} \mu_{ex})/(R_o^2 (3 \mu_{ex} - 8 \mu_{in}))&\\
\ C_{2i}=&(20 (R_o X_1 - 2 \gamma_{nu1}) \mu_{in} + C_{9i} R_o^2 (3 \mu_{ex} + 12 \mu_{in}))/(30 \mu_{ex} \mu_{in})&\\
\ C_{3i}=&(-C_{10i} R_o^6 + 4 R_o^2 s_2 \gamma_u + 2 R_o^3 \gamma_{nu2} + 24 s_2 \kappa_b - R_o^4 (X_2 - 3 C_{4i} \mu_{ex} - 2 C_{4i} \mu_{in}))/& \nonumber \\
   &(R_o^6 (-\mu_{ex} - 
   6 \mu_{in}))& \\
\ C_{4i}=&(C_{10i} R_o^6 (\mu_{ex} - \mu_{in}) + 7 (-R_o^4 X_2 + 4 R_o^2 s_2 \gamma_u + 2 R_o^3 \gamma_{nu2} + 24 s_2 \kappa_b) \mu_{in})/& \nonumber \\
   &(7 R_o^4 (-3 \mu_{ex} - 2 \mu_{in}) \mu_{in})&\\
\ C_{5i}=&(-4 C_{11i} R_o^7 - 4 R_o^4 X_3 + 40 R_o^2 s_3 \gamma_u + 8 R_o^3 \gamma_{nu3} + 240 s_3 \kappa_b + C_{6i} R_o^5 (19 \mu_{ex} + 16 \mu_{in}))/& \nonumber \\
   &(R_o^7 (-13 \mu_{ex} - 32 \mu_{in}))&\\
\ C_{6i}=&(-13 C_{11i} R_o^7 \mu_{ex} - 8 (C_{11i} R_o^7 - 3 R_o^4 X_3 + 30 R_o^2 s_3 \gamma_u + 6 R_o^3 \gamma_{nu3} + 180 s_3 \kappa_b) \mu_{in})/& \nonumber \\
   &(6 R_o^5 \mu_{in} (19 \mu_{ex} + 16 \mu_{in}))&\\
\ C_{7i}=&(-5 C_{12i} R_o^8 - 5 R_o^4 X_4 + 90 R_o^2 s_4 \gamma_u + 10 R_o^3 \gamma_{nu4} + 540 s_4 \kappa_b + C_{8i} R_o^6 (33 \mu_{ex} + 30 \mu_{in}))/& \nonumber \\
   &(R_o^8 (-27 \mu_{ex} - 50 \mu_{in}))&\\
\ C_{8i}=&(C_{12i} R_o^8 (-54 \mu_{ex} - 45 \mu_{in}) + 55 (R_o^4 X_4 - 18 R_o^2 s_4 \gamma_u - 2 R_o^3 \gamma_{nu4} - 
108 s_4 \kappa_b) \mu_{in})/& \nonumber \\
   &(11 R_o^6 \mu_{in} (33 \mu_{ex} + 
   30 \mu_{in}))&\\
\ C_{9i}=&(5 (C_{11i} R_o^7 (-25 \mu_{ex} - 24 \mu_{in}) + 24 (-R_o^4 X_3 + 10 R_o^2 s_3 \gamma_u + 2 R_o^3 \gamma_{nu3} + 60 s_3 \kappa_b) \mu_{in}))/& \nonumber \\
   &(12 R_o^5 (-19 \mu_{ex} - 16 \mu_{in}))& \\
\ C_{10i}=&(7 (-C_{12i} R_o^8 (-168 \mu_{ex} - 165 \mu_{in}) (-3 \mu_{ex} - 2 \mu_{in}) + 11 \mu_{in} (R_o^4 ((-264 X_2 - 45 X_4) \mu_{ex} & \nonumber \\
   &- 30 (8 X_2 + X_4) \mu_{in}) + R_o^3 (2 (264 \gamma_{nu2} + 45 \gamma_{nu4}) \mu_{ex} + 60 (8 \gamma_{nu2} + \gamma_{nu4}) \mu_{in}) & \nonumber \\
      &+ 2 R_o^2 \gamma_u (135 s_4 (3 \mu_{ex} + 2 \mu_{in}) + 
16 s_2 (33 \mu_{ex} + 30 \mu_{in})) + 12 \kappa_b (135 s_4 (3 \mu_{ex} + 2 \mu_{in})& \nonumber \\
   & +  16 s_2 (33 \mu_{ex} + 30 \mu_{in})))))/(88 R_o^6 (-33 \mu_{ex} - 30 \mu_{in}) (-8 \mu_{ex} - 7 \mu_{in}))&\\
\ C_{11i}=&(24 (R_o^4 X_3 - 10 R_o^2 s_3 \gamma_u - 2 R_o^3 \gamma_{nu3} - 60 s_3 \kappa_b) \mu_{in})/(R_o^7 (-25 \mu_{ex} - 24 \mu_{in}))&\\
\ C_{12i}=&(165 (R_o^4 X_4 - 18 R_o^2 s_4 \gamma_u - 2 R_o^3 \gamma_{nu4} - 108 s_4 \kappa_b) \mu_{in})/(R_o^8 (-168 \mu_{ex} - 165 \mu_{in}))&\\
\ C_{1e}=&(R_o^2 (-C_{9e} - 2 C_{2e} \mu_{ex} + R_o (-2 \gamma_{nu1} + R_o (C_{9i} R_o + X_1 - 4 C_{1i} R_o \mu_{in}))))/(6 \mu_{ex})&\\
\ C_{2e}=&(C_{9e} + R_o (2 (\gamma_{nu1} + 3 C_{2i} \mu_{ex}) - 
R_o (C_{9i} R_o + X_1 - 2 C_{1i} R_o (3 \mu_{ex} + 2 \mu_{in}))))/(4 \mu_{ex})&\\
\ C_{3e}=&-C_{4e} R_o^2 + C_{4i} R_o^5 + C_{3i} R_o^7&\\
\ C_{4e}=&1/2 Ro^3 (5 C_{4i} + 7 C_{3i} R_o^2)&\\
\ C_{5e}=&-(1/(10 \mu_{ex}))
 R_o^2 (C_{11e} - C_{11i} R_o^7 - R_o^4 X_3 + 10 R_o^2 s_3 \gamma_u + 2 R_o^3 \gamma_{nu3} + 60 s_3 \kappa_b + 6 C_{6e} \mu_{ex} & \nonumber \\
    &+ 4 C_{6i} R_o^5 \mu_{in} + 8 C_{5i} R_o^7 \mu_{in})&\\
\ C_{6e}=&(C_{11e} - C_{11i} R_o^7 - R_o^4 X_3 + 10 R_o^2 s_3 \gamma_u + 2 R_o^3 \gamma_{nu3} + 60 s_3 \kappa_b + 2 C_{6i} R_o^5 (5 \mu_{ex} + 2 \mu_{in}) & \nonumber \\
    &+ 2 C_{5i} R_o^7 (5 \mu_{ex} + 4 \mu_{in}))/(4 \mu_{ex})&\\
\ C_{7e}=&-R_o^2 (C_{12e} - C_{12i} R_o^9 - R_o^5 X_4 + 18 R_o^3 s_4 \gamma_u + 2 R_o^4 \gamma_{nu4} + 108 R_o s_4 \kappa_b + 8 C_{8e} \mu_{ex} & \nonumber \\
    &+ 6 C_{8i} R_o^7 \mu_{in} + 10 C_{7i} R_o^9 \mu_{in})/(12 \mu_{ex}))&\\   
\ C_{8e}=&(C_{12e} - C_{12i} R_o^9 - R_o^5 X_4 + 18 R_o^3 s_4 \gamma_u + 2 R_o^4 \gamma_{nu4} + 108 R_o s_4 \kappa_b + 
  2 C_{8i} R_o^7 (6 \mu_{ex} + 3 \mu_{in}) & \nonumber \\
      &+ 2 C_{7i} R_o^9 (6 \mu_{ex} + 5 \mu_{in}))/(4 \mu_{ex})&\\
\ C_{9e}=&R_o (-2 \gamma_{nu1} + R_o (C_{9i} R_o + X_1 + 4 C_{1i} R_o (\mu_{ex} - \mu_{in})))&\\
\ C_{10e}=&(C_{10i} R_o^7 - 8 (C_{4i} R_o^5 + C_{3i} R_o^7 - 1/2 R_o^5 (5 C_{4i} + 7 C_{3i} R_o^2)) \mu_{ex} - R_o (4 R_o^2 s_2 \gamma_u + 2 R_o^3 \gamma_{nu2}& \nonumber \\
    & + 24 s_2 \kappa_b +  2 R_o^4 (5 C_{4i} + 7 C_{3i} R_o^2) \mu_{ex} + 6 C_{3i} R_o^6 \mu_{in} -  R_o^4 (X_2 - 2 C_{4i} \mu_{in})))/R_o^2&\\
\ C_{11e}=&C_{11i} R_o^7 + R_o^4 X_3 - 10 R_o^2 s_3 \gamma_u - 2 R_o^3 \gamma_{nu3} - 60 s_3 \kappa_b + 4 C_{5i} R_o^7 (2 \mu_{ex} - 2 \mu_{in}) + 4 C_{6i} R_o^5 (\mu_{ex}  & \nonumber \\
    &- \mu_{in})&\\
\ C_{12e}=&C_{12i} R_o^9 + R_o^5 X_4 - 18 R_o^3 s_4 \gamma_u - 2 R_o^4 \gamma_{nu4} - 108 R_o s_4 \kappa_b + 2 C_{7i} R_o^9 (5 \mu_{ex} - 5 \mu_{in}) & \nonumber \\
    &+ 2 C_{8i} R_o^7 (3 \mu_{ex} - 3 \mu_{in})&\\ \nonumber
 \end{align}    
\section{Clausius Mossotti Factor}
\begin{align}
Re[CMF]=&(-2 \sigma_r^2 + (-2 (\bar{C}_m + \epsilon_r)^2 + \bar{C}_m^2 \sigma_r + (-2 + \bar{C}_m + \bar{C}_m^2) \sigma_r^2) \bar{\omega}^2 + (\bar{C}_m (-1 + \epsilon_r) &  \nonumber \\
&- \epsilon_r) (2 \epsilon_r + \bar{C}_m (2 + \epsilon_r)) \bar{\omega}^4 + \bar{G}_m^2 (-2 + \sigma_r + \sigma_r^2 + (-2 + \epsilon_r + \epsilon_r^2) \bar{\omega}^2)  & \nonumber \\
&+ \bar{G}_m (\sigma_r^2 + \epsilon_r^2 \bar{\omega}^2 - 
 4 \sigma_r (1 + \bar{\omega}^2)))/((2 \sigma_r + 
\bar{G}_m (2 + \sigma_r))^2+ (\bar{G}_m^2 (2 + \epsilon_r)^2 & \nonumber \\
& + \bar{C}_m^2 (2 + \sigma_r)^2 + 4 \bar{G}_m (\epsilon_r^2 + 2 \sigma_r) + 4 \bar{C}_m (2 \epsilon_r + \sigma_r^2) + 4 (\epsilon_r^2 + \sigma_r^2)) \bar{\omega}^2 + (2 \epsilon_r& \nonumber \\
& + \bar{C}_m (2 + \epsilon_r))^2 \bar{\omega}^4) &
\end{align}

\newpage
\section*{\LARGE Supplementary material}

\textbf{Dielectrophoretic force and translational velocity of a vesicle in quadrupole field}\\

  A vesicle subjected to a non-uniform AC electric field, experiences a DEP force due to up-down asymmetric electric stresses. The electric traction acting on a unit area of a vesicle in the z-direction can be written as
  $F_z^E= \tau_r^E \cos\theta-\tau_\theta^E \sin\theta$.
  The total z-directional DEP force is determined by integrating the electric traction over the vesicle surface as
   \begin{align}
   \ F_{DEP}=\int_{0}^{2 \pi} \int_{0}^{\pi} F_z^E R_o^2 \sin\theta d\theta d\phi=2\pi \epsilon_{ex} R_o^3 Re[CMF] \nabla E_{\infty}^2 \label{Fdep}
   \end{align}
   Equation \ref{Fdep} represents two approaches to estimate the DEP force. The integral expression is obtained from a more fundamental approach of Maxwell stress as employed in this work. The right hand side is the classical expression for DEP force obtained by the dipole moment approach. Here Re[CMF] is the real part of Clausius Mossotti factor which depends upon the frequency of the applied electric field as well as on the electric properties of the particle and the suspending media\cite{jones1990} (Appendix-H). A positive value of Re[CMF] implies that a vesicle experiences positive DEP force and moves towards region of high electric field whereas negative Re[CMF] leads to  negative DEP which moves a vesicle to a region of low electric field.\\
   
  Classical theory \cite{jones1990} based on dipole moments suggests positive or negative DEP (movement towards region of higher or lower electric fields respectively) if  Re[CMF] is positive or negative respectively. In this work,  Re[CMF] is calculated by equating the classical expression for dielectrophoretic force to the dielectrophoretic force calculated by the Maxwell stress approach. 
   A cross over frequency for transition from negative to positive DEP is obtained by setting $Re[CMF]=0$, which gives a lower cross over (LCO) frequency
   \begin{align}
   \bar{\omega}_{LCO}=\frac{2 \sqrt{\sigma_r}}{\sqrt{3} \sqrt{
    \bar{C}_m (2 \sigma_r + \bar{C}_m (2 + \sigma_r))}}
    \label{LCO}
   \end{align}
  valid for $\sigma_r>1$ (expression for $\sigma_r<1$ is complicated and not provided here)
  whereas the upper cross over (UCO) frequency for transition from positive to negative DEP is given by
   \begin{align}
   \bar{\omega}_{UCO}=\frac{
    \sqrt{3} \sqrt{
     \bar{C}_m (2 \epsilon_r^3 + 
        \bar{C}_m^2 (\epsilon_r - \sigma_r) (2 + \sigma_r) + 
        2 \bar{C}_m (\epsilon_r - \sigma_r) (2 \epsilon_r + \
   \sigma_r))}}{\sqrt{(\bar{C}_m (-1 + \epsilon_r) - \epsilon_r) (2 \
   \epsilon_r + \bar{C}_m (2 + \epsilon_r))^2}}
   \label{UCO}
   \end{align}
    The expression for the CMF (as provided by Jones \cite{jones1990}) is identical to that obtained in the present work by the Maxwell stress approach \cite{desai2009}, although some discrepancy in the reported values of the  CMF  in literature is noticed \cite{jones1990,victoria2009}.  \\

   The solution of kinematic condition, on integration with respect to the first Legendre polynomial, over a vesicle surface, gives steady dielectrophoretic velocity of the vesicle.  
   \begin{align}
   \ U_{DEP}=&\frac{ds_1}{dt}=\frac{2 R_o (X_1 Y+6 R_o (-5 Y_0 + 3 Y_2) \epsilon_{ex} \Lambda_o)}{9 Y \mu_{ex}} &\label{VELdim}
   \end{align} 
  Substitution of electric stress coefficients (all $X's$ and $Y's$) and non-dimensionalization of equation \ref{VELdim} shows that the DEP velocity varies linearly with $\bar{f}$ (a dimensionless factor described in the section below). The DEP velocity shows dependency on conductivity as well as permittivity ratio along with the membrane properties $C_m, G_m$. \\  
   
\textbf{Dielectrophoresis}\\
  
  \begin{figure}[tp]
            \centering
             \hspace{0.12cm}
           \begin{subfigure}[b]{0.46\linewidth}       \includegraphics[width=\linewidth]{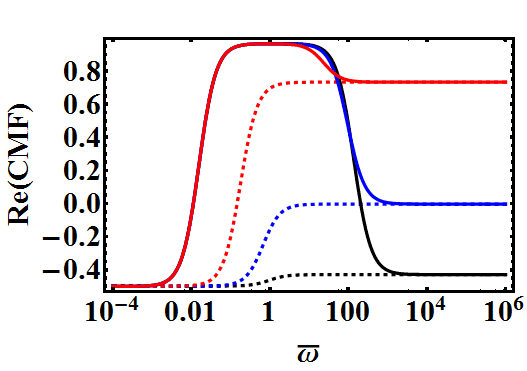}
                \caption{}
              \end{subfigure}
             \hspace{0.12cm}
         \begin{subfigure}[b]{0.46\linewidth}
         \includegraphics[width=\linewidth]{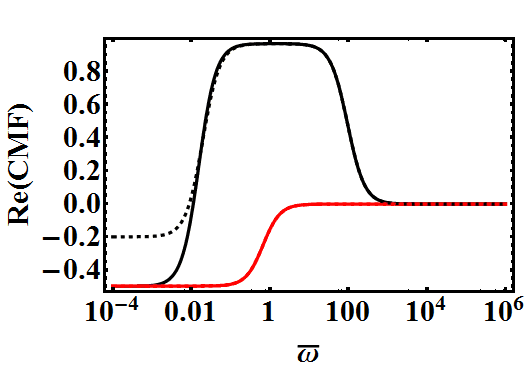}
            \caption{}
             \end{subfigure}          
   \caption{Re(CMF) vs frequency plot: (a) Role of permittivity ratio when $\sigma_r=280$(solid) and  $\sigma_r=0.003$(dashed) ($\bar{C}_m=125, \bar{G}_m=0$), \textemdash ($\epsilon_r=0.1$), \textcolor{blue}{\textemdash} ($\epsilon_r=1$), \textcolor{red}{\textemdash} ($\epsilon_r=10$) (b) Role of membrane conductance ($\bar{C}_m=125,\epsilon_r=1, \bar{G}_m=0$(solid),$\bar{G}_m=0.5$(dashed)), \textemdash ($\sigma_r=280$), \textcolor{red}{\textemdash} ($\sigma_r=0.003$)} 
             \label{CMF}
  \end{figure}  
  \begin{figure}[tp]
            \centering
        \begin{subfigure}[b]{0.45\linewidth}
        \includegraphics[width=\linewidth]{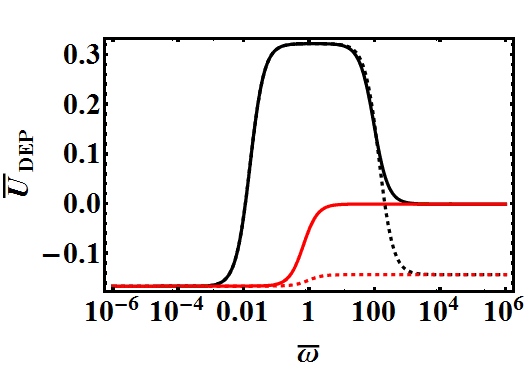}
                    \caption{}
           \end{subfigure}
          \hspace{0.12cm}
       \begin{subfigure}[b]{0.45\linewidth}
       \includegraphics[width=\linewidth]{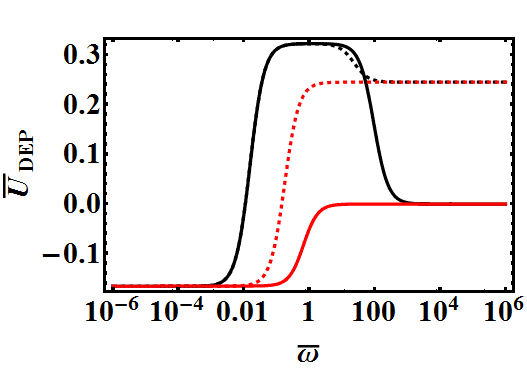}
                 \caption{}
        \end{subfigure}
             \hspace{0.12cm}
      \begin{subfigure}[b]{0.45\linewidth}
        \includegraphics[width=\linewidth]{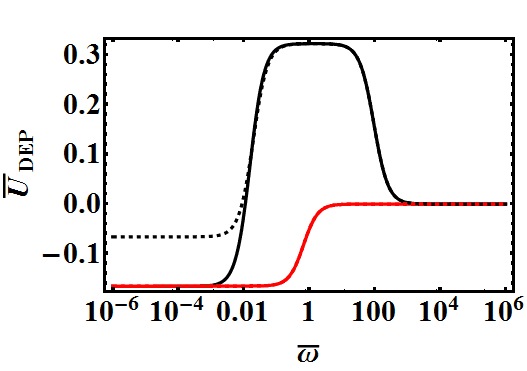}
                 \caption{}
             \end{subfigure} 
  \caption{DEP velocity vs frequency:(a) Role of low permittivity ratio when $\sigma_r$=280 (black) and $\sigma_r$=0.003 (red) ($\epsilon_r=0.1$ (dashed), $\epsilon_r=1$ (solid), $\bar{C}_m=125,\bar{f}=0.5,\bar{G}_m=0$) (b) Role of high permittivity ratio when $\sigma_r$=280 (black) and $\sigma_r$=0.003 (red) ($\epsilon_r=10$ (dashed), $\epsilon_r=1$ (solid), $\bar{C}_m=125,\bar{f}=0.5,\bar{G}_m=0$) (c) Role of membrane conductance when $\sigma_r$=280 (black) and $\sigma_r$=0.003 (red) ($\bar{C}_m=125,\epsilon_r=1, \bar{f}=0.5, \bar{G}_m=0$(solid), $\bar{G}_m=0.5$(dashed))} 
             \label{VEL}
  \end{figure}
   The expression for Re[CMF] indicates that in the low frequency regime, a vesicle with a non-conducting membrane shows a  negative value essentially due to a vesicle acting as a dielectric drop in a conducting fluid. In the intermediate frequency regime, the vesicle acts as a leaky dielectric drop when $t_{cap}^{-1}<\omega<t_{MW}^{-1}$ or $t_{ex}^{-1}$(which is =1), whichever greater. In this range, when $\sigma_r<1$, that is outer medium more conducting, a drop shows negative dielectrophoresis, whereas for $\sigma_r>1$ positive dielectrophoresis is seen.  At very high frequencies, the membrane is uncharged, and the vesicle behaves as a perfect dielectric inner fluid in a perfect dielectric outer fluid. The response is then governed by $\epsilon_r$. When the outer medium has more permittivity, it shows negative dielectrophoresis, whereas when the inner fluid has more permittivity, positive dielectrophoresis can be seen. 
  These features are demonstrated in figure \ref{CMF}. \\

  Figure \ref{CMF}a shows the effect of dielectric constant ratio $\epsilon_r$, on Re[CMF]. It is seen that for $\sigma_r>1$ there is an upper and a lower cross over frequency for $\epsilon_r<1$. However, for $\epsilon_r>1$ only lower cross over frequency is observed. This  shows that a vesicle undergoes a transition from negative to postive DEP  at low frequencies, whereas at very high frequencies the Re[CMF] and thereby the dielectrophoretic velocity tend to zero (figure \ref{VEL}a). While LCO frequency is independent of $\epsilon_r$ when $\sigma_r>1$, and accurately predicted by equation \ref{LCO} the LCO frequency has $\epsilon_r$ dependency  for $\sigma_r < 1$. The UCO frequency has $\epsilon_r$ dependency for all values of $\sigma_r$ and this is predicted well by equation \ref{UCO}. When $\sigma_r<1$ (figure \ref{CMF}a) there is no upper and lower cross over frequency for $\epsilon_r\le 1$ and the vesicle always shows negative DEP in the entire frequency range (figure \ref{VEL}a). For $\epsilon_r>1$ there is a LCO frequency which results in negative to positive DEP with an increase in frequency. \\

  Figure \ref{CMF}b shows the effect of membrane conductivity on the Re[CMF]. It indicates that the membrane conductivity predominantly affects the low frequency behavior. For sufficiently high membrane conductance, a vesicle can behave like a drop in DC field in the low frequency regime, thereby exhibiting negative dielectrophoresis for $\sigma_r<1$ and positive dielectrophoresis for $\sigma_r>1$. \\
  
  \begin{figure}[tbp]
          \centering
           \begin{subfigure}[b]{0.45\linewidth}
     \includegraphics[width=\linewidth]{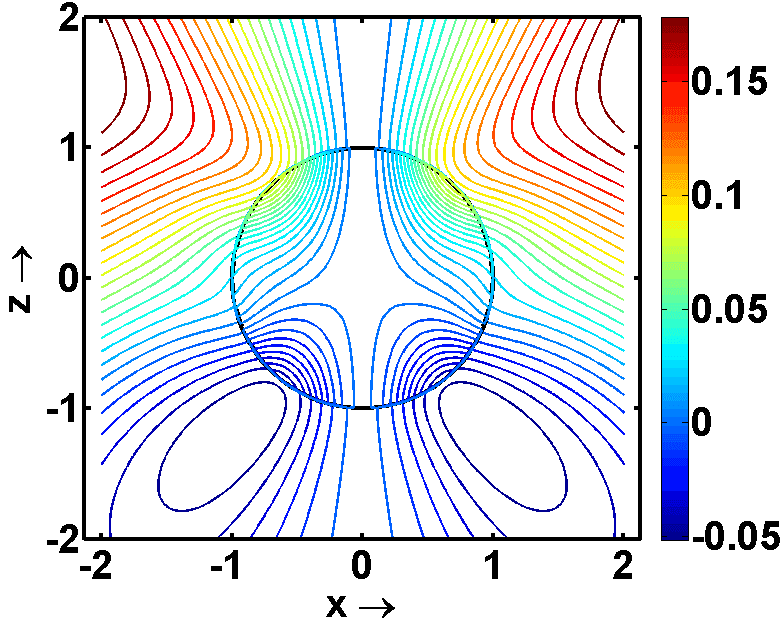}
       \caption{}
     \end{subfigure} 
           \hspace{0.12cm}
    \begin{subfigure}[b]{0.45\linewidth}
      \includegraphics[width=\linewidth]{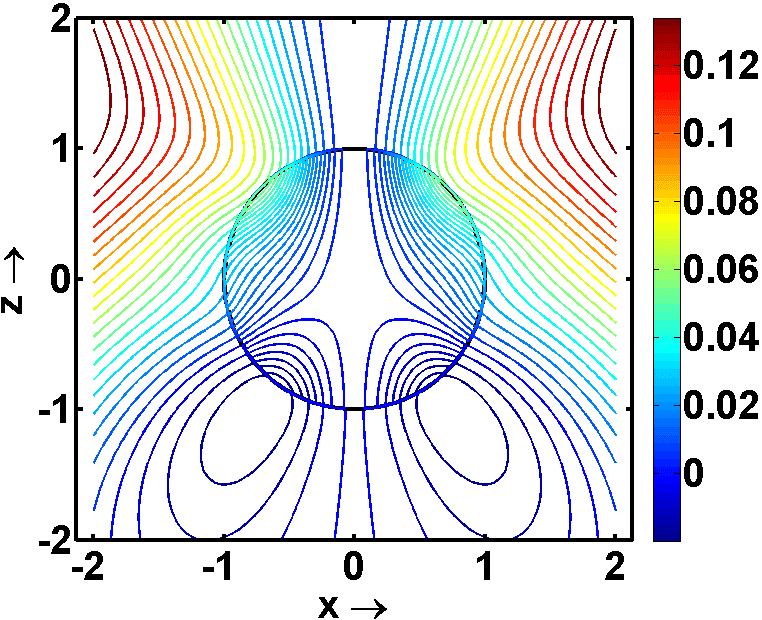}
       \caption{}
    \end{subfigure}
    \hspace{0.12cm}
     \begin{subfigure}[b]{0.45\linewidth}
     \includegraphics[width=\linewidth]{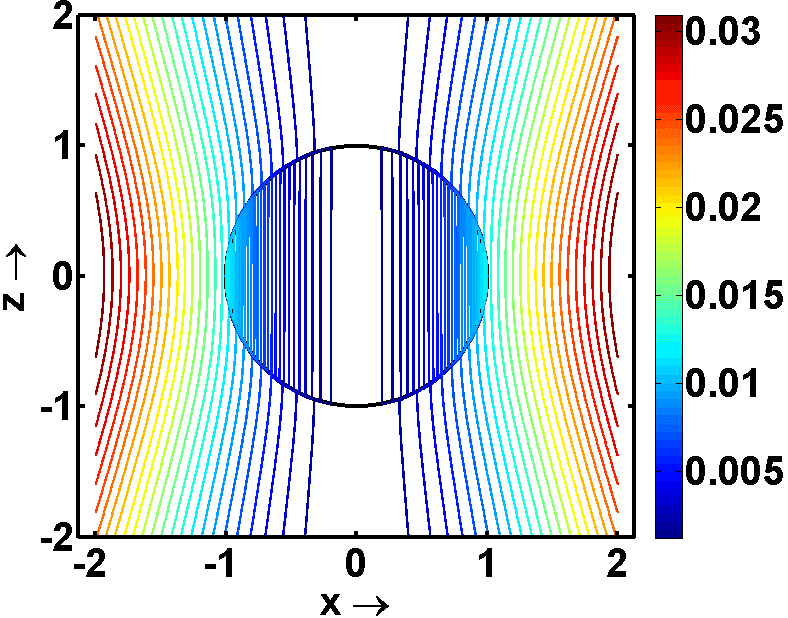}
      \caption{}
     \end{subfigure}                        
  \caption{Effect of viscosity ratio on the velocity profile for a spherical vesicle undergoing dielectrophoresis (a) $\mu_r=0.001$, (b) $\mu_r=1$, (c) $\mu_r=1000$ ($\bar{\omega}=1, \bar{f}=0.1$)} 
    \label{CONTOUR}
  \end{figure} 
  
 \textbf{Non-dependency of dielectrophoretic velocity on viscosity ratio and Bousinesq number}\\
   
\begin{figure}[tbp]
    \centering
  \includegraphics[width=0.65\linewidth]{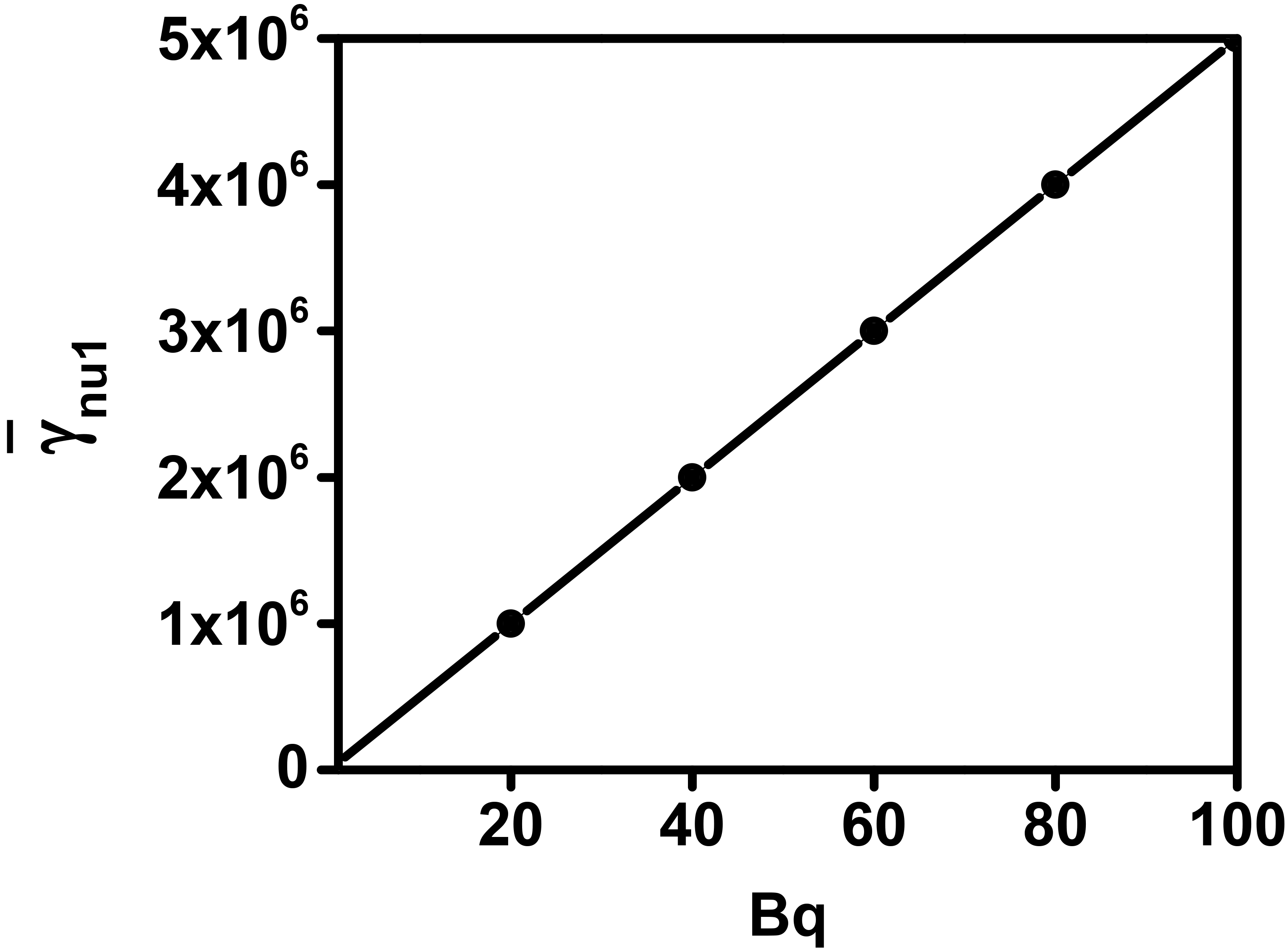}
   \caption{Dependency of non-uniform membrane tension on Boussinesq number ($\bar{C}_m=125, \bar{G}_m=0, \epsilon_r=1, \sigma_r=280, Ca=71033, \bar{f}=0.5, \bar{\omega}=1$)} 
   \label{gamma1}
  \end{figure}

It is important to self consistently solve the electrohydrodynamics problem using the Maxwell-stress and low Re approach, even for an undeformed sphere, if one is interested in knowing the dielectrophoretic velocity of a deformable particle such as a vesicle. This is essentially due to the coupling between the non-uniform tensions associated with different Legendre modes. Thus, for an undeformed sphere, the $P_2$ and $P_4$ electric stresses generated by the electric fields, need to be satisfied for an undeformed sphere. Although the normal stresses associated with the  non-uniform tensions are decoupled, the non-uniform tensions get coupled in the tangential stress balance. In an undeformed sphere, the electric stresses  are balanced  by the non-uniform tension, leading to fluid flow associated with the $3^{rd}-5^{th}$ Gegenbauer streamfunctions (these flows eventually lead to deformation which is discussed next). Although these do not contribute to the  dielectrophoretic velocity, they give rise to fluid flow, which has a dependence on the viscosity ratio (figure \ref{CONTOUR}). The non-uniform tension associated with the translational mode $\gamma_{nu1}$, leads to a rigid body like motion.  Therefore the drag resisting the dielectrophoretic motion is the same as a rigid sphere ($6 \pi \mu_{ex} R_o U$), and is independent of the viscosity ratio, although it can be calculated only after solving the coupled problem. Therefore the dielectrophoretic velocities calculated by the Maxwell stress approach in this work are in agreement with that calculated by the dipole moment method with a "rigid body" drag (equation \ref{VELdim}).  Despite the dielectrophoretic velocity showing a rigid body like drag, the velocity streamlines are quite different than that caused by translation of a rigid spherical particle, and show multiple rolls associated with the $P_2$ and $P_4$ electric stresses associated with the applied field (figure \ref{CONTOUR}). Similarly for a  membrane with surface viscosity, the membrane viscous stress is absorbed by the non-uniform tension such that the dielectrophoretic velocity remains independent of the Boussinesq number too.  The non-uniform tension though depends linearly on the Boussinesq number (figure \ref{gamma1}).\\
  
\begin{figure}[tp]
      \centering
    \includegraphics[width=0.65\linewidth]{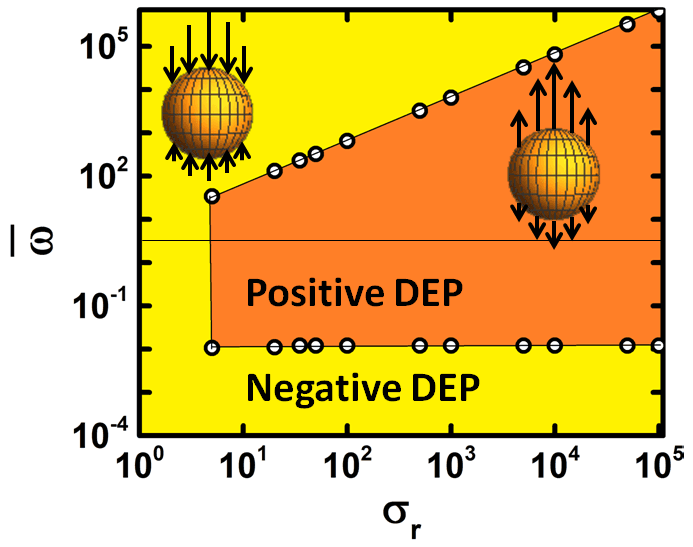}
  \caption{Phase diagram for vesicle DEP at $ \epsilon_r=\mu_r$=1, $\bar{C}_m$=125, $\bar{G}_m$=0 in mixed quadrupole AC electric field} 
    \label{PD}
  \end{figure}

  These results can be presented in a phase diagram (figure \ref{PD}) which shows a negative dielectrophoretic region at low frequencies  for all conductivity ratios (due to vesicle acting as a dielectric drop in a perfect conductor) and at all frequencies at low conductivity ratios (due to negative dielectrophoresis in the intermediate frequency range (since the polarization vector is opposite to the applied field). At moderate and high conductivity ratios, positive dielectrophoresis is seen in an intermediate range of frequencies, essentially in agreement with liquid drops of same conductivity ratio (having polarization vector in the direction of the applied field).
  The above discussion indicates that the dipole moment method with a rigid particle Stokes drag assumption should suffice to describe the dielectrophoretic motion of a vesicle in non-uniform electric field. The experimental results in \cite{victoria2009} are in qualitative agreement with the present model, although quantitative studies cannot be done due to the axisymmetric electrode configuration considered in this work. It should be mentioned here that experiments are typically conducted in pure quadrupole field. In such a field, no dielectrophoresis will be seen for a vesicle  kept exactly at the center of the electrode system. However, an off-center vesicle, say in the $z$ direction will experience a uniform field given by $E_o=-2 \Lambda_o z_o$ and a quadrupole potential of $\Lambda_o$, thereby enabling the use of the results derived in this work. 

\end{document}